\documentclass[12pt,preprint]{aastex}
\usepackage{psfig}
\def\HI {H\kern0.1em{\sc i}} 
\def\radm {rad m$^{-2}$} 

\def\dg{$^{\circ}$}

\begin{document}
\title{~~\\ ~~\\ A View through Faraday's fog:\\
Parsec scale rotation measures in AGN}
\shorttitle{Faraday's Fog}
\shortauthors{Zavala \& Taylor}
\author{R. T. Zavala\altaffilmark{1,2} \& G. B. Taylor\altaffilmark{1}}
\email{rzavala@nrao.edu, gtaylor@nrao.edu}
\altaffiltext{1}{National Radio Astronomy Observatory, P.O. Box 0, 
Socorro, NM 87801}
\altaffiltext{2}{Department of Astronomy, New Mexico State University, 
MSC 4500 P.O. Box 30001, Las Cruces, NM 88003-8001}


\slugcomment{Accepted to the Astrophysical Journal}

\begin{abstract}

Rotation measure observations of 9 quasars, 4 BL Lacertae objects, 
and 3 radio galaxies are presented. The rest frame rotation 
measures in the cores of the quasars and the jets of the radio 
galaxy M87 are several thousand \radm. The BL Lacertae objects 
and the jets of the quasars have rest frame rotation measures of 
a few hundred \radm. A nuclear rotation measure of 500 \radm\ in the rest
frame is suggested as the dividing line between quasar and BL 
Lacertae objects. The substantial rotation measures of the BL 
Lacertae objects and quasars 
cast doubt on the previous polarization position angle investigations 
of these objects at frequencies of 15 GHz or less. BL Lacertae 
itself has a rotation measure that varies in time, similar to the 
behavior observed for the quasars 3C\,273 and 3C\,279. 
A simple model with magnetic fields of 40 $\mu$G or less can account 
for the observed rotation measures.

\end{abstract}

\keywords{galaxies: active -- galaxies: ISM -- 
 galaxies: jets -- galaxies: nuclei -- radio continuum: galaxies}

\section{Introduction}

Polarization observations using Very Long Baseline Interferometry (VLBI) 
have been probing magnetic field orientations at high 
angular resolution for almost 20 years \citep{cot84}.
Details of magnetic vector orientation are required to understand 
the relativistic jets of AGN, but the presence
of a magnetized plasma in the centers of these objects complicates 
the analysis. Faraday rotation \citep{farad}, or the rotation of the plane of 
polarization of an electromagnetic wave as it traverses a magnetized 
medium, changes the observed polarization orientation from that 
intrinsic to a source. The intrinsic polarization angle 
$\chi_0$ is related to the observed polarization angle $\chi$ by 

\begin{equation}
\chi = \chi_{0} + \rm{RM}{\lambda}^{2}
\end{equation}

\noindent where $\lambda$ is the observed wavelength.  The 
linear relationship to $\lambda^2$ is the characteristic signature 
of Faraday rotation. The slope of the line is known as the      
Rotation Measure (RM) and depends linearly on the electron 
density $\rm{n_{e}}$, the net line of sight magnetic field 
$\rm{B_{\parallel}}$, and path length $\rm{dl}$ through the plasma. 
Using units of cm$^{-3}$, mG and parsecs the rotation measure 
is given by:

\begin{equation}
\rm{RM = 812\int{n_{e}B_{\parallel}\,dl}\quad rad\ m^{-2}}.
\end{equation}

Simultaneous, multi-frequency observations 
allow the rotation measure to be determined, and the recovery of the 
intrinsic polarization orientation. Faraday rotation towards an 
extragalactic source was first observed by \citet{cp62}, who found a
rotation measure of $-$60 \radm\ towards the center of Centaurus A. 
Without knowledge of the rotation measure errors in the intrinsic 
polarization angle can be substantial. Rotation measures of 250 \radm\ 
can change the intrinsic polarization angle by 25\dg\ at 8 GHz. 
Thus, parsec scale rotation measures are required if polarization 
observations are to yield the true magnetic vector
orientation in the relativistic jets of AGN. Faraday rotation
probes the density weighted magnetic field along the line of sight, 
so an added benefit is the ability to probe the physical conditions
in the Faraday screen itself. 

We present rotation measures for 16 quasars, radio galaxies, and 
BL Lacertae objects obtained with the Very Long Baseline Array
(VLBA)\footnote{The National Radio Astronomy Observatory is operated by 
Associated Universities, Inc., under cooperative agreement with the National 
Science Foundation.}. Observations and data reduction are described in 
\S2. In \S3 results for individual objects with maps of the rotation 
measure, magnetic vector orientation corrected for Faraday rotation, 
and spectral indices are shown. In \S4 we discuss our results, which 
are followed by our conclusions. We assume H$_0 = 50$ km s$^{-1}$ 
Mpc$^{-1}$, q$_0$=0.5, and spectral index definition $S_{\nu} 
\propto \nu^\alpha$ throughout.

\section{Observations and Data Reduction}
The observations, performed on 2000 June 27 (2000.40), were carried out at 
seven widely separated frequencies between 8.1 and 15.2 GHz using the 10 
element VLBA. This 24 hour observation targeted the sources listed in 
Table 1. Prior to self-calibration all processing was performed in the  
Astronomical Image Processing System \citep[AIPS;][]{van96}. 
Amplitude calibration using APCAL was performed using the recorded system 
temperatures and gain tables. The task FRING was used on two minutes of
data from 2200+420 to remove errors due to clock and correlator model 
inaccuracies. A global fringe fit was run on all the data to remove 
the remaining delay and rate errors. The R$-$L delay offset was removed
using the procedure CROSSPOL. A bandpass correction table was made
with BPASS using 3C\,279 as a bandpass calibrator. The data were then 
averaged in frequency across the bandwidths shown in Table 2. 

Self-calibration was done using DIFMAP \citep{shep,spt} and AIPS in 
combination. Polarization leakage of the antennas (D-terms) were 
determined using the AIPS task LPCAL. This calibration was performed 
twice, first using 
0133$+$476 with an accompanying CLEAN model for this polarized source. 
Plots of the real versus imaginary crosshand polarization data indicated 
that a satisfactory D-term solution was obtained. After this calibration 
was performed the radio galaxy 2021$+$614 was found to be unpolarized. 
The D-terms were solved for again using 2021$+$614 as an unpolarized 
calibrator as a consistency check. Both sources produced nearly 
identical corrections for the D-terms, and 2021$+$614 was used as the 
final D-term calibrator.
   
Absolute electric vector position angle (EVPA)
calibration was determined by using the EVPA's of 0851$+$202, 0923$+$392,
1308$+$326 and 2200$+$420 listed in the VLA Monitoring 
Program\footnote{http://www.aoc.nrao.edu/$\sim$smyers/calibration/} 
\citep{tmy00}. The EVPA calibration at 8 GHz was readily
obtained from the polarization calibration website. The  observations
were interpolated in time as necessary. Polarization monitoring 
observations at 8 and 22 GHz were interpolated to produce position 
angles at 12 and 15 GHz. Figure 1 of \citet{zt02} shows calibrated 
position angle results for these four calibrators with the associated 
VLA data from the website. As a further test of our EVPA calibration we 
examined the rotation measure of component C4 of 3C\,279 after the EVPA 
calibration was completed. In \citet{zt01} C4 was shown to have a relatively 
low RM and polarization angle $\chi = -84^{\circ}$ at 8 GHz to 
$-87^{\circ}$ at 15 GHz on 2000 
January 27. After the EVPA calibration for this experiment was completed 
we measured C4 and the EVPA was $-$87\dg\ for all frequencies. Although 
C4 does vary its position angle and RM with time the agreement between 
the observations of 2000 January and 2000 July suggests that the 
polarization calibration was performed correctly.  

To perform the rotation measure analysis data cubes in $\lambda^2$ were 
constructed. This provides adequate short and long spacings in 
$\lambda^2$ to properly recover RMs between $\pm$ 30000 \radm. The 
12 and 15 GHz images used to produce the polarization angle maps were 
tapered to approximate the 8 GHz resolution, and a restoring beam matched 
to the 8 GHz beam was used. 

\section{Results}
Results for individual sources are presented. There are usually 
three images for most sources. 
These images display the rotation measure, E vectors corrected for
Faraday rotation, and spectral indices for the sources. Rotation 
measure results for the radio galaxies 3C\,111, 3C\,120 and M87 were 
presented in \citet{zt02}, thus only spectral index and E vector maps 
are shown in this paper. Vector lengths shown are proportional 
to the polarized flux density, and the scale is given for each 
figure. These plots were made with the AIPS task COMB with PCUT 
set to 5$\sigma$ derived from the Stokes Q and U maps, and ICUT 
set to 3$\sigma$ in total intensity. No polarization data for the 
radio galaxy 2021+614 are provided as this source was unpolarized. 

The rotation measure maps consist of a color plot of the observed RM 
results overlaid on 15 GHz contours. The 15 GHz contours begin at the 
3$\sigma$ level in the Stokes I map
and increase by factors of two. Insets of the plots of polarization 
angle $\chi$ versus $\lambda^2$ are shown, with the linear fits to 
the data provided. Values of the RM which result from the fits are 
placed in the inset plots in units of \radm. The error bars on the 
$\chi$ data points are determined by adding in quadrature the noise
from the Stokes Q and U maps with the errors determined from the 
EVPA calibration technique previously reported in \citet{zt01}. 
These are $\pm$ 2\dg\ at 8 GHz and $\pm$ 4\dg\ at 12 and 15 GHz.

Spectral index maps were made with the AIPS task COMB, with pixels less
than 5$\sigma$ in the Stokes I images blanked. The spectral index 
plots were made for the $\alpha^{12.1}_{8.1}$ interval.

\subsection{B0133+476}

This quasar exhibits a relatively high RM of $-1420 \pm 56$ \radm\ 
in the core. 
The percent polarization increases at the location of the fit 
from 1.3\% at 15 GHz to 1.9\% at 8.1 GHz. The extension NW of 
the core with a positive rotation measure of $40 \pm 90$ \radm\ 
is consistent with an RM of zero. There is 
therefore no convincing evidence for a change of sign of the B field 
in this region. This region does coincide with an optically thin spectrum 
from 8$-$12 GHz, as indicated in Fig.~\ref{0133si}. The core has 
a flat spectrum across these frequencies.

\subsection{B0212+735}

The quasar 0212+735 shows an RM of $-542 \pm 55$ \radm\ at the center, 
and $+119 \pm 64$ \radm\ in the jet (Fig.~\ref{0212rm}).  The center fit 
exhibits a flat spectrum, with $\alpha \approx -0.15$. Percent 
polarization at the peak in the I map is approximately 6\% from 12$-$15 
GHz and decreases to 3\% at 8 GHz. The optically thin jet maintains a 
percent polarization $\sim$ 13\%. The thin (width about 1 pixel) region 
with an indicated RM of $\approx$ 1000 \radm\ separates two areas where the 
slope of the rotation measure changes sign. The polarization vectors 
also change direction across this boundary. The fits to a $\lambda^2$ 
law show different slopes across 8$-$12 GHz as compared to 12$-$15 GHz. 
The RMs of order 1000 are then not physically real, but result from the 
superposition of two regions of differing polarization angles. 
Unfortunately, it is not possible to reject these fits easily in preparing 
the RM maps.
        
\subsection{3C\,111}

The rotation measure image for this radio galaxy was published in 
\citet{zt02}. No reliable rotation measures could be determined for 
the core due to a weak to non-existent polarization at 12 and 8 GHz. 
The jet had a rotation measure of $-750\pm 61$ \radm\ 3 mas east 
of the core, which decreased to $-200 \pm 56$ \radm\ 5 mas east of 
the core. A positive spectral index coincides with the depolarized 
core of this radio galaxy. The E fieldvectors are well ordered 
along the jet. 

\subsection{B0420-014}

No reliable fits to a $\lambda^2$ law were obtained for this quasar. 
The plot of polarization angle versus $\lambda^2$ (inset to 
Fig.~\ref{0420rmsi}a) is representative of the results obtained
for 0420$-$014. Two different RM slopes are implied; a steep
positive slope from 15-12 GHz, and a shallower, negative slope
across the 8 GHz position angle data. The spectral index plot 
(Fig.~\ref{0420rmsi}b) shows that a flat spectrum  
characterizes most emission from this quasar across 8$-$12 GHz. 
Percent polarizations $m$ integrated over the source are 
0.9\%, 0.1\% and 0.4\% at 8.1, 12.5 and 15 GHz, respectively. 
We set an upper limit to $m$ of 0.2\% at 12.1 GHz. The upper limit 
for 12.1 GHz is due to noisier polarization data at this frequency. 
Rotation measure fits using all but the 12.1 GHz data create inconsistent 
slopes of position angle versus $\lambda^2$. The 8 GHz points are 
consistent with a rotation measure of zero, while the two points 
at 12.5 and 15.1 GHz suggest a much higher rotation measure.   

\subsection{3C\,120}

As for 3C\,111, the RM maps for this radio galaxy were presented 
in \citet{zt02}. The core rotation measure was $2080 \pm 100$ \radm. 
This decreased to approximately $100 \pm 60$ \radm\ after a projected 
distance of 1 parsec, which is similar to what is observed 
in quasars. The jet, where the percent polarization is $\sim$ 
18\%, was shown to have good agreement to a $\lambda^2$ law.
The spectral index in the jet is  
$\alpha^{12.1}_{8.1} = -0.5$ or less. Electric field vectors 
are perpendicular to the jet axis. The E vectors in the core 
are coincident with rotation measures that may show a deviation 
from a $\lambda^2$ law \citep[Fig.~4]{zt02}.   

\subsection{B0528+134}

This quasar shows a decrease in RM as one proceeds along the jet 
from the core (Fig.~\ref{0528rm}). The inset plots in Fig.~\ref{0528rm} 
have RMs of $325 \pm 58$ and $64 \pm 59$ \radm. 
The percent polarization in the 
core is 1\% at 15 GHz, decreasing to 0.5\% at 8 GHz. The spectral index 
$\alpha^{12.1}_{8.1}$ at the location of the core RM fit has turned 
over and is slightly positive (Fig.~\ref{0528si}. 
A small steeper spectrum region separates the flat spectrum core and
jet. The jet fractional polarization varies from 2\% to 3.5\% from 15 GHz to 
8 GHz respectively. 

\subsection{B0923+392}

Both RM fits displayed for this quasar are consistent 
with an RM of zero. Both inset RM fits in Fig.~\ref{0923rm}a have 
errors in the rotation measure of $\pm 60$ \radm.
A flat 8-12 GHz spectrum characterizes most 
of the emission from this quasar, as shown in Fig~\ref{0923si}. 

\subsection{M87}

The rotation measure map for this radio galaxy was presented in
\citet{zt02}. The rotation measures are confined to a region 
from 18 to 27 mas west of the core. The rotation measure
varied from $\approx$ $-$5000 \radm\ to almost 10,000 \radm. 
The errors in the rotation measures are approximately $\pm 150$ 
\radm.
A sign change in the slope of the RM is seen over a projected 
distance of 0.3 pc. The electric field vectors in the area 20 mas west 
the core are perpendicular to the jet boundary. The cutoff used 
to create the spectral index maps (\S3) prevents the spectral index 
from being mapped in this area. 

\subsection{3C\,279}

Both the flat spectrum core and steeper spectrum jet (see 
Fig.~\ref{3c279si})  of this quasar show good agreement to a 
$\lambda^2$ law. The low rotation measure of the core 
($-97 \pm 56$ \radm) is similar to that seen in \citet{tay00} 
and \citet{zt01}. \citet{tay98} found a much higher core 
rotation measure of $-$1280 \radm.  

\subsection{B1308+326}

The low RM of this BL Lac is contrasted with a turn 
in the E fieldof almost 35\dg\ from core to the edge of the polarized 
emission west of the core. The fits give rotation measures of 
$113 \pm 55$ and $-28 \pm 57$ \radm. 
Percent polarization in the core is 
$\sim$ 3\% while in the jet it increases to almost 6\%. The core has 
a flat spectrum. Electric field vectors in the core are perpendicular 
to the jet direction, and turn almost parallel to the jet at 2
mas west of the core. There is an unpolarized jet component 
9 mas west of the core which is not shown.      

\subsection{B1611+343}

As seems typical for quasars the RM in the core and jet both seem 
to be in good agreement with the $\lambda^2$ law. Similar to 
the general results of \citet{tay98,tay00} the rotation measure is highest 
in the core ($-518 \pm 55$ \radm), and decreases to $-45 \pm 55$ \radm\ 
within 10 parsecs of the core. This follows the change in 
spectral index from a flat spectrum core to a steeper spectrum jet. 
The electric field vectors are essentially parallel to the jet 
throughout this quasar. 

\subsection{1803+784}

This BL Lac object is optically thick at 8 GHz at the core, and 
has an fractional polarization of 5\%, while maintaining 
agreement to the $\lambda^2$ law. The jet has a flat RM (14 
$\pm$ 73 \radm), is optically thin, and has approximately 
20$-$30\% polarization. In a similar fashion 
as the quasars the core shows a higher rotation measure ($-201 
\pm 55$ \radm) as compared to the jet.  

\subsection{1823+568}

Breaking with the trend seen for other objects reported herein, 
the jet 5 mas south of the core of this BL Lac object displays 
an RM ($-200 \pm 88$ \radm) comparable to that in the core 
($-128 \pm 55$ \radm). These two regions are 
separated by a nearly flat RM 2 mas south of the core. The B 
vectors remain perpendicular to the jet axis, and the core and 
jet have the usual flat and steeper spectra respectively. 

\subsection{2005+403}

At first glance this quasar seems to exhibit a sign change in the RM, 
but the quality of the core RM fit is poor. The core has an 
RM of $668 \pm 58$ \radm, and the jet $-200 \pm 57$ \radm. 
The 8 GHz core polarization angles suggest a flat RM, while the 12
and 15 GHz data suggest the steeper, positive RM of 668 \radm. The 
core has an inverted spectrum from 8 to 15 GHz. The location of 
the RM fit 1.5 mas east of the core has an inverted spectrum 
from 8$-$12 GHz, but is optically thin at 15 GHz.  

\subsection{B2200+420 (BL Lac)}

This prototype of the BL Lac object class \citep{BL} has previously been 
observed to have a core rotation measure of $-$550 \radm\ and a 
jet rotation measure of $-$110 \radm\  with good agreement to 
a $\lambda^2$ law \citep{rey}. However, Fig.~\ref{blrm} clearly shows 
a breakdown in the $\lambda^2$ law for the core. The rotation measure in 
the jet has increased to $-287 \pm 56$ \radm\ as compared to the 
\citet{rey} result. Our data show the jet has a rotation measure 
comparable to the jets in quasars, and has a higher polarization 
than the core (m$_{core}$ = 3\%, m$_{jet}$ = 18\%). 

\subsection{2251+158}
This bright quasar exhibits an interesting structure in both the 
RM and B vector image. As the inset plots of Fig.~\ref{2251rm} show, 
agreement to a $\lambda^2$ law is only evident at the peak in the I image. 
The RM at this point is $-263 \pm 56$ \radm.
Percent polarizations for the core range from 1\% at 8 GHz to 0.7\% 
at 15 GHz. Rotation measures are not given in the remaining insets 
due to the poor fits to a Faraday rotation law. The reliability of the 
indicated E fielddirections in Fig.~\ref{2251rm}b are suspect due 
to the unreliable RM fits. The spectral index is flat 
within 3 mas of the core (Fig.~\ref{2251si}), and becomes steep 
beyond that point.

\section{Discussion} 

Some general rotation measure properties of the sample 
will be discussed, and then properties of separate classes 
of AGN. Data used to prepare the figures for this 
section are listed in Table 3.  

\subsection{Rotation measure vs. core dominance}

In an effort to examine the cause of the observed rotation 
measures a plot of the absolute value of the maximum RM versus 
15 GHz core dominance (R$_{c}$) was created (Fig.~\ref{rcrm}). 
The core dominance is defined as the ratio of the peak flux density 
in the map divided by the sum of the CLEAN components in the map. The 
rotation measure has been corrected for redshift by a factor of 
(1 + z)$^2$. Quasars are plotted as dots, radio galaxies as 
triangles, and BL Lac objects as X's. Multi-epoch 
rotation measure data for the quasar 
3C\,279 are plotted as diamonds 
connected by a dashed line. The plotted rotation measure is 
for the core, except for the radio galaxies 3C\,111 and M87. 
For these two radio galaxies we plot the maximum rotation measure 
observed in the jet as depolarization in these galaxies 
prevented the determination of a rotation measure in their
cores.   

At first glance Fig.~\ref{rcrm} shows that on parsec scales
AGN rotation measures are independent of core dominance. 
Closer examination shows that the BL Lacertae objects have a 
systematically smaller maximum RM as compared to quasars and 
radio galaxies. This suggests that a moderate ($\approx$ 500 \radm\ 
or less) rotation measure in the rest frame could serve as 
a dividing line between quasars and BL Lac objects. This 
division may not be robust, as clearly 3C\,279 has crossed below
this threshold. The small number of BL Lac objects in 
Fig.~\ref{rcrm} represent snapshots in time, and future 
monitoring will be needed to confirm that BL Lac objects 
are restricted to relatively low rotation measures. 
\citet{ver95} have shown that BL Lac itself can have optical 
emission lines with equivalent widths greater than 5 angstroms, 
and thus not meet the criteria for classification as a BL Lac
object as set by \citet{sfk93}. Therefore, it is not inconceivable 
that the RM variation in the core of a BL Lac could yield a rotation 
measure more consistent with a quasar core. As the jets 
of BL Lacs have rotation measures consistent with those seen for quasar 
jets similar Faraday depths are present on parsec scales for these
two optical AGN classes. 
 
The path of 3C\,279 through Fig.~\ref{rcrm} shows that at least 
one quasar can move vertically and horizontally through a core 
dominance$-$RM space. Future monitoring may show that all quasars 
have a variable RM independent of core dominance. The cores can also 
brighten or fade, so some horizontal movement in Fig.~\ref{rcrm} 
occurs. 3C279 maintains a constant magnetic field orientation across the three
year span of RM monitoring of 145$-$150\dg, with the exception 
of epoch 1998.59 when the magnetic vector orientation was 120\dg.
The magnetic field orientation was determined from the RM corrected intrinsic 
E field vector rotated by 90\dg.

Small number statistics make comparisons between the object 
classes difficult. However, the systematic offset between the 
BL Lac's and quasars in Fig.~\ref{rcrm} may make a comparison 
worthwhile. The quasars in Fig.~\ref{rcrm} have an average 
(absolute value) rotation measure of 2400 \radm, and a median 
of 1500 \radm. The BL Lacs have an average of 450 \radm, and a median 
of 440 \radm. These data suggest that BL Lac cores have a 
low, but not insignificant, rotation measure as compared to 
quasar cores. Such a rotation measure property separating
BL Lacs from quasar cores must be tested over time and for a 
larger sample before it is to be considered a defining characteristic. 
  
As shown in \citet{zt02} the radio galaxies M87, 3C\,120  
and 3C\,111 have relatively high rotation measures; equal to or
greater than the rest frame RMs in the quasar cores presented here. 
A similar result was found for the radio galaxy 3C\,166 which has a 
rest frame RM of $-$2300 \radm\ for its inner jet and 
has a depolarized core \citep{thv}.
The highest parsec scale rotation measure in the rest frame 
(20$-$40,000 \radm) still belongs to the quasar OQ 172 \citep{udom}, 
so rotation measures of order 10,000 \radm\ do not belong exclusively 
to the radio galaxies. The depolarization of radio galaxy cores in 
\citet{zt02} may be due to Faraday depolarization across an 
observing band as the nuclear torus surrounding the black hole 
may provide the requisite Faraday depth to account for the 
observed depolarization \citep{rjf}.

\subsection{Rotation measure vs. percent polarization}

A high core rotation measure (RM greater than 500 \radm) is clearly 
anti-correlated with core percent polarization (Fig.~\ref{mcrm}). 
The symbols in Fig.~\ref{mcrm} are the same as for Fig.~\ref{rcrm}. 
This depolarization for high rotation measures is not due to Faraday 
depolarization across the observing bandwidth. For Faraday rotation 
to depolarize the signal across the 32 MHz bandwidth 
at 15 GHz requires an RM of more than 600,000 \radm. This is 
clearly inconsistent with the observed RMs of order 1000 \radm.

A possible explanation for the depolarization is a gradient
in the Faraday screen across the source, or beamwidth  
depolarization \citep{gw66}. 
A useful rule of thumb is that a turn of one
radian in polarization angle across the beam is sufficient
to cause substantial depolarization. The required RM gradient
is then

\begin{equation}
\nabla \rm{RM} = \frac{\Delta \chi}{\lambda^2~\rm{Beam}}
\end{equation}

\noindent where the RM gradient $\nabla$RM equals the change in 
polarization angle at the given wavelength squared across a beam. 
At 15.1 and 8.1 GHz gradients of 2540 and 770 \radm\ beam$^{-1}$, 
respectively, will depolarize a source across a beamwidth. Invoking 
an RM gradient to depolarize the 15 GHz signal provides a gradient 
more than three times what is needed to depolarize the 8 GHz signal. 
Obviously, a gradient can depolarize a source at 8 GHz, and 
have no appreciable effect at a higher frequency. The changes 
in the observed rotation measure are attributed to a foreground
screen sampled on a changing sightline as components move along
the jets in the AGN \citep{tay00}. Once a component has moved 
beyond a projected distance of $\sim$ 10 parsecs from the core 
the screen changes to a lower and more uniform Faraday depth. 

The multi-epoch data for 3C\,279 can also be used to look for any 
correlation of percent polarization with rotation measure. 
This is shown in Fig.~\ref{279time}, where the rotation
measure in the core and percent polarization at 8 and 15 GHz are 
plotted versus observation epoch for 3C\,279. The initial high 
rotation measure at epoch 1997.07 occurs with a relatively low percent 
polarization at both frequencies. As the magnitude of the RM 
decreases the percent polarization rises. Interestingly, the 8 
GHz percent polarization lags the rise in the 15 GHz percent 
polarization at epoch 1998.59. This is consistent with a 
Faraday screen with a gradient that decreases in magnitude
as components move along the jet. 

Tracking the motion of model components in the region of high 
Faraday depth will set an upper limit to the scale size of the 
foreground Faraday screen. For example, epochs 1997.07 and
1998.59 could set such a limit if the motion of a model 
component near the core could be accurately measured and its 
motion tracked. Unfortunately, while we can measure the motion 
of component C4 (the component 3.5 mas west of the core) of 
3C\,279 across the four epochs, we were unable to unambiguously 
identify components within approximately a milliarcsecond of the 
core. These may be stationary components 
1.2 mas or less from the core of 3C\,279 as reported by \citet{weh01}.
Another factor is that these observations were not designed 
with the (u,v) coverage required to properly conduct a motion study. 

No correlation was found between the rest frame rotation measures 
and spectral index of the cores of the objects presented here.

\subsection{Implications for BL Lac objects}

The unified scheme for AGN describes BL Lacertae objects as strongly 
Doppler boosted sources whose relativistic jet is viewed close to the 
observer's line of sight \citep{zen}. In such a scenario the relativistic jet 
can sweep away magnetized thermal gas which is probably intrinsic 
to the central parsecs of an AGN. Therefore, the foreground 
Faraday screen would be removed. An observer should then see 
little or no Faraday rotation towards a BL Lac object on parsec scales. 

Surprisingly, \citet{rey} found a relatively high quasar-like value 
for the core RM of BL Lac. The BL Lac objects 
presented here have core rest frame rotation measures of 500 
\radm\ or less. While BL Lac itself has a high core RM, the agreement to a 
$\lambda^2$ law is not good, in contrast to the fits 
obtained by \citet{rey}. Failure to follow the Faraday rotation law 
could  indicate an optical depth effect. BL Lac has a flat spectrum 
($\alpha^{12.1}_{8.1}$=0.33) core and the combination of several 
components of varying, but physically unrelated, electric vector 
position angles can explain the lack of agreement to a $\lambda^2$ law. 
The blending of components of unrelated polarization angles is also
hinted at by the core percent polarization of 2.8\%. \citet{denn} 
reported that core polarizations of BL Lac in excess of 1\% 
indicated the emergence of a new polarized component from the core.
Full resolution 15 GHz maps with a 1.1 by 0.6 mas beam of BL Lac do not
resolve the hypothetical component, but an extension to the south 
of the core may be an emerging component.

The jets of the BL Lac 1823+568 and BL Lac itself have rotation 
measures comparable to those found for quasar jets. Compared to 
\citet{rey} the RM of the jet of BL Lac has increased by 170 \radm, 
and this time-variability argues for a Faraday screen intrinsic 
to the central regions of this object. 

The non$-$negligible jet rotation measures for BL Lacertae
objects has implications for interpretations of intrinsic
magnetic field orientations in these objects. \citet{gab00} 
analyzed magnetic field orientations in the jets of BL Lac objects 
at 5 GHz and found that the majority of jets have
magnetic vectors transverse to the jet axis. A correction 
using an integrated rotation measure of approximately 50 \radm\ 
or less was applied for most objects in their analysis. 
At 5 GHz the parsec scale rotation measure of 250 \radm\ changes 
the vector orientation by 52\dg. Therefore, any attempt to 
make conclusions on magnetic or electric vector orientations
in BL Lacertae objects at frequencies 
less than 15 GHz without parsec scale rotation 
measures must be suspect. 

Our rotation measure results allow for the determination of the intrinsic
EVPA orientation in the BL Lacertae objects presented here. To determine
the degree of alignment of the electric vector with the jet direction 
we used the MODELFIT feature in DIFMAP to represent the BL Lac objects 
with Gaussian components, and derived a jet orientation axis $\theta$. 
This angle $\theta$, along with the difference $|\theta - \chi|$ for the 
core and brightest jet components, are listed in Table 4. There are no intrinsic
electric vector orientations perpendicular to the jets or cores for the four 
BL Lacs. All the BL Lac objects presented here are optically thin at 15 GHz 
in both their cores and jets. The intrinsic magnetic field 
orientation is therefore determined by rotating the intrinsic electric vector 
by 90 \dg. None of the four BL Lac objects presented here shows an 
intrinsic magnetic field orientation parallel to the jet. With 
the obvious problems of small number statistics two out of four 
of the cores and jets of the BL Lacs have intrinsic magnetic vectors 
oriented perpendicular to the jet axes. The flat/inverted spectrum 
cores of the BL Lac objects do make interpretations of the magnetic
vector orientations problematic at frequencies of 8 GHz or less. 
The averaging of components within our beam also complicates 
the analysis. Direct comparison to the extensive compilation of \citet{gab00}
is difficult as 10 years or more separate our observations. During this 
time much can change on the small scales probed by VLBI. 

\subsection{Depolarization in 0420$-$014}

As discussed in \citet{tay00} quasar cores are expected to be viewed 
through a line of sight likely to produce high (several 1000 \radm)
rotation measures. The quasar 0420$-$014 is weakly
polarized (Fig.~\ref{0420rmsi}), and no reliable fits to a $\lambda^2$ law 
were obtained. 0420$-$014 is a high optical polarization quasar \citep{hpq}
and is one of the EGRET detected blazars \citep{egret}. \citet{jor01} show that 
this quasar has structure on sub-milliarcsecond scales which includes
stationary and moving components. \citet{mar02} showed 0420$-$014 
that has two polarization components in their 22 and 43 GHz maps. Their 
1997.58 data show the core is 0.5\% polarized at 22 GHz, and 2.0\% 
polarized at 43 GHz. 

A relatively simple mechanism for our observed low polarization for 
0420$-$014 is a superposition of components with differing polarization 
angles as described by \citet{pg}. They describe how spectral components of 
differing Stokes parameters combine to produce a flat-spectrum, depolarized 
image. This can result in several extrema appearing in a plot of
percent polarization versus frequency. This is similar to what is 
presented here for 0420$-$014, which has nearly equal percent 
polarizations at 8 and 15 GHz, but a minimum at 12 GHz. Untapered, 
full-resolution 15 GHz data (Fig.~\ref{2pol}) clearly resolves 
two polarized components similar to those seen by \citet{mar02} 
. The peak percent polarization in this image is 1.3\%.   

\subsection{Magnetic fields and the Faraday screen} 

To determine a magnetic field strength from a rotation measure
two free parameters must be specified: the electron density and 
the path length through the Faraday rotating medium. This requires
knowledge of the location and nature of the foreground Faraday screen. 

\citet{asada} suggest the jet of 3C\,273 itself as a possible 
Faraday rotating medium. They report a rotation measure gradient 
across the jet of 3C\,273 and attribute this to a helical magnetic 
field which wraps around the jet. This explains the change in sign of 
the rotation measure across the jet as the magnetic field points towards 
or away from the observer due to the twisting of the helix.   
There are only two beamwidths across the region 
where the RM gradient is determined, so it is difficult to conclusively
resolve such a feature. The quasar 1611+343 (\S3.11) is the only AGN 
presented here where there is just sufficient resolution across the 
jet to search for such a rotation measure gradient. No rotation 
measure gradient is apparent transverse to the jet axis in 
Fig.~\ref{1611rm}. If the helical field does serve as a Faraday screen
then this requires thermal material to be mixed with the relativistic
jet itself. If the helical field is also a source of synchrotron 
radiation then internal Faraday depolarization should also be apparent
\citep{burn}. Faraday rotation measures across a jet of sufficient 
width to resolve any gradient in rotation measure are therefore required.  

The line of sight integrated magnetic field topology can be investigated 
using the change of slope of the rotation measure plots. A change in sign 
like that observed for M87 \citep{zt02} requires the line of sight magnetic 
field to change by 180\dg. A similar sign change was observed for the core
of 3C\,273 \citep{zt01}. The net magnetic field in the Faraday rotating 
medium must change across the plane of the sky on projected distances
of a few tenths of a parsec to around 5 parsecs. 

We now put forth a simple model to explain the observed rotation measures.
Upper limits to electron densities 
are set by the requirement that the optical depth for free-free 
absorption be less than unity. We assume the path length through the Faraday 
screen is the same as that seen for the plane of the sky. A temperature 
of 10$^4$ K, suitable for the narrow line region of an AGN, is assumed. 
Then the electron density (cm$^{-3}$) required to give an optical depth 
of one as a function of path length L (cm) is:

\begin{equation}  
{\rm n_{e} = \frac{1}{\sqrt{2.7\times10^{-28}L}}}. 
\end{equation}

\noindent This sets an upper of a few times 10$^4$ cm$^{-3}$ to maintain 
an optical depth less than one, with the Gaunt factor set to unity. 
This is in line with the densities calculated for narrow line regions 
in AGN \citep{ost,ss} of 10$^3$ cm$^{-3}$. Setting the electron density 
to 1000 cm$^{-3}$ is then a reasonable estimate. 

Rotation measures vary from 1000s of \radm\ in quasar cores and the jet 
of M87, to a few hundred \radm\ for quasar jets and the BL Lac objects. 
These screens may be very different entities, and can vary their 
electron densities as the level of the ionizing continuum varies. 
With these caveats in mind let the depth of the Faraday screen 
be approximately a parsec, and consider RMs in the observer's 
frame of 2000 \radm\ and 200 \radm\ for the two extremes. Fields 
of 3 $\mu$G and 0.3 $\mu$G, respectively, are needed to account
for these rotation measures. Even for M87 fields of less than 
40 $\mu$G \citep{zt02} can cause the observed RMs under these 
assumptions. The fields are much less than required for equipartition 
with the thermal gas surrounding the jet, and cannot confine 
the jet. 

Other than observational conjecture, is there any theoretical 
basis for a Faraday screen with a depth of a few parsecs?
\citet{hr02} assign the role of the Faraday screen to the jet 
of 3C\,345, albeit on kiloparsec scales. A layer several jet 
radii thick, which preserves the helical orientation in the jet, 
is found to exist over these distances in their simulations. 
Such a layer would be a few parsecs thick if it existed on 
milliarcsecond scales. This would require a high-amplitude 
helical twisting, which does not seem to be present on parsec 
scales and would in any event destabilize the jet (Hardee, 
private communication). A mixing layer caused by 
Kelvin-Helmholtz instabilities \citep{ddy} has been suggested for the 
screen put forth in \citet{zt02} for the rotation measures in 
M87 (de Young, private communication). This mixing layer can 
account for the observed change in the RM slope by the turbulence 
in such a layer. 

The simulations shown in \citet{bss} and \citet{bick}, which 
attempt to explain Compact Symmetric Objects (CSO) may also 
prove useful. The CSOs are considered frustrated radio 
galaxies whose jets are disrupted by collisions with dense 
clouds in their cores. These clouds become ionized as a result 
of the shocks from the collision, and they can serve as a Faraday 
screen. The densities in the clouds need to be at least 1000 
cm$^{-3}$ in order to be stable against tidal forces from a 
supermassive black hole \citep{bss}, which agrees with the 
simple estimates above. These shocked clouds should exhibit 
free-free absorption at lower frequencies, which should be 
examined in future observations \citep{bss}. A screen of 
shocked clouds may not be present in every case, but M87 and 
3C\,120 are two examples where the model appears useful.

Attributing the observed rotation measures in M87 as arising 
in a shocked cloud gives special meaning to the intrinsic 
magnetic field vectors. Such a collision is not very disruptive 
and is likely to be oblique. The B field vectors would then be
more or less aligned with the jet flow (Bicknell, private 
communication). Rotating the E vectors in Fig.~\ref{M87} by 90\dg\ 
does produce vectors parallel to the local jet flow. An explanation 
is still required for the lack of polarized emission in M87 
outside of the area of the RM data. Invoking a high Faraday 
depth towards the core of a radio galaxy \citep{rjf} accounts 
for the observed RM without invoking the shocked cloud scenario.
A heterogeneous Faraday screen of several 10$^5$ \radm\ would
need a region of a lower Faraday depth to allow the rotation 
measure to be observed. This should require significant free$-$free
absorption against the jet where the Faraday depth is high enough 
to depolarize the jet.  

As components are ejected along a jet they should exhibit similar
Faraday rotation effects as they pass behind such shocked clouds. 
The region 2 mas west of the core of 3C\,120 \citep{jlg} may 
be such a cloud. Components exhibit variations in their light 
curve and polarization properties when they pass this feature 
\citep{jlg}. Polarization monitoring of 3C\,120 could 
conclusively identify this feature as a persistent source 
of Faraday rotation via a long-lived, dense cloud. G{\'o}mez et al. 
did report a high rotation measure towards this component, but it was 
unfortunately based on only two frequencies. 

Finally, we consider the hot gas which confines the narrow line region 
clouds \citep{elvis} as a potential Faraday screen. This turbulent 
hot gas will cause the radiation traversing the region to experience a 
random walk magnetic field 
orientation This will cause the RM derived magnetic field 
strength to underestimate the true field by a factor of $\sqrt{3\rm{N}}$ 
where N is the number of cells \citep{odea}. There are two predictions 
which can be used to test this scenario. First, the rotation measure distribution 
should have a Gaussian distribution with zero mean and a width given by 
a standard deviation $\sigma$. This is expected for any random walk process. 
Second, should a counter jet be detected in polarized flux it should 
have a higher rotation measure than the jet oriented towards the observer. 
The higher RM results from the longer path length through the Faraday 
screen which the counter jet is seen through. Unfortunately, we still 
await the detection of a polarized counter jet on parsec scales. 

The RM distribution can be measured and examined for a sign 
of the random walk Gaussian signature. Most of our sources have 
their rotation measure extent covered by just a few beams, so 
it is difficult to place conclusive results on the histograms. The  
quasar 1611+343 does have a broad jet, and provides the best sampling
of the RM maps shown here. The histogram is shown in Fig.~\ref{1611dist}, 
and is distinctly non-Gaussian. 

Rotation measure changes in 3C 279 over 6 months to 1.5 years also argue 
against the intercloud region as a Faraday screen. It is difficult 
to reconcile changes on such short timescales for a medium 100's of 
parsecs across. 

Although we invoke a mixing layer/thin skin Faraday screen for the source
of the observed rotation measures, a turbulent medium may play a role in 
some cases. M87 could conceivably have its rotation measure distribution 
derived from such a screen. This is especially true if we consider the 
observed polarization in the jet to be seen through a window in an otherwise
very depolarizing screen \citep{rjf}.
 
\section{Conclusions}

Parsec-scale rotation measures are presented for 9 quasars, 
4 BL Lacertae objects, and 3 radio galaxies. Rotation measures in the 
cores of these AGN are independent of core dominance. The four 
BL Lac objects have a median core rotation measure of 440 \radm, 
systematically less than the median rotation measure of 1500 \radm\ 
for quasar cores. A core rotation measure could be determined for 
one radio galaxy only due to depolarization at 8 and 12 GHz
for these galaxies. An anti-correlation of core rotation measure with 
percent polarization is consistent with depolarization by a gradient 
in rotation measure in AGN cores. This is also in agreement with the 
observation that quasar rotation measures decrease rapidly beyond a 
projected distance of 10 parsecs from the core. A decreasing rotation 
measure gradient proceeding from the core along the jet also explains 
the time variation on rotation measure and percent polarization for the 
quasar 3C\,279.

BL Lacertae is shown to have a variable rotation measure in its core 
and jet, consistent with a Faraday screen intrinsic to this AGN. 
The rotation measures in the jets of BL Lacertae and the BL Lac 
object 1823+568 result in substantial turns in polarization angles 
at frequencies less than 15 GHz. Therefore, results of magnetic 
vector orientations in the jets of BL Lac objects determined at these 
frequencies without VLBI-scale rotation measures are called into 
question. Rotation measures in the jets of BL Lac objects are 
comparable to those seen in the jets of quasars. This further blurs 
the distinction between BL Lacertae objects and quasars. 

Depolarization in the quasar 0420$-$014 is attributed to polarized 
components which just begin to be resolved at 15 GHz. The combination
of different polarization angles in these components results in the 
observed depolarization and lack of agreement to a $\lambda^2$ law for
this quasar. 

Magnetic fields of 40 $\mu$G or less can account for the observed 
rotation measures for path lengths through the Faraday rotating 
medium of 1 parsec and electron densities $\approx 10^3$ 
cm$^{-3}$. The line of sight magnetic field reverses
across projected distances on the sky of a few parsecs or less. 

\acknowledgments
We thank Kathleen LeFebre, NRAO Socorro Librarian, for assistance
in locating the reference to Faraday's discovery of Faraday rotation.
R.T.Z gratefully acknowledges support from a pre-doctoral research 
fellowship from NRAO and from the New Mexico Alliance for Graduate 
Education and the Professiorate through NSF grant HRD-0086701. 
This research has made use of the NASA/IPAC Extragalactic 
Database (NED) which is operated by the Jet Propulsion Laboratory, 
Caltech, under contract with NASA, and NASA's Astrophysics Data System 
Abstract Service.



\begin{deluxetable}{lccccr}
\tabletypesize{\scriptsize}
\tablecaption{T{\sc arget} S{\sc ources} \label{tbl-1}}
\tablewidth{0pt}
\tablehead{
\colhead{Source} & \colhead{Name}   & \colhead{Identification}   &
\colhead{Magnitude\tablenotemark{a}} &
\colhead{z}  & \colhead{$S_{15}$} \\
\colhead{(1)} & \colhead{(2)} & \colhead{(3)} & \colhead{(4)} & 
\colhead{(5)} & \colhead{(6)} }

\startdata
0133+476   &DA 55   &Q &18.0 &0.86 &2.22 \\
0212+735   &        &Q &19.0 &2.37 &2.69 \\
0415+379   &3C\,111  &G &18.0 &0.05 &5.98 \\
0420$-$014 &        &Q &17.8 &0.92 &4.20 \\
0430+052   &3C\,120 &G  &14.2 &0.03 &3.01 \\
0528+134   &        &Q  &20.0 &2.06 &7.95 \\
0923+392   &4C\,39.25 &Q &17.9 &0.70 &10.84 \\
1228+126 &M87       &G &9.6 &0.00 &2.40 \\
1253$-$055 &3C\,279  &Q &17.8 &0.54 &21.56 \\
1308+326   &        &BL &19.0 &1.00 &3.31 \\
1611+343    &DA 406  &Q  &17.5 &1.40 &4.05 \\
1803+784   &        &BL &17.0 &0.68 &2.05 \\
1823+568   &        &BL &18.4 &0.66 &2.31 \\
2005+403   &        &Q  &19.5 &1.74 &2.51 \\
2021+614   &        &G  &19.5 &0.23 &2.21 \\
2200+420   &BL Lac  &BL &14.5 &0.07 &3.23 \\
2251+158   &3C\,454.3 &Q &16.1 &0.86 &8.86 \\
\enddata


\tablenotetext{a}{Note that many sources are highly variable.}

\tablecomments{Col. (1): B1950 source name. Col. (2): Alternate
common name. Col. (3): Optical identification from the literature 
(NED) with  Q = quasar, G = radio galaxy, BL = BL Lac object.  
Col. (4): Optical magnitude from NED. Col. (5) Redshift. Col. (6): 
Total flux density at 15 GHz measured by \citet{kel98}. }

\end{deluxetable}


\clearpage

\begin{deluxetable}{cc}
\tabletypesize{\scriptsize}
\tablecaption{O{\sc bservational} P{\sc arameters} \label{tbl-2}}
\tablewidth{0pt}
\tablehead{
\colhead{Frequency} & \colhead{Bandwidth}}

\startdata
8.114, 8.209, 8.369, 8.594   &8   \\
12.115, 12.591& 16       \\
15.165 &32 \\
\enddata


\tablecomments{Frequencies in GHz, bandwidths in MHz}

\end{deluxetable}

\clearpage

\begin{deluxetable}{lrrrrrrrr}
\tabletypesize{\scriptsize}
\tablecaption{R{\sc esults} \label{tbl-3}}
\tablewidth{0pt}
\tablehead{
\colhead{Source} & \colhead{Peak} & \colhead{Integ} & \colhead{PeakPOL}  & \colhead{$RM_{0}$} 
& \colhead{$R_{c}$} & \colhead{$m_{c}$} & \colhead{$RM_{i}$} 
& \colhead{$\alpha^{12.1}_{8.1}$} \\
\colhead{(1)} & \colhead{(2)} & \colhead{(3)} & \colhead{(4)} & 
\colhead{(5)} & \colhead{(6)} & \colhead{(7)} & \colhead{(8)} & 
\colhead{(9)} }

\startdata
0133+476 & 3736 & 3802 & 50 & -1410 & 0.983 & 1.32 & -4878 & 0.68  \\ 
0212+735 & 1844 & 2445 & 41 & -542 & 0.754 & 2.22 & -6155 & -0.14  \\ 
3C111 & 1537 & 2263 &$<$1.8  & -730 & 0.679 & $<$0.1 & -804 & 1.21  \\ 
0420$-$014 & 2644 & 2872 & 6 & \nodata & 0.921 & 0.23 & \nodata & 0.59  \\ 
3C120 & 797 & 2519 & 4 & 2082 & 0.316 & 0.48 & 2209 & -0.26  \\ 
0528+134 & 3100 & 3439 & 32 & -163 & 0.901 & 1.03 & -1526 & 0.46  \\ 
0923+392 & 7179 & 8959 & 175 & -218 & 0.801 & 2.44 & -630 & 0.003  \\ 
M87 & 1029 & 1920 & $<$1.8 & 9563 & 0.536 & $<$0.2 & 9640 & 0.24  \\ 
3C279 & 14650 & 21335 & 1457 & -91.00 & 0.687 & 9.95 & -216 & 0.42  \\ 
3C279\tablenotemark{a} & 14070 & 21980 & 1238 & -140 & 0.641 & 8.80 & 
-332 & 0.31  \\ 
3C279\tablenotemark{b} & 19895 & 24766 & 1344 & -310 & 0.803 & 6.76 & 
-735 & 0.58  \\ 
3C279\tablenotemark{c} & 17500 & 22336 & 700 & -1280 & 0.783 & 4.00 & -3035 
& 1.10  \\ 
1308+326 & 667 & 1079 & 20 & 113 & 0.618 & 3.00 & 452 & -0.14  \\ 
1611+343 & 2692 & 3880 & 58 & -519 & 0.694 & 2.17 & -2989 & 0.29  \\ 
1803+784 & 1716 & 2179 & 86 & -201 & 0.788 & 5.03 & -567 & 0.19  \\ 
1823+568 & 647 & 829 & 45 & -128 & 0.780 & 6.97 & -353 & 0.32  \\ 
2005+403 & 1403 & 2327 & 21 & 668 & 0.603 & 1.50 & 5015 & 1.38  \\ 
2021+614 & 1557 & 2187 & $<$1.5 & \nodata & 0.712 & $<$0.1 & \nodata 
& 0.60  \\ 
BL Lac & 2029 & 2982 & 56 & -376 & 0.680 & 2.78 & -430 & 0.33  \\ 
2251+158 & 5509 & 8442 & 40 & -263 & 0.653 & 0.73 & -910 & 0.48  \\ 
\enddata

\tablenotetext{a}{\citet{tay98}}
\tablenotetext{b}{\citet{tay00}}
\tablenotetext{c}{\citet{zt01}}

\tablecomments{Col. (1): Source name. Col. (2): Peak flux density (mJy beam$^{-1}$) Col.  
(3): Sum of CLEAN components (mJy) Col. (4): Polarized flux density (mJy beam$^{-1}$) at 
location of peak Col. (5) Core or maximum jet RM (\radm) Col. (6): 
Core dominance Col. (7): Percent polarization Col. (8): Intrinsic RM
Col. (9): 8.1$-$12.1 GHz spectral index }

\end{deluxetable}

\clearpage

\begin{deluxetable}{lccccc}
\tabletypesize{\scriptsize}
\tablecaption{BL L{\sc ac} O{\sc bject} EVPA A{\sc lignments} \label{tbl-4}}
\tablewidth{0pt}
\tablehead{
\colhead{Source} & \colhead{$\theta$} & \colhead{$\chi_{core}$} & 
\colhead{$\theta_{jet}$} & \colhead{$|\theta - \chi_{core}|$} & 
\colhead{$|\theta - \chi_{jet}|$} }

\startdata
1308+326 & $-66^{\circ}$ & 161$^{\circ}$ & 158$^{\circ}$ & 47$^{\circ}$ & 44$^{\circ}$  \\
1803+476 & $-85^{\circ}$ & 136$^{\circ}$ & 115$^{\circ}$ & 41$^{\circ}$ & 20$^{\circ}$  \\
1823+568 & $-162^{\circ}$ & 33$^{\circ}$ & 25$^{\circ}$ & 15$^{\circ}$ & 7$^{\circ}$  \\
BL Lac & $-161^{\circ}$ & 174$^{\circ}$ & 56$^{\circ}$ & 25$^{\circ}$ & 37$^{\circ}$ \\
\enddata

\tablecomments{$\theta$ is position angle of the jet, $\chi$ is the
EVPA of the core or jet component.}

\end{deluxetable}

\clearpage

\begin{figure}
\vspace{19.2cm}
\includegraphics{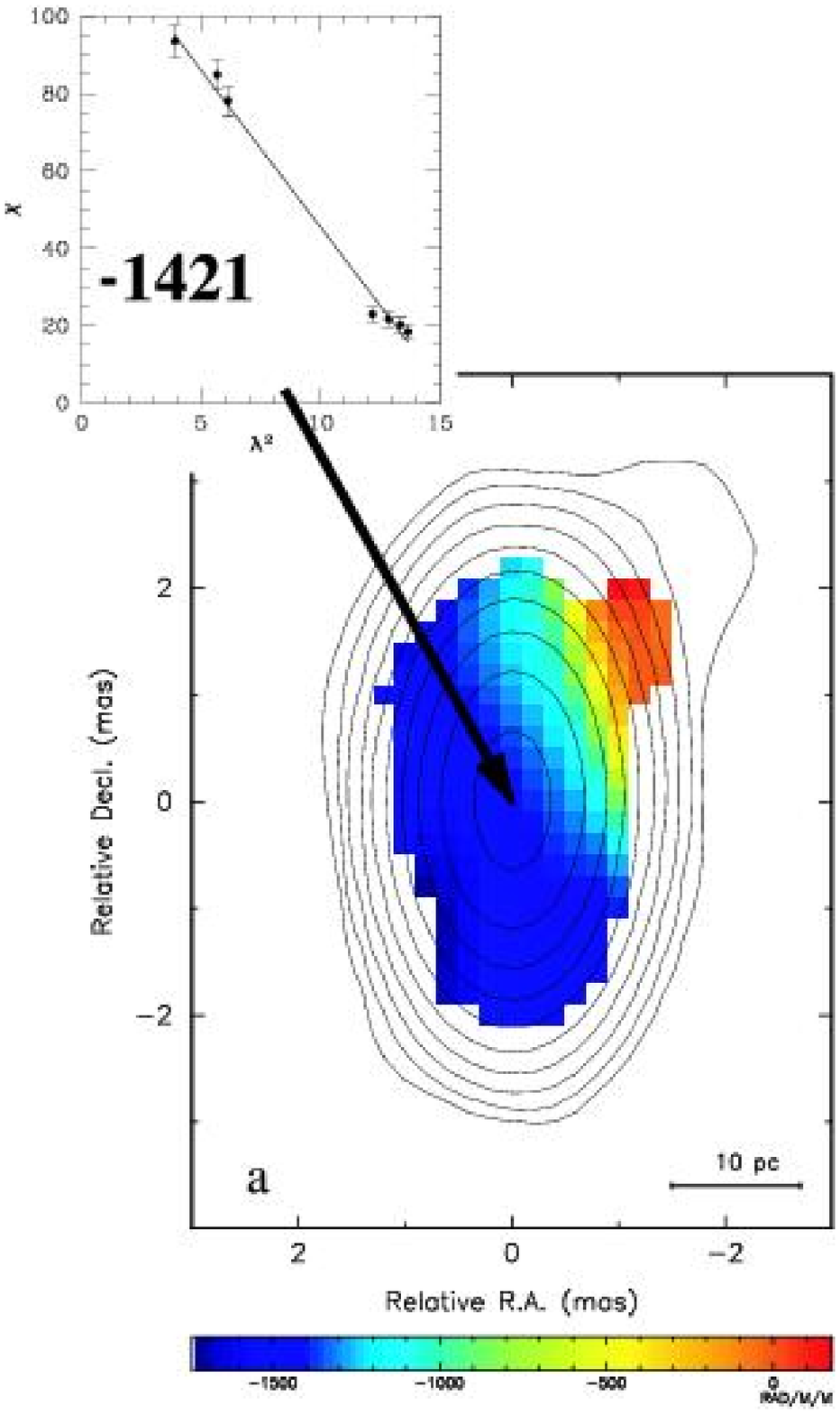}
\includegraphics{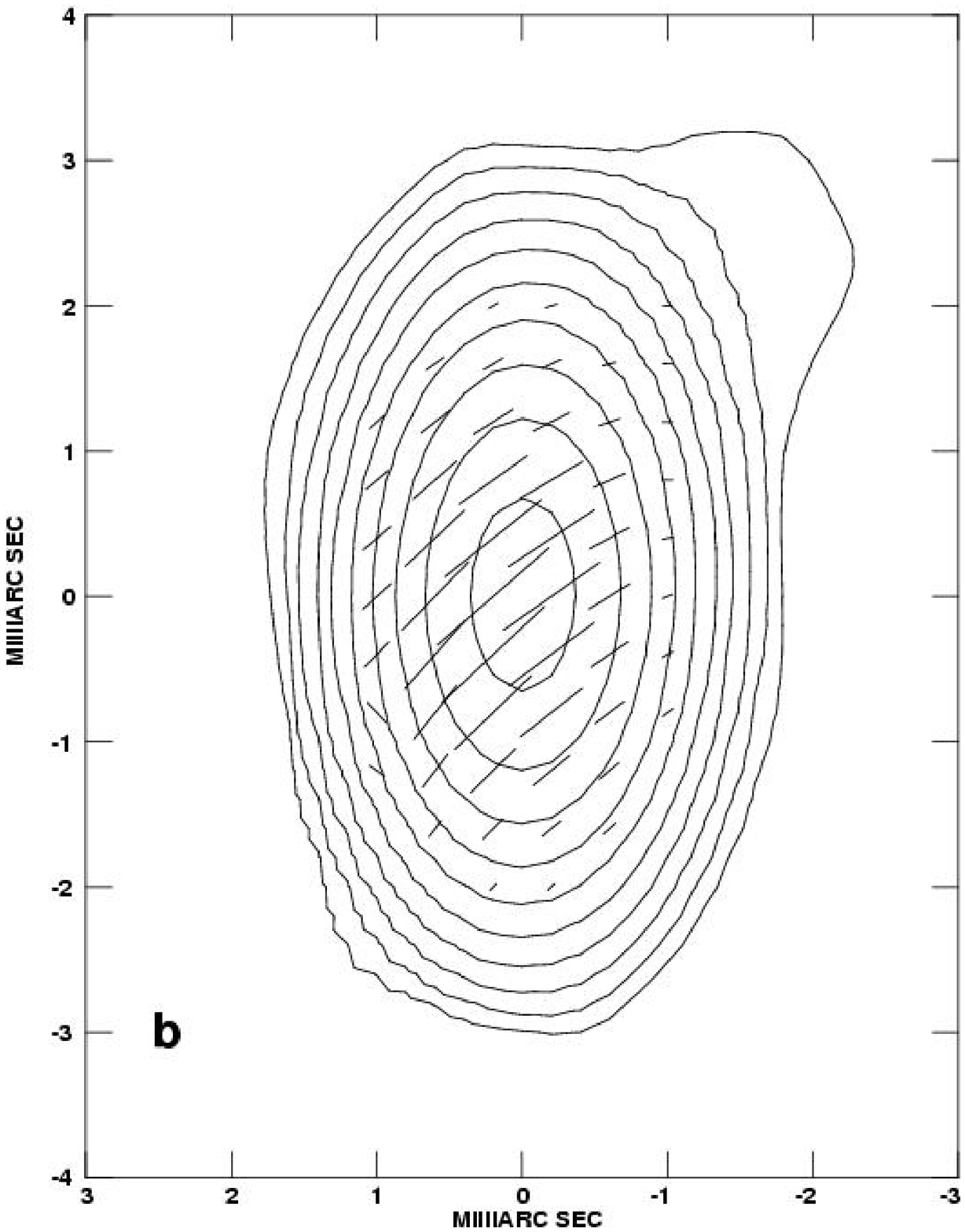}
\caption{(a) Rotation measure image (color) for 0133+476 overlaid on 
Stokes I contours at 15 GHz. The inset is a plot of EVPA $\chi$ 
(deg) versus $\lambda^2$ (cm$^2$). (b) Electric vectors (1 mas = 
50 mJy beam$^{-1}$ polarized flux density) corrected for Faraday 
Rotation overlaid on Stokes I contours. Contours start at 5.4 mJy 
beam$^{-1}$ and increase by factors of two.}
\label{0133rm}
\end{figure}
\clearpage

\begin{figure}
\vspace{19.2cm}
\includegraphics{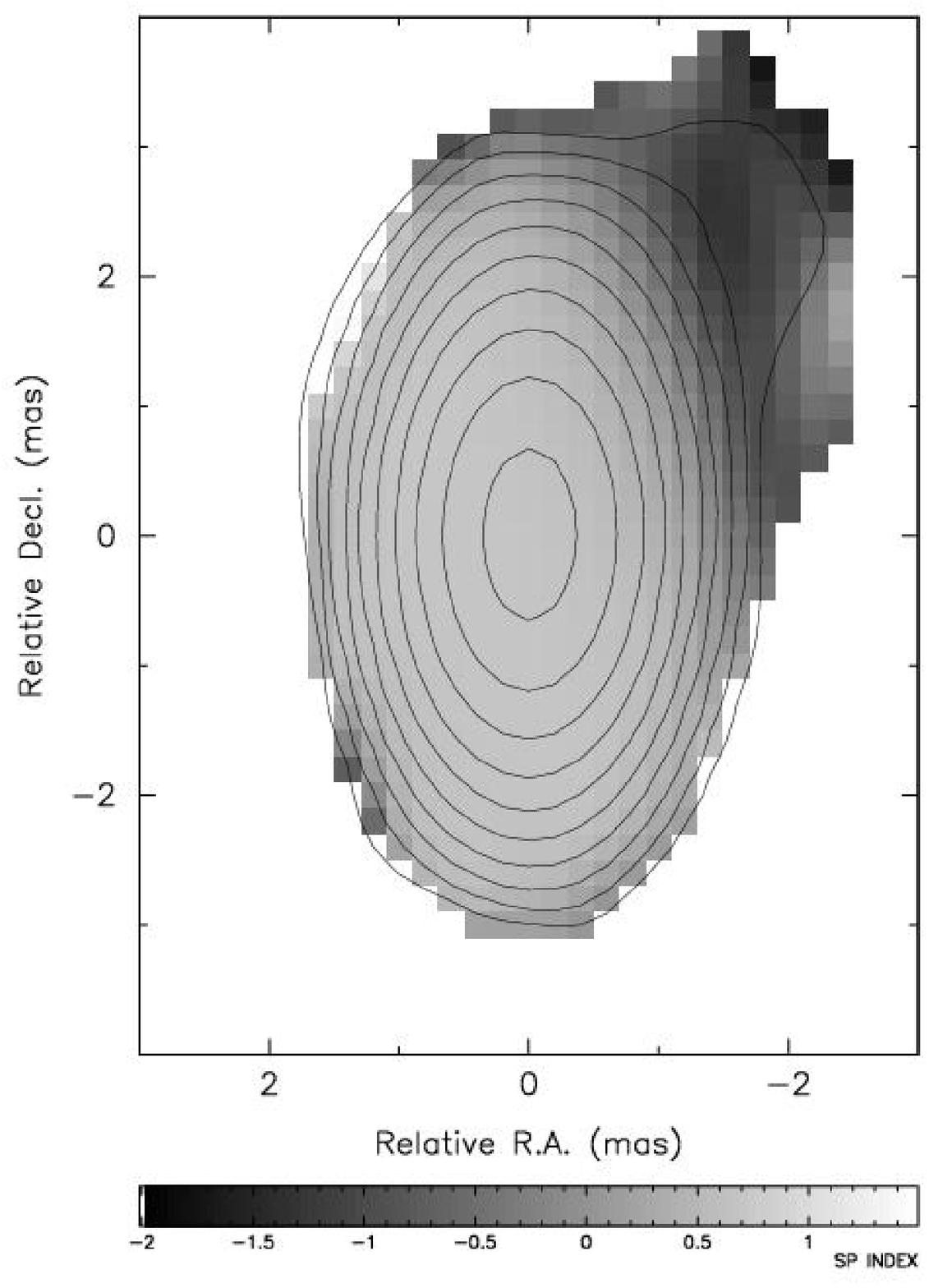}
\caption{spectral index $\alpha^{12.1}_{8.1}$ plot for 0133+476 overlaid on 
Stokes I contours at 15 GHz. Contours start at 5.4 mJy beam$^{-1}$ and 
increase by factors of two.}
\label{0133si}
\end{figure}

\begin{figure}
\vspace{19.2cm}
\includegraphics{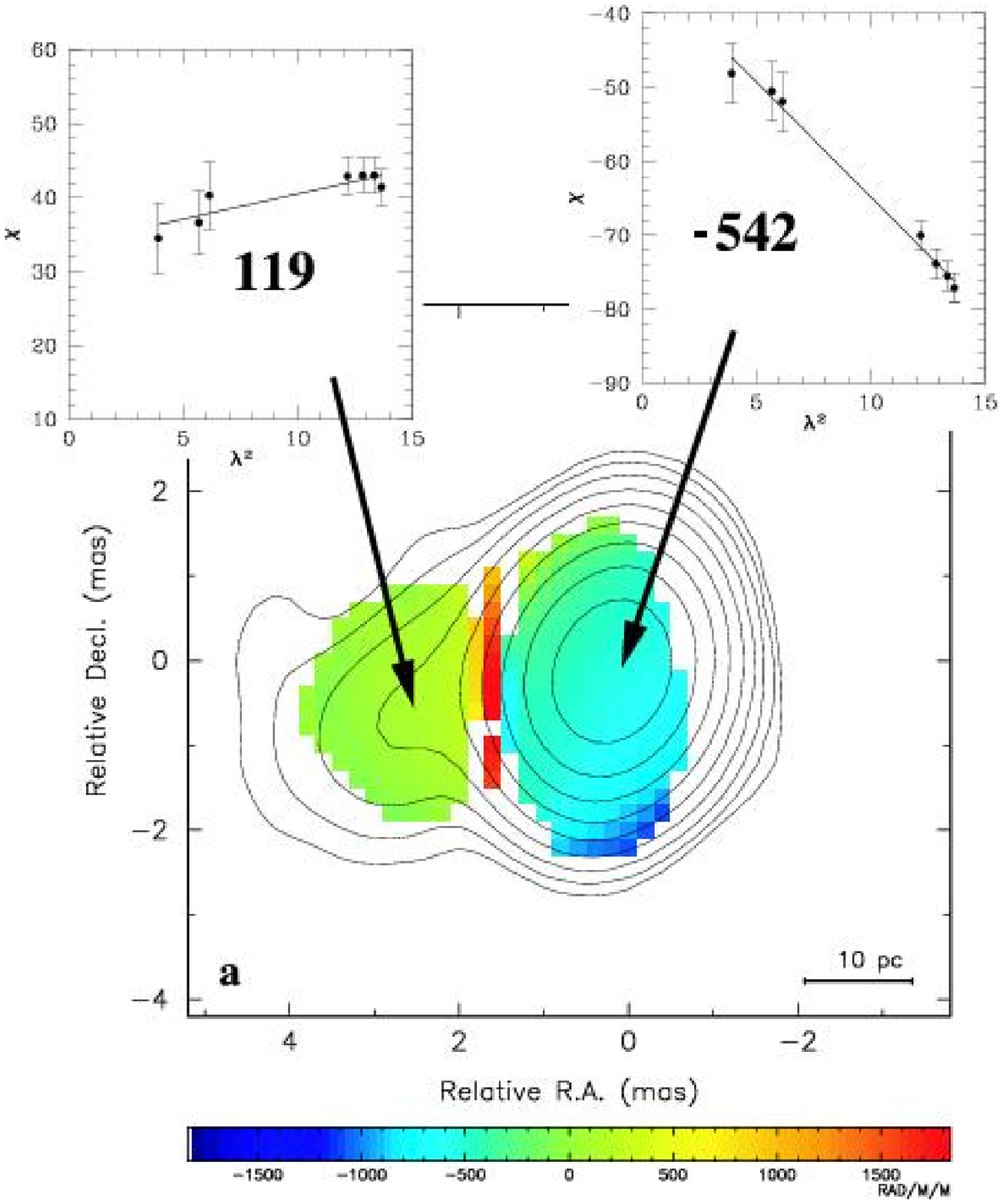}
\includegraphics{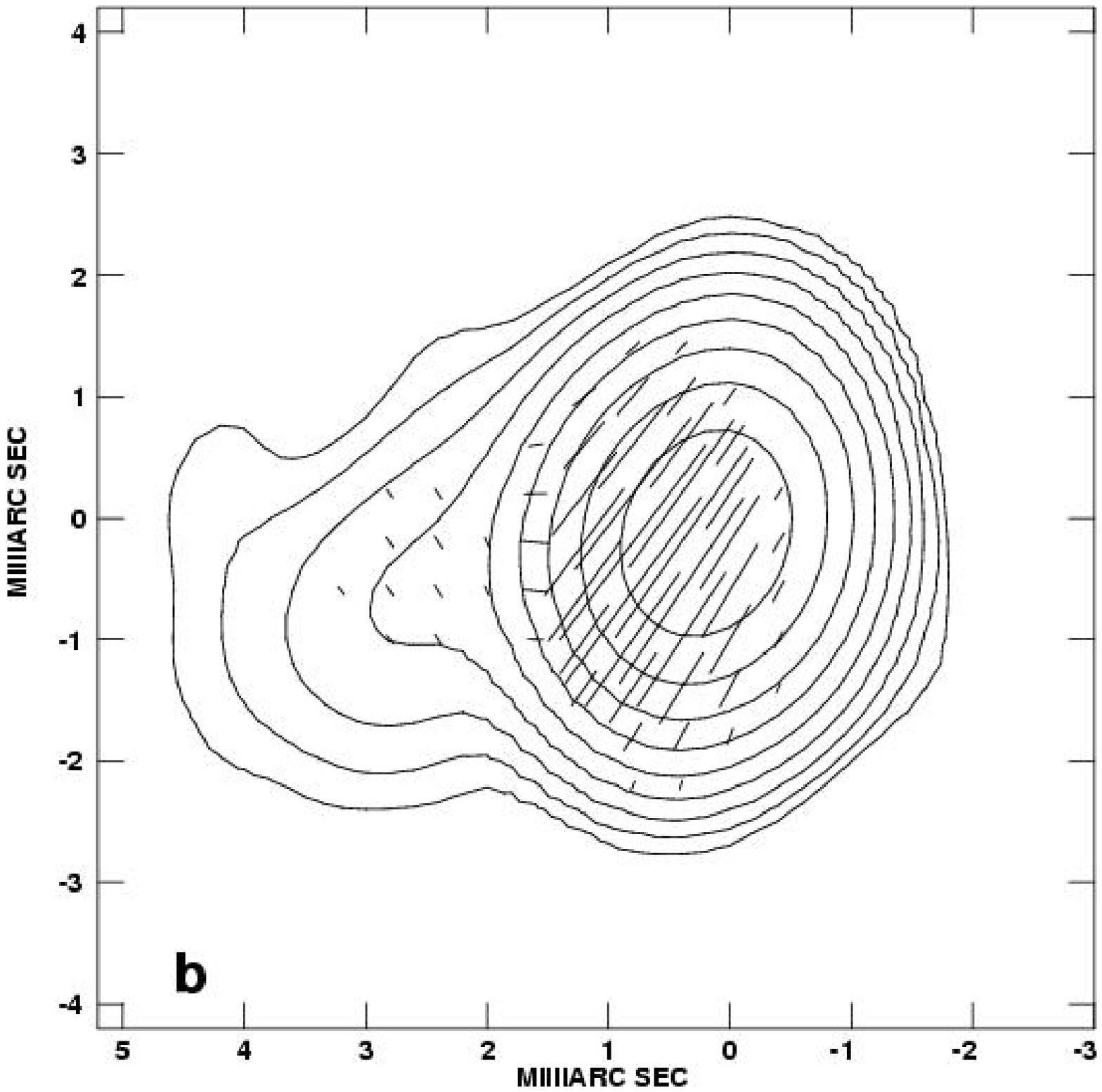}
\caption{(a) Rotation measure image (color) for 0212+735 overlaid on 
Stokes I contours at 15 GHz. The insets show plots of EVPA $\chi$ 
(deg) versus $\lambda^2$ (cm$^2$). (b) Electric vectors (1 mas = 
50 mJy beam$^{-1}$ polarized flux density) corrected for Faraday 
Rotation overlaid on Stokes I contours. Contours start at 3.9 mJy 
beam$^{-1}$ and increase by factors of two.}
\label{0212rm}
\end{figure}
\clearpage

\begin{figure}
\vspace{19.2cm}
\includegraphics{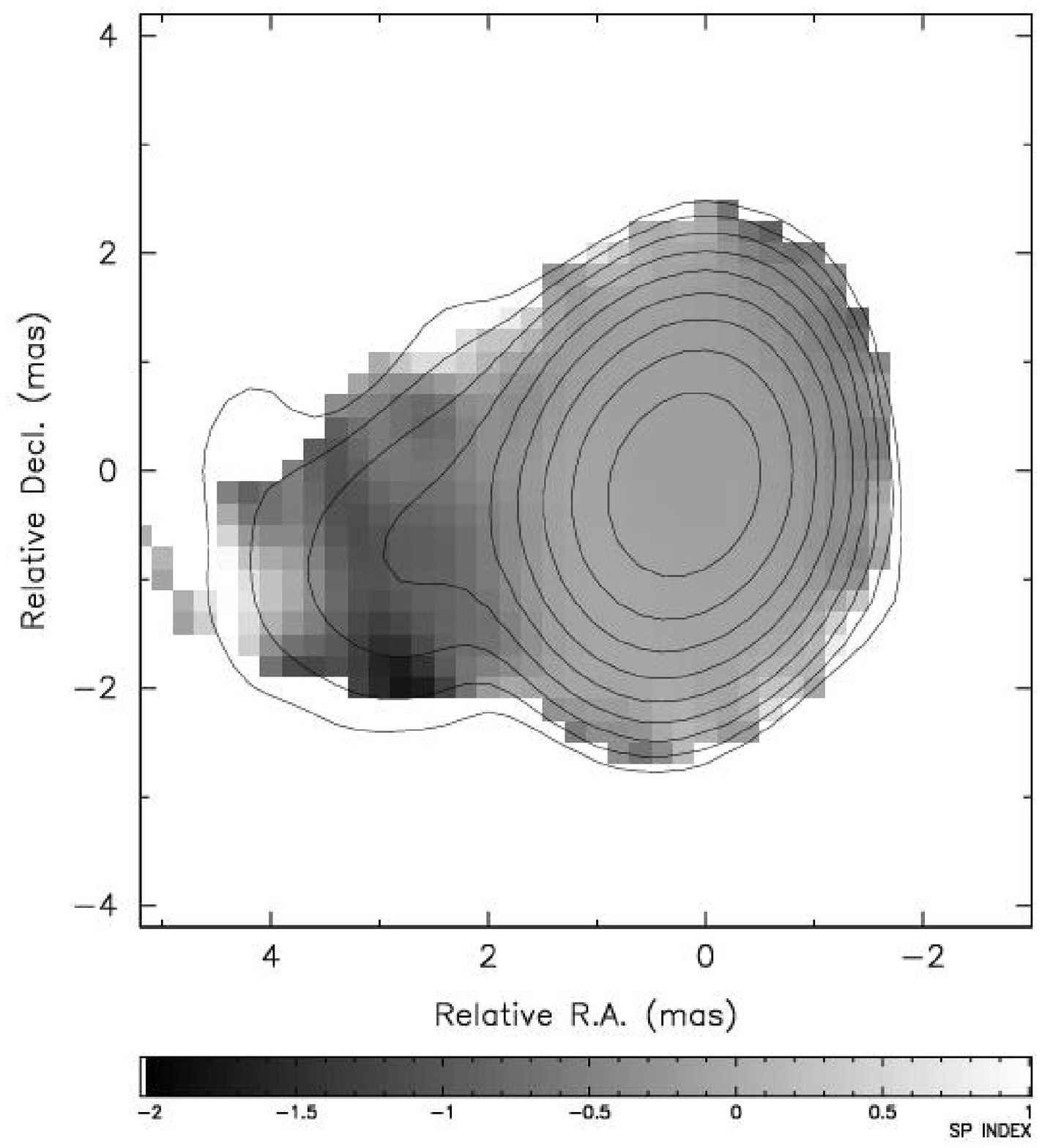}
\caption{Spectral index $\alpha^{12.1}_{8.1}$ plot for 0212+735 
overlaid on Stokes I contours at 15 GHz.  Contours start at 3.9
mJy beam$^{-1}$ and increase by factors of two.}
\label{0212si}
\end{figure}
\clearpage

\begin{figure}
\vspace{19.2cm}
\includegraphics{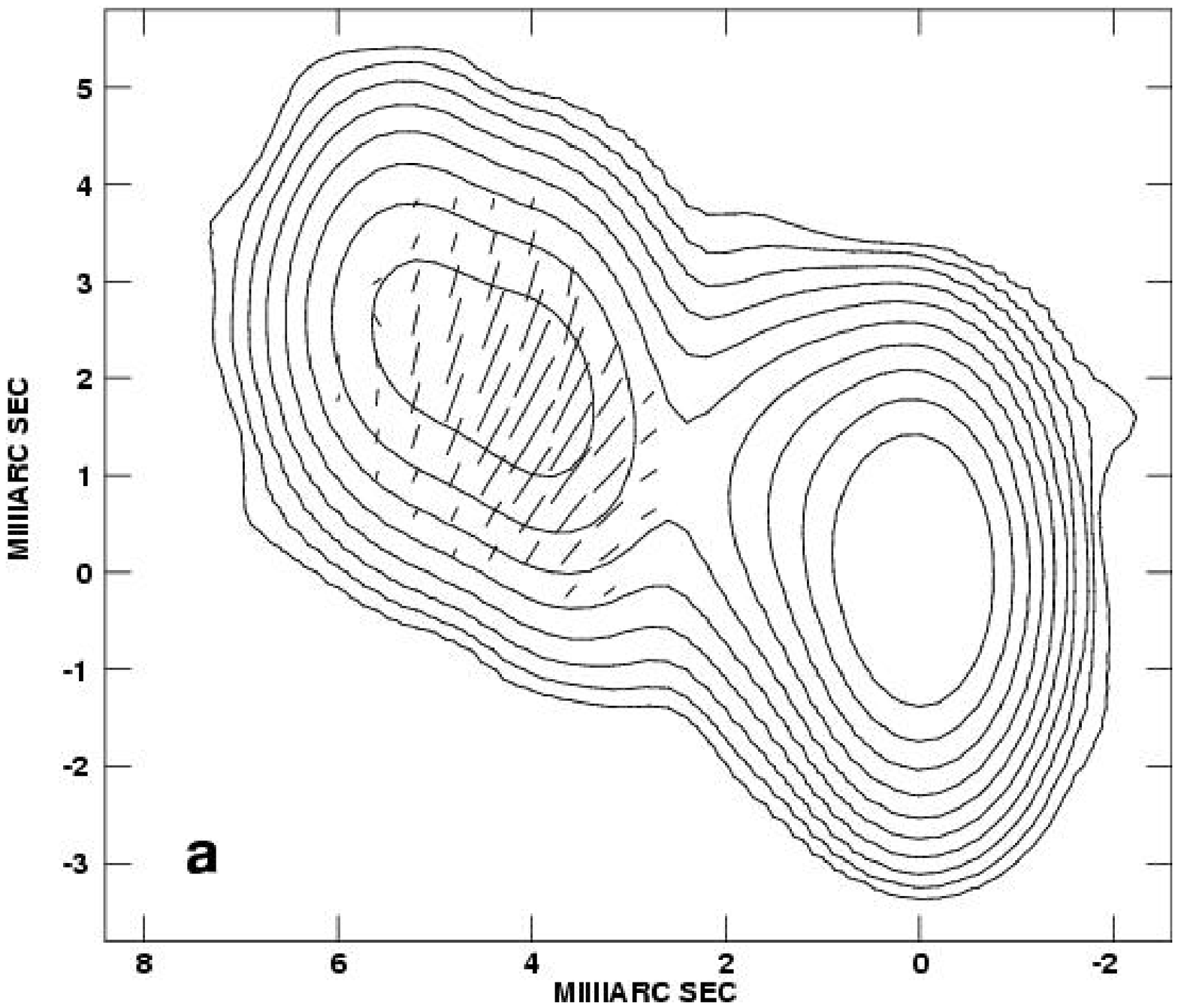}
\includegraphics{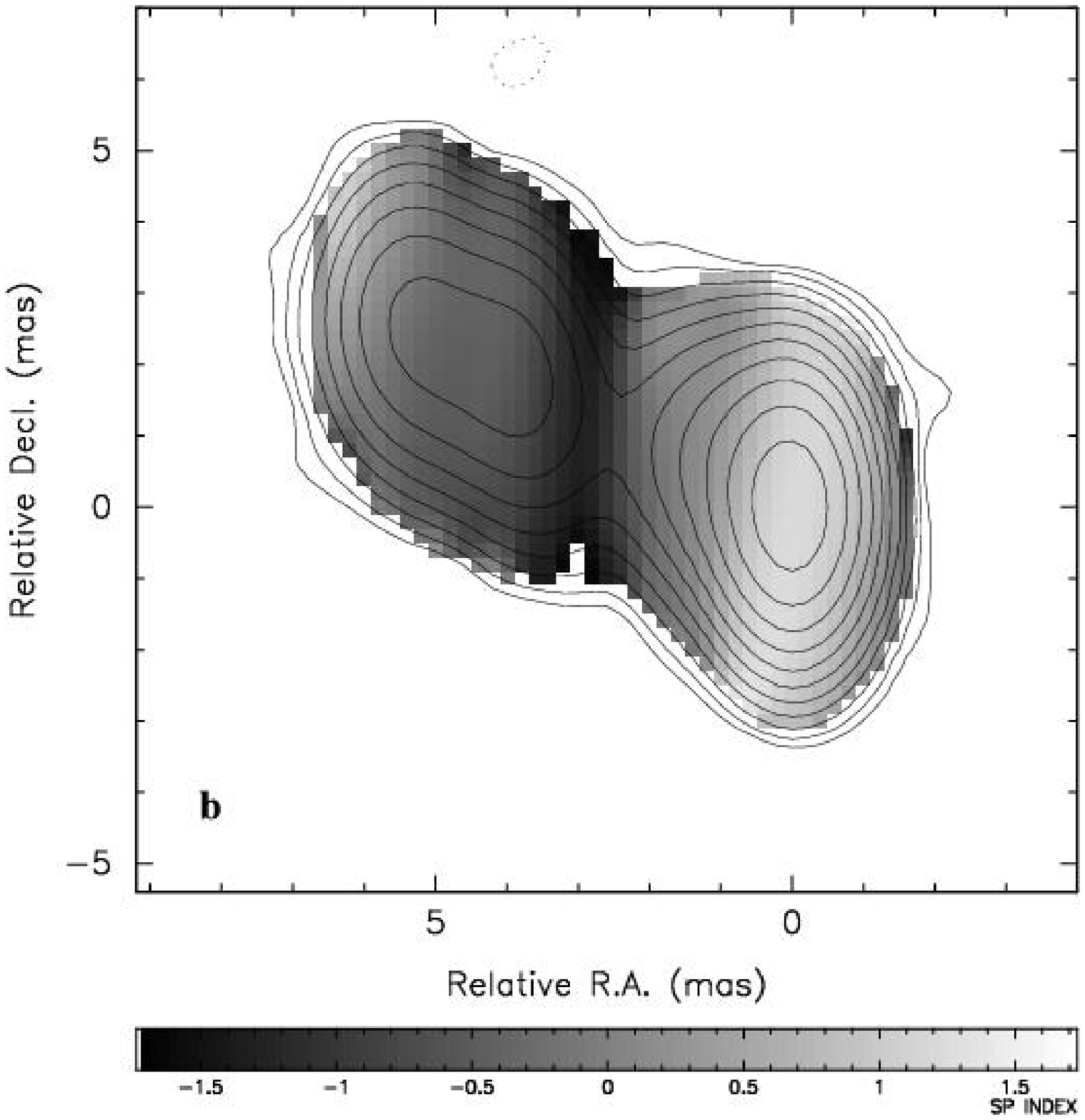}
\caption{(a) Electric vectors (1 mas = 25 mJy beam$^{-1}$ polarized flux 
density) corrected for Faraday Rotation overlaid on 15 GHz Stokes I contours. 
(b).Spectral index $\alpha^{12.1}_{8.1}$ plot for 3C\,111 overlaid on Stokes 
I contours at 15 GHz. Contours start at 0.9 mJy beam$^{-1}$ and increase by 
factors of two.}
\label{3c111}
\end{figure}
\clearpage

\begin{figure}
\vspace{19.2cm}
\includegraphics{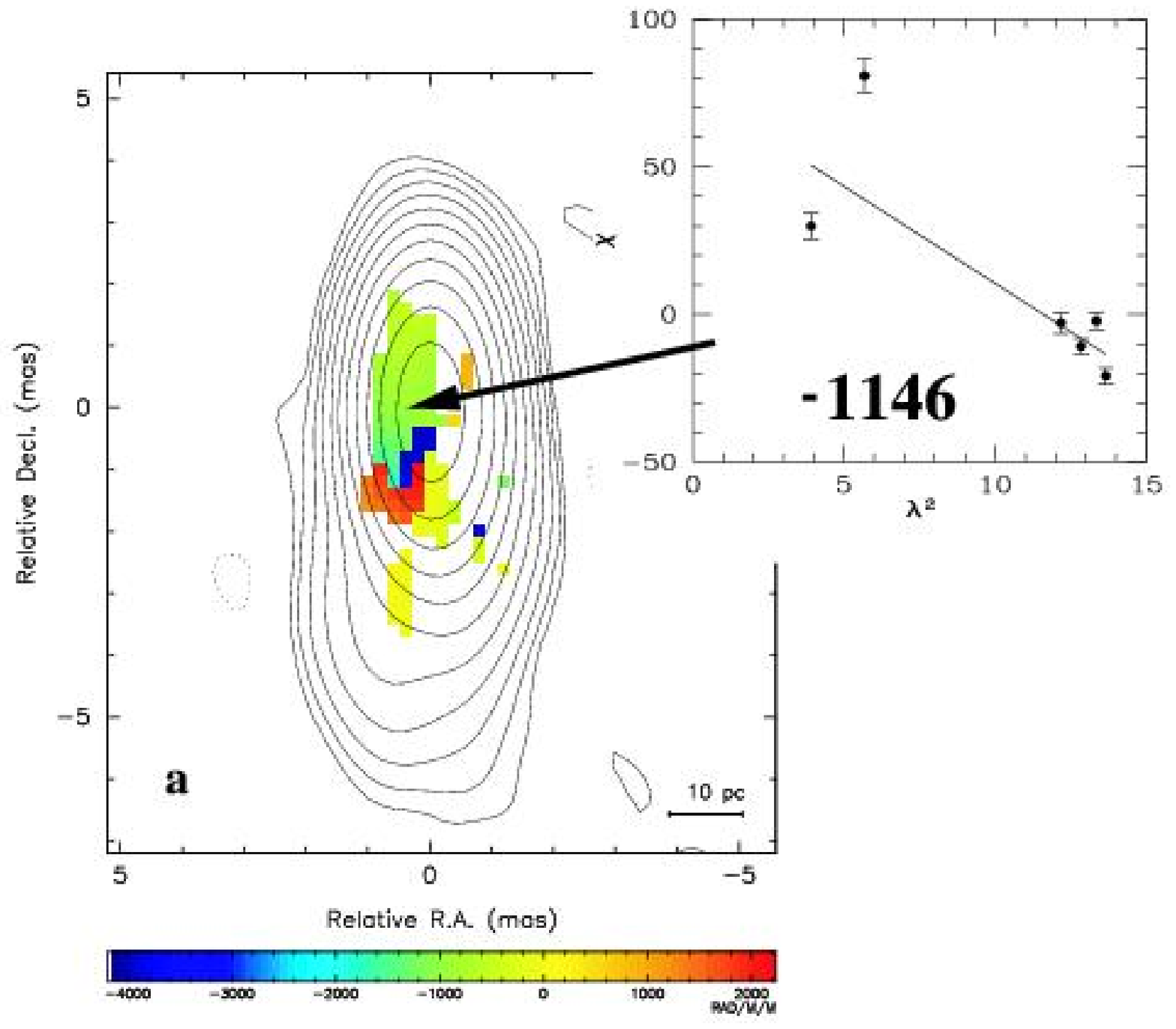}
\includegraphics{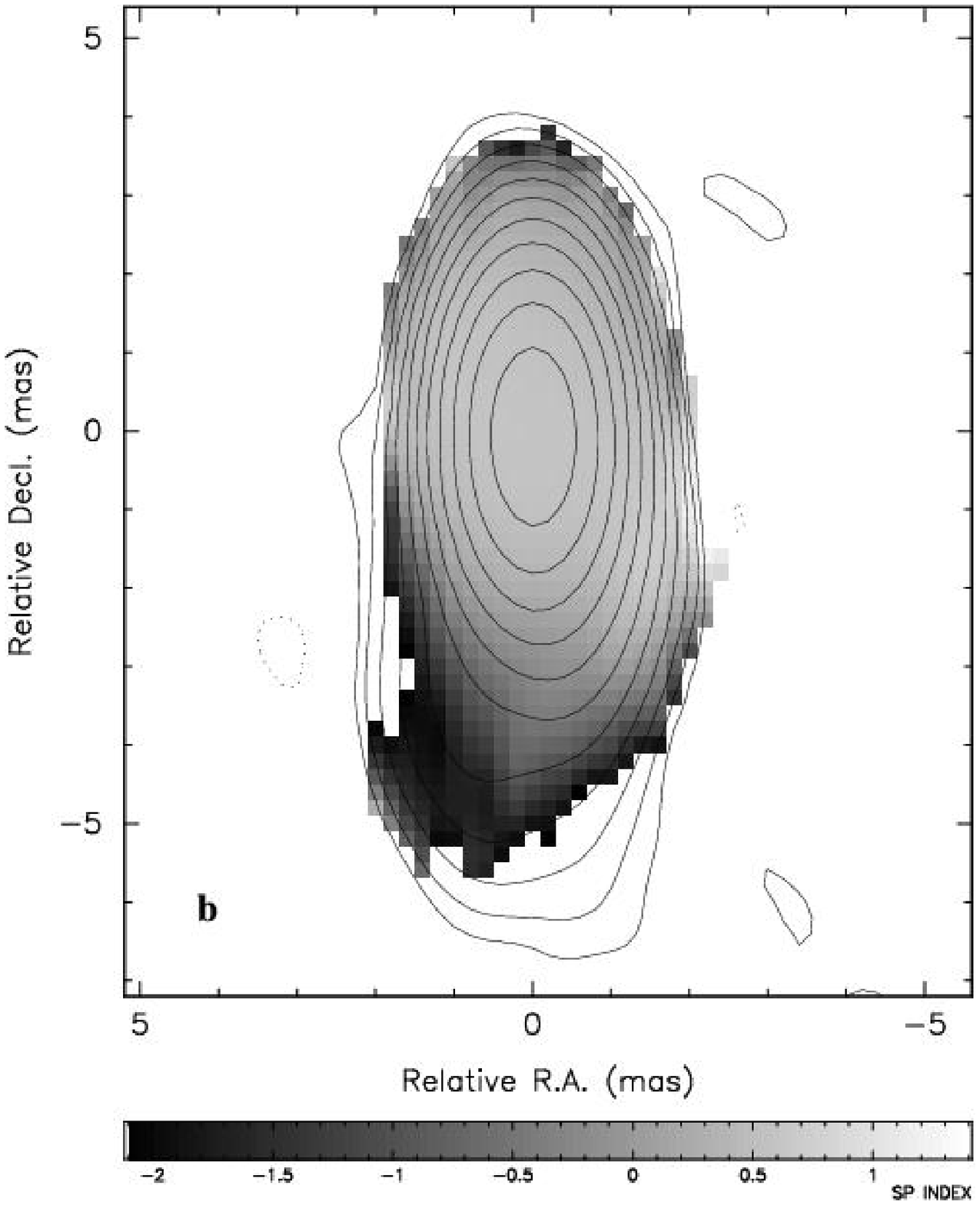}
\caption{(a) Rotation measure image (color) for the quasar 0420$-$014 overlaid on 
Stokes I contours at 15 GHz. The inset is a plot of EVPA $\chi$ (deg)
versus $\lambda^2$ (cm$^2$). (b)Plot of spectral index
$\alpha_{12.1}^{8.1}$ overlaid on 15 GHz Stokes I contours. Contours
start at 1.5 mJy beam$^{-1}$ and increase by factors of two.}
\label{0420rmsi}
\end{figure}
\clearpage


\begin{figure}
\vspace{19.2cm}
\includegraphics{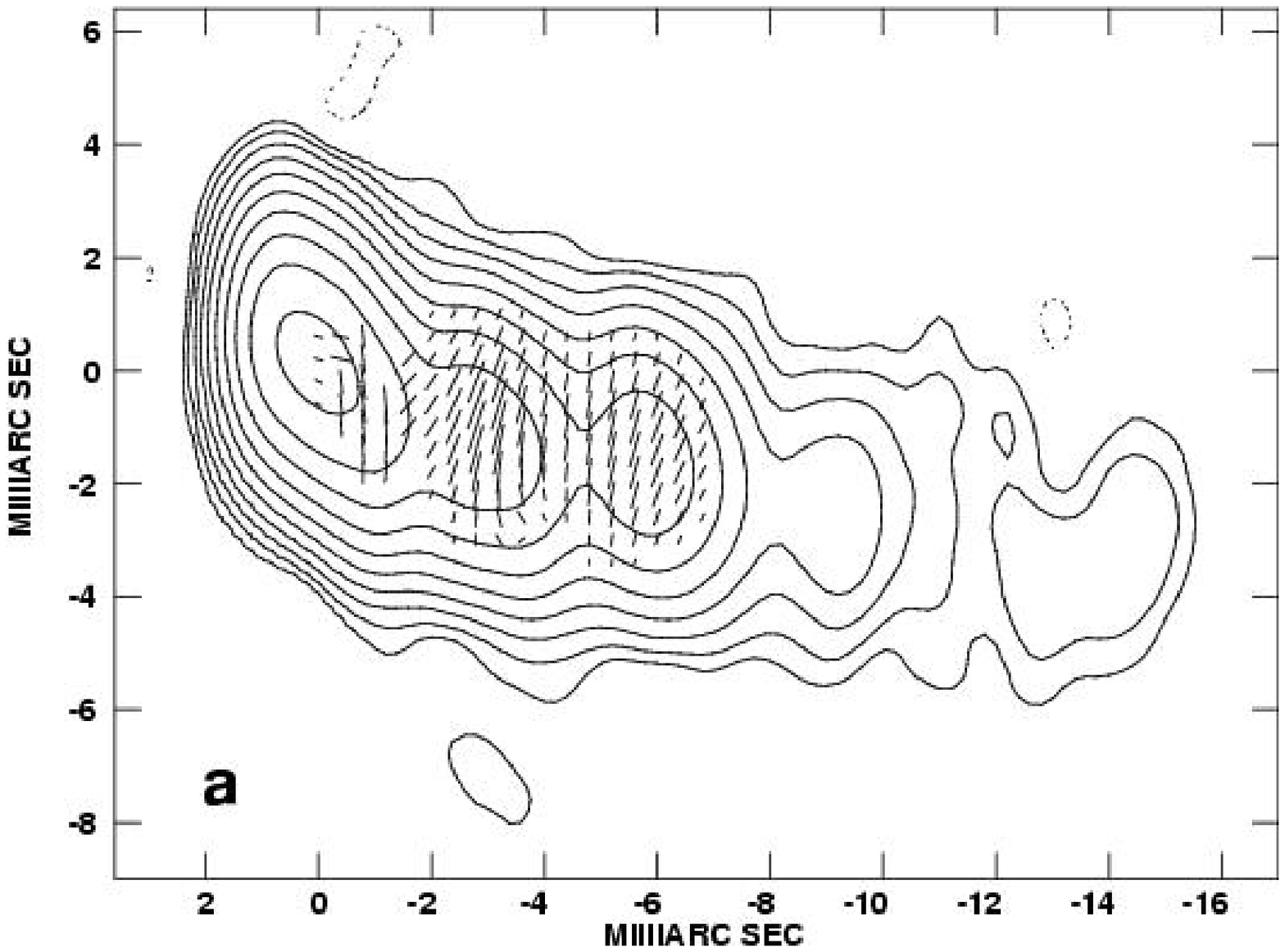}
\includegraphics{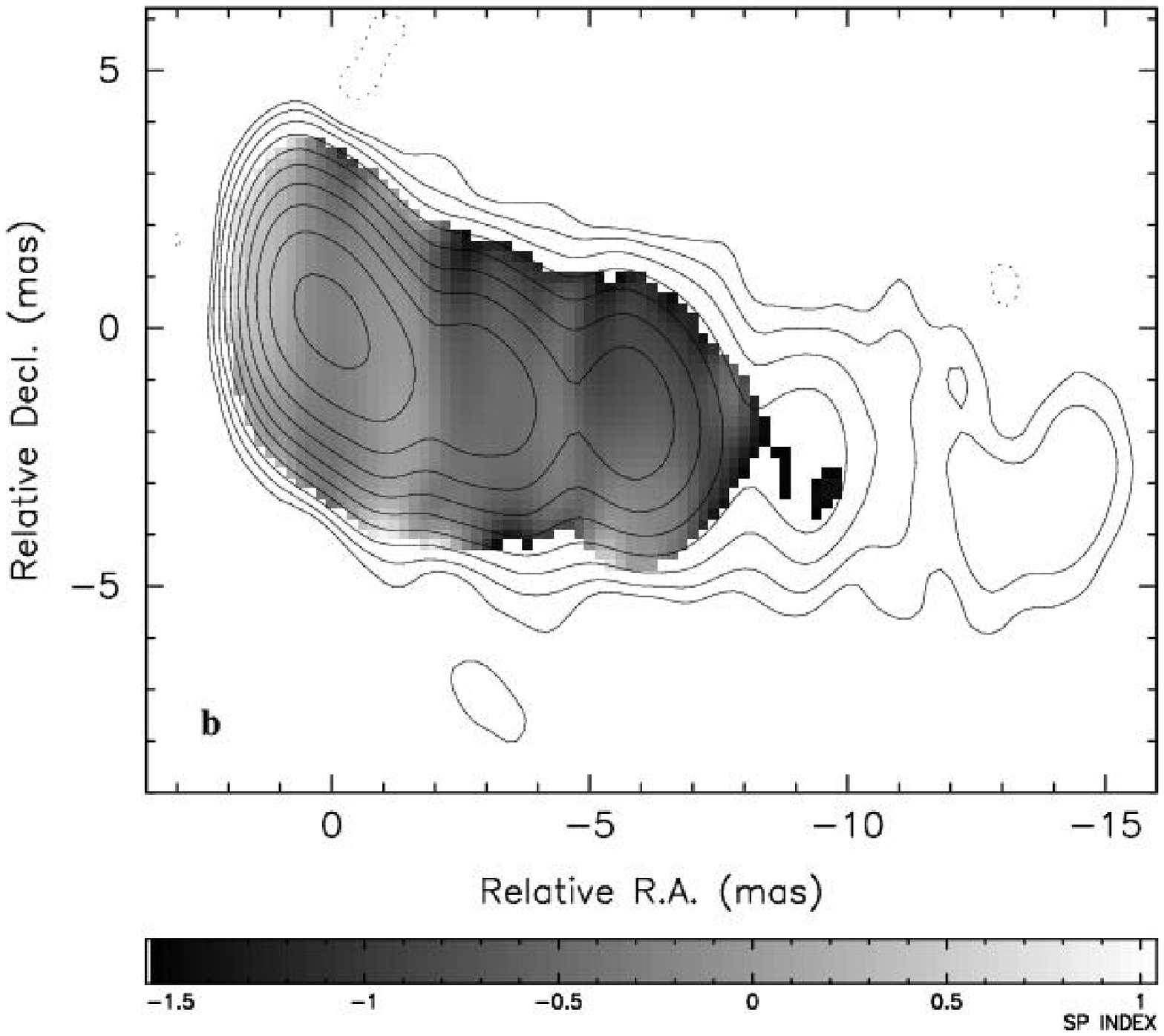}
\caption{(a) Electric vectors (1 mas = 25 mJy beam$^{-1}$ polarized flux 
density) for the radio galaxy 3C\,120 corrected for Faraday Rotation 
overlaid on 15 GHz Stokes I contours. b). Plot of spectral 
index $\alpha^{12.1}_{8.1}$ of the radio galaxy 3C\,120 overlaid on 
15 GHz Stokes I contours. Contours start at 1.22 mJy beam$^{-1}$ and 
increase by factors of two.}
\label{3c120}
\end{figure}
\clearpage

\begin{figure}
\vspace{19.2cm}
\includegraphics{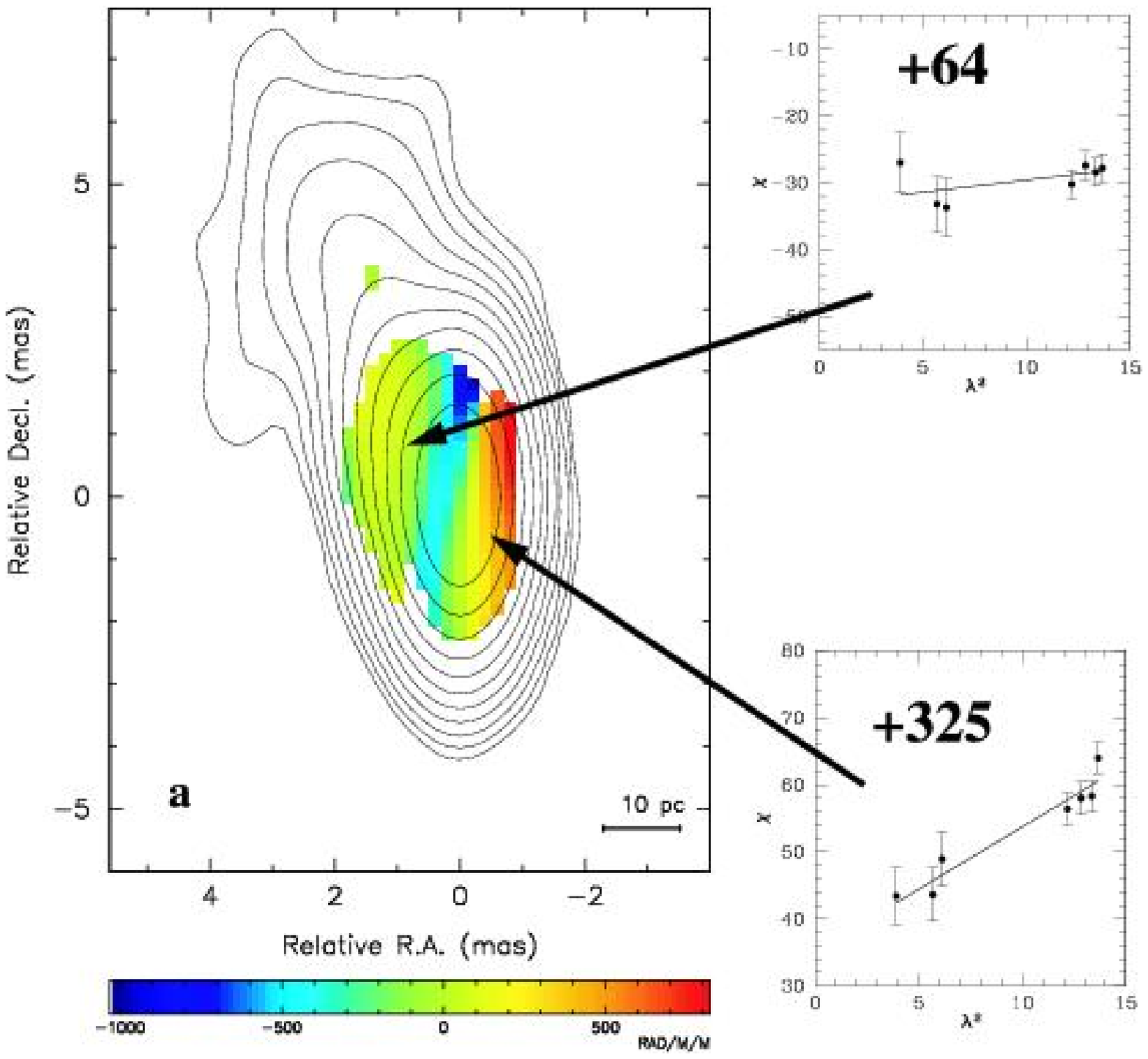}
\includegraphics{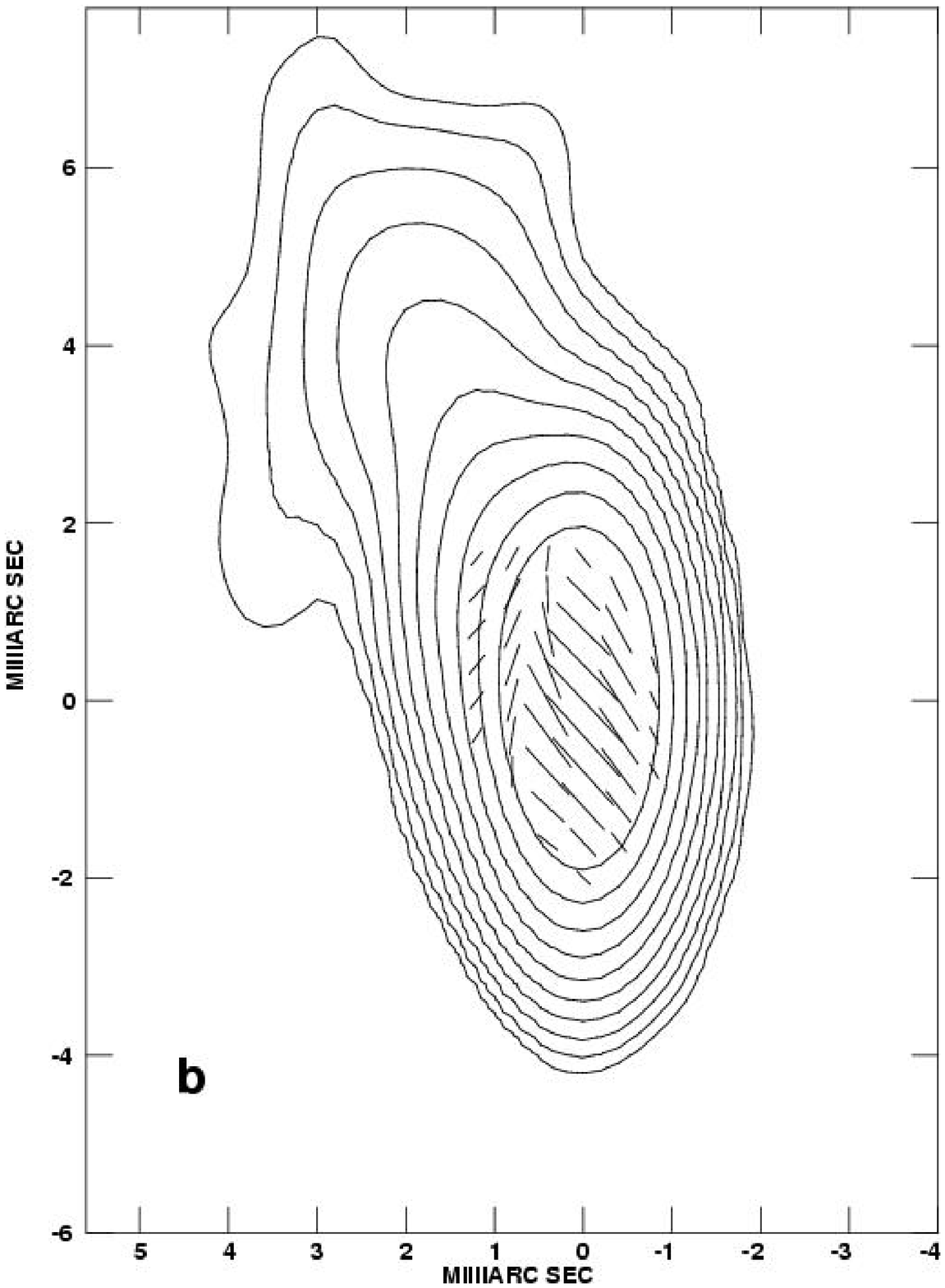}
\caption{(a) Rotation measure image (color) for 0528+134 overlaid on 
Stokes I contours at 15 GHz. The insets show plots of EVPA $\chi$ 
(deg) versus $\lambda^2$ (cm$^2$). (b) Electric vectors (1 mas = 
25 mJy beam$^{-1}$ polarized flux density) corrected for Faraday Rotation overlaid on
Stokes I contours. Contours start at 1.2 mJy beam$^{-1}$ and 
increase by factors of two.}
\label{0528rm}
\end{figure}
\clearpage

\begin{figure}
\vspace{19.2cm}
\includegraphics{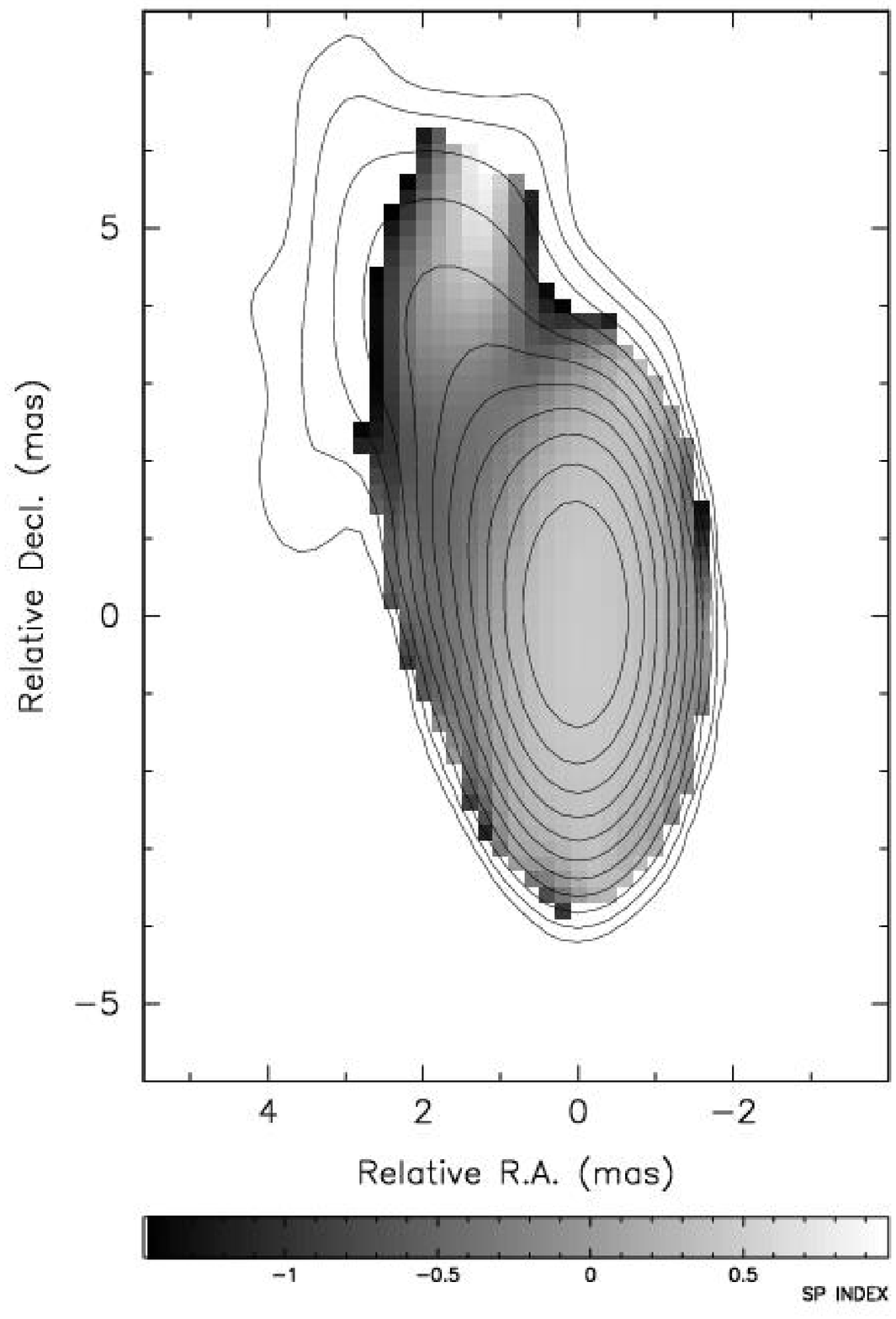}
\caption{Plot of spectral index $\alpha^{12.1}_{8.1}$ of the quasar 
0528+134 overlaid on 15 GHz Stokes I contours. Contours start at 
1.2 mJy beam$^{-1}$ and increase by factors of two.}
\label{0528si}
\end{figure}
\clearpage

\begin{figure}
\vspace{19.2cm}
\includegraphics{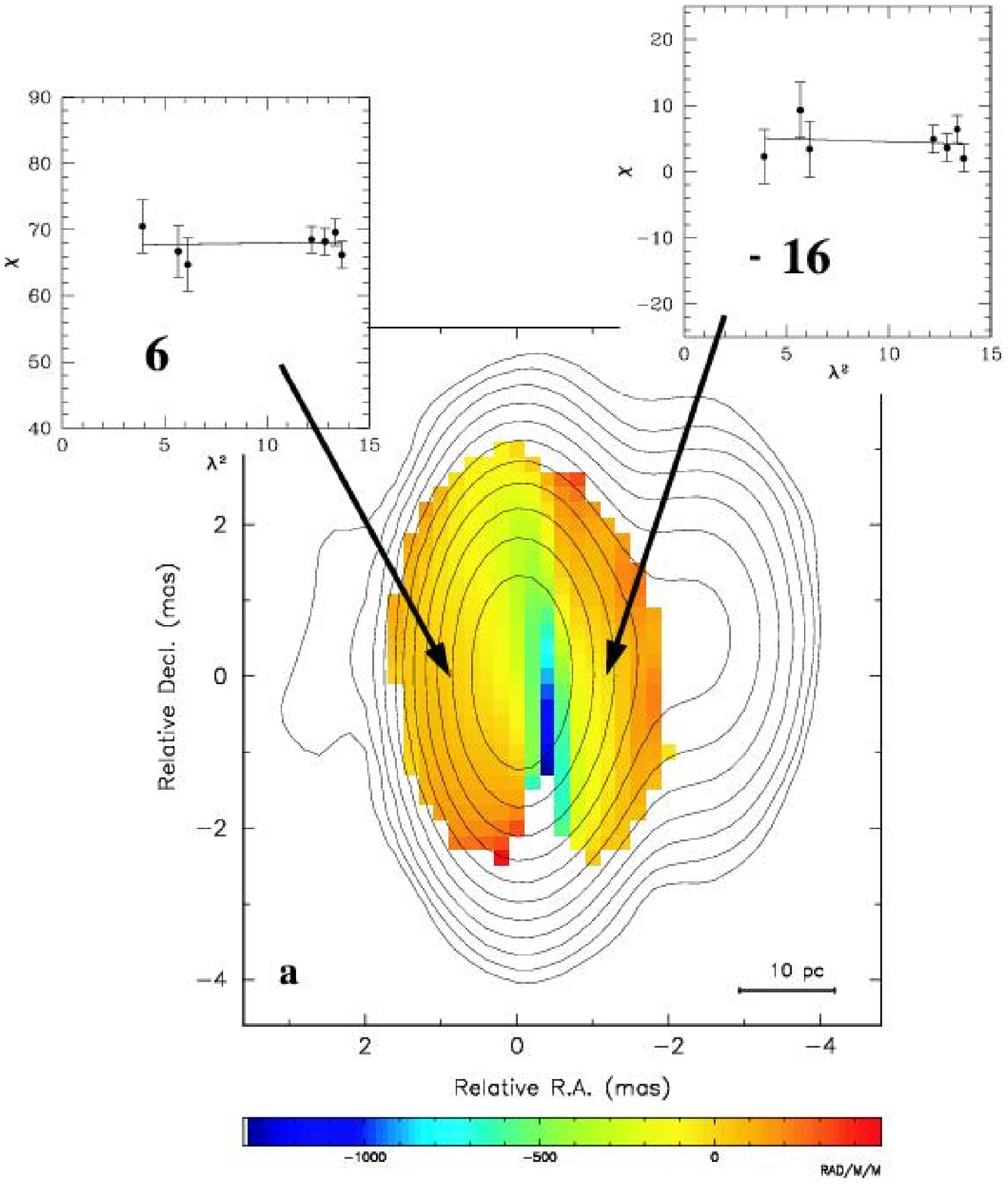}
\includegraphics{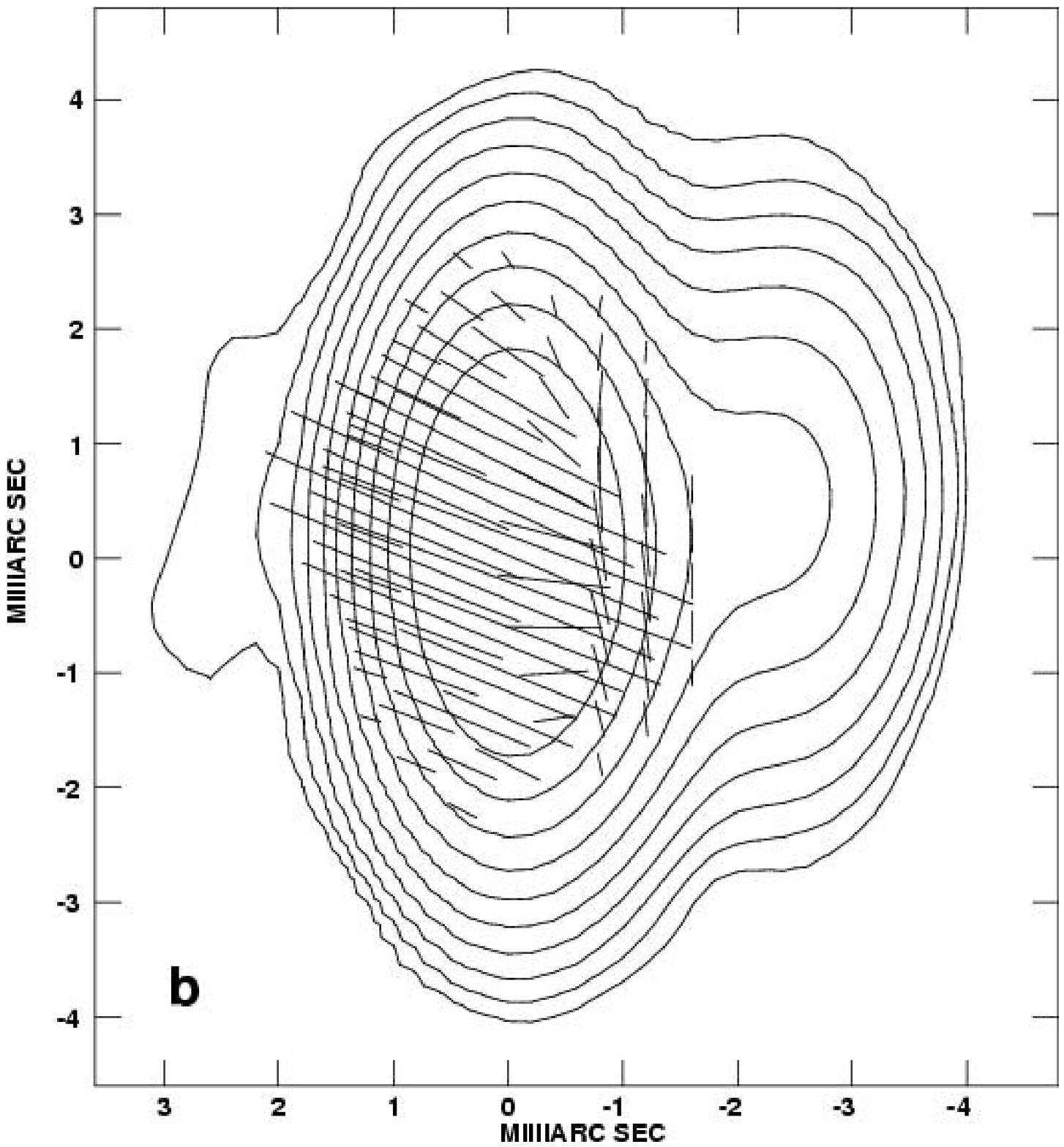}
\caption{(a) Rotation measure image (color) for 0923+392 overlaid on 
Stokes I contours at 15 GHz. The insets show plots of EVPA $\chi$ 
(deg) versus $\lambda^2$ (cm$^2$). (b) Electric vectors (1 mas = 
50 mJy beam$^{-1}$ polarized flux density) corrected for Faraday Rotation overlaid on
Stokes I contours. Contours start at 3.3 mJy beam$^{-1}$ and 
increase by factors of two.}
\label{0923rm}
\end{figure}
\clearpage

\begin{figure}
\vspace{19.2cm}
\includegraphics{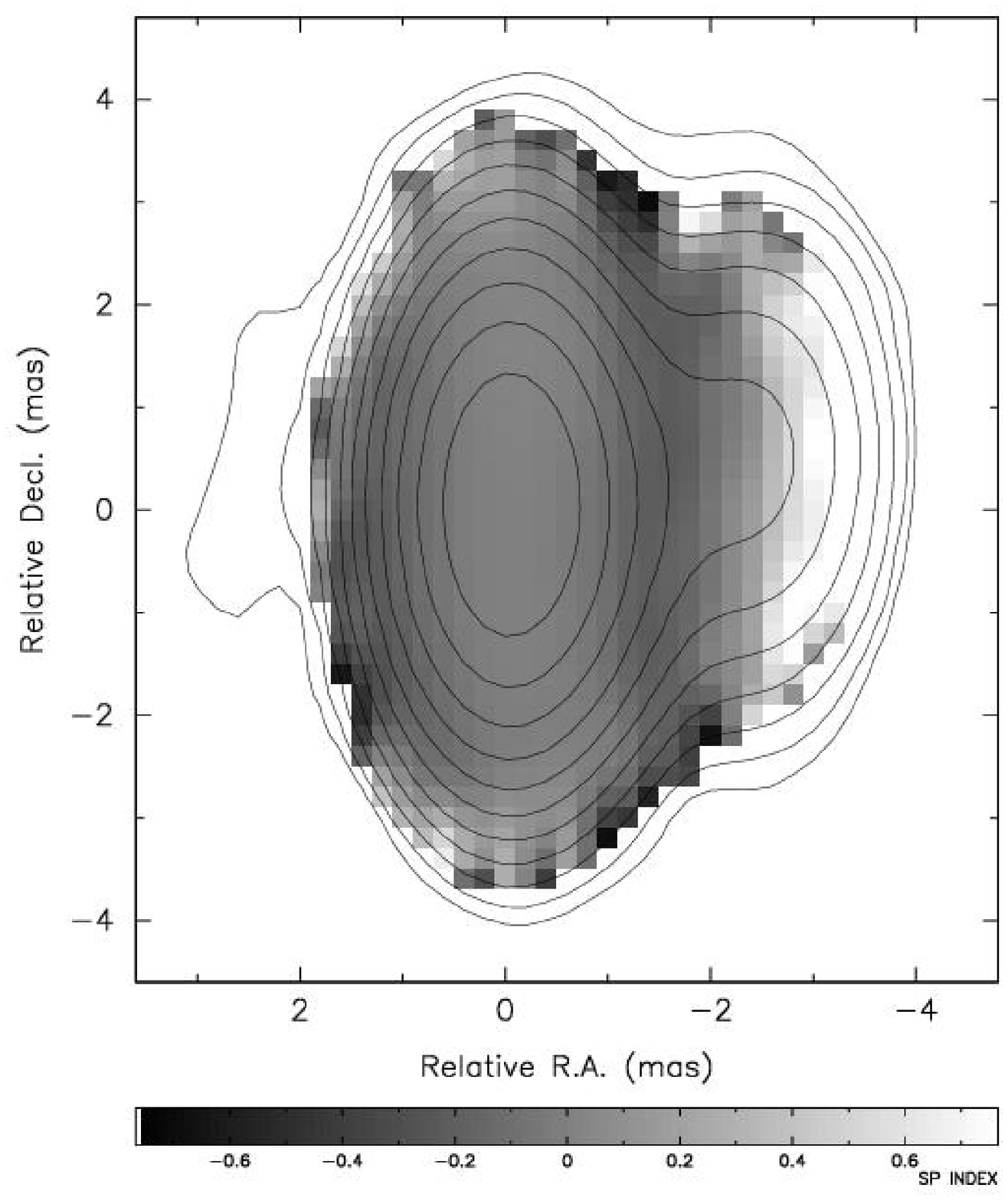}
\caption{Plot of spectral index $\alpha^{12.1}_{8.1}$ of the quasar 
0923+392 overlaid on 15 GHz Stokes I contours. Contours start at 
3.3 mJy beam$^{-1}$ and increase by factors of two.}
\label{0923si}
\end{figure}
\clearpage

\begin{figure}
\vspace{19.2cm}
\includegraphics{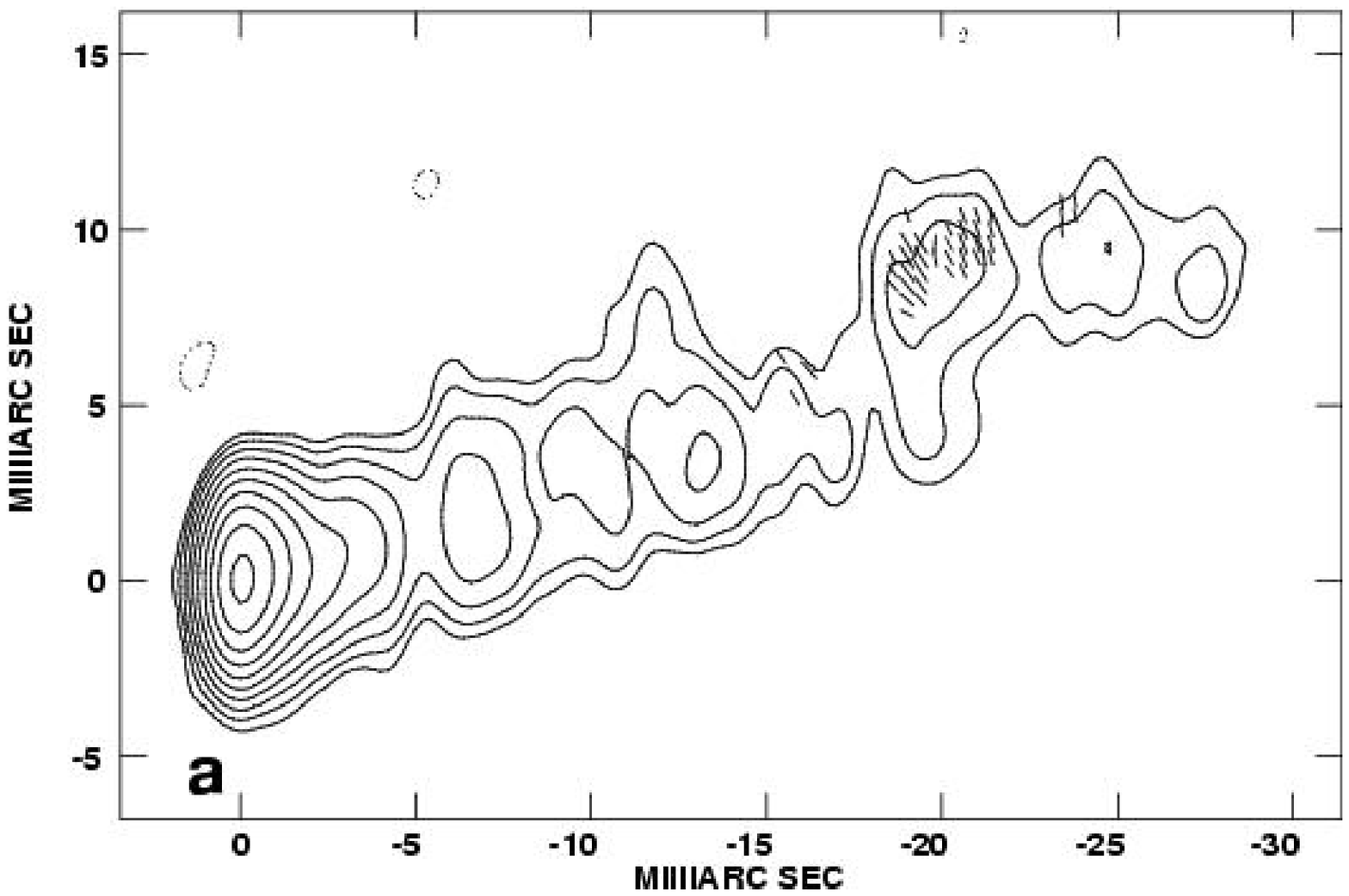}
\includegraphics{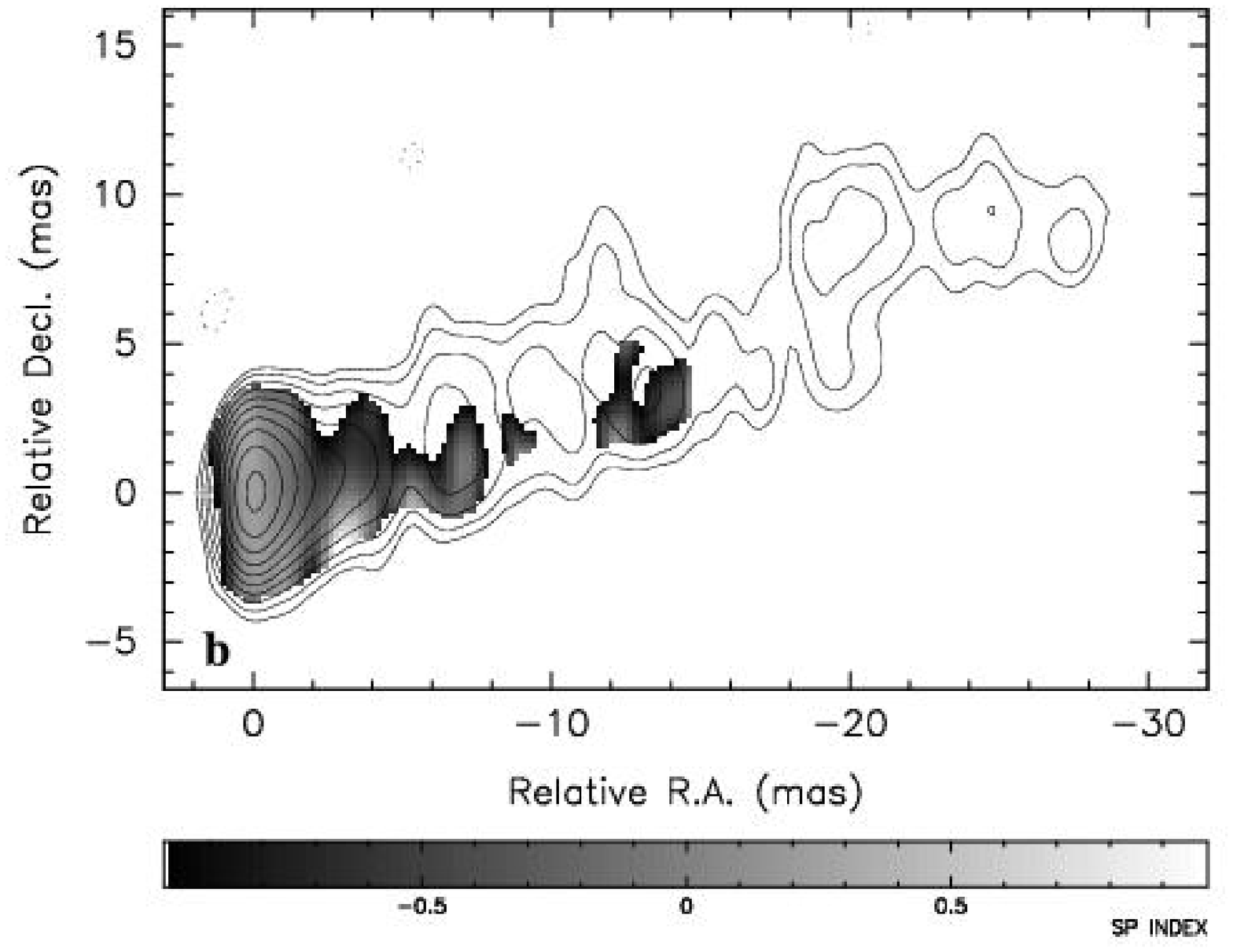}
\caption{(a) Electric vectors (1 mas = 3.3 mJy beam$^{-1}$ polarized flux density) 
for M87 corrected for Faraday Rotation overlaid on 15 GHz Stokes I contours. 
b). Plot of spectral index $\alpha^{12.1}_{8.1}$ overlaid on 15 GHz 
Stokes I contours. Contours start at 1.7 mJy beam$^{-1}$ and increase by 
factors of two.}
\label{M87}
\end{figure}
\clearpage

\begin{figure}
\vspace{19.2cm}
\includegraphics{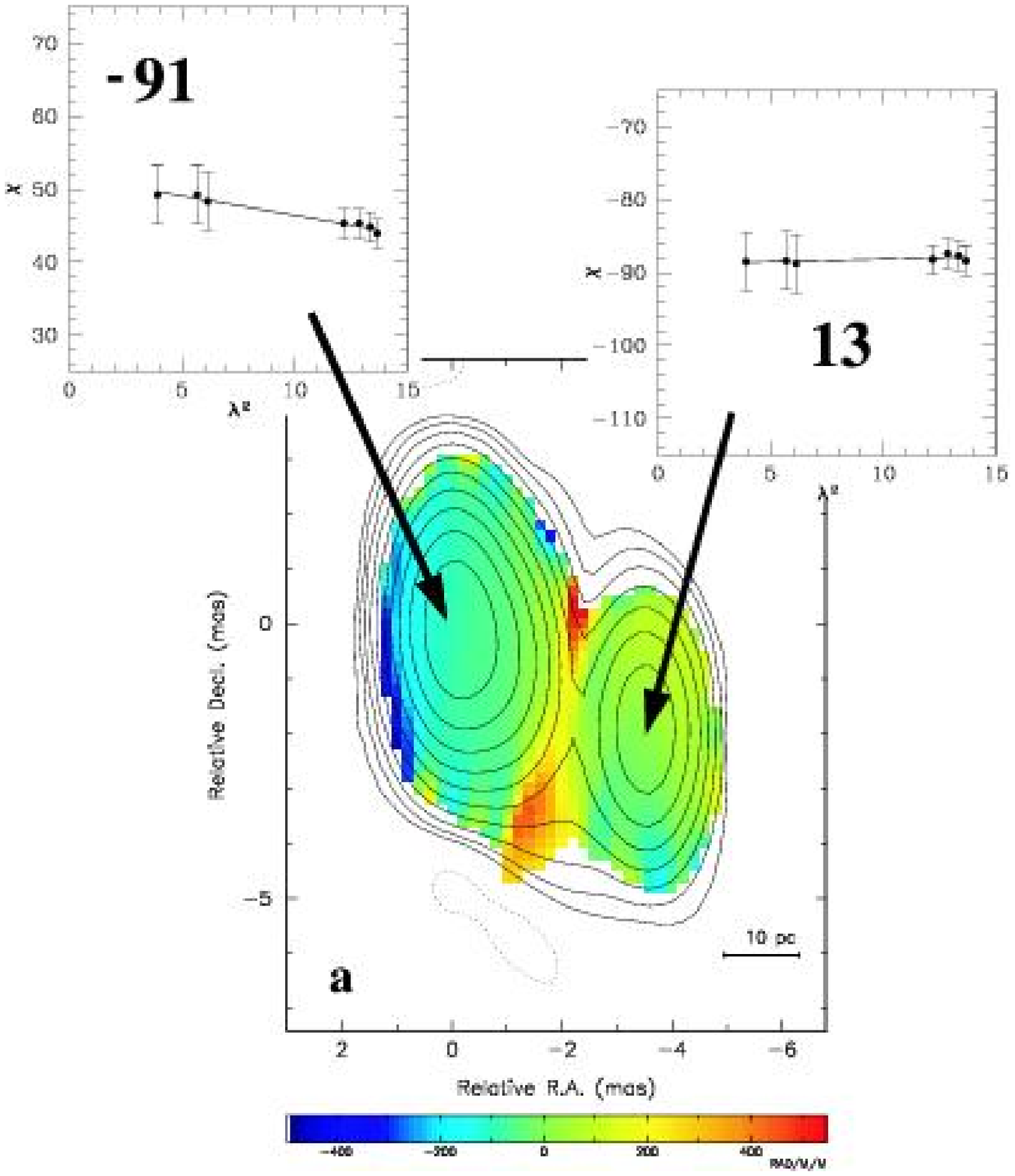}
\includegraphics{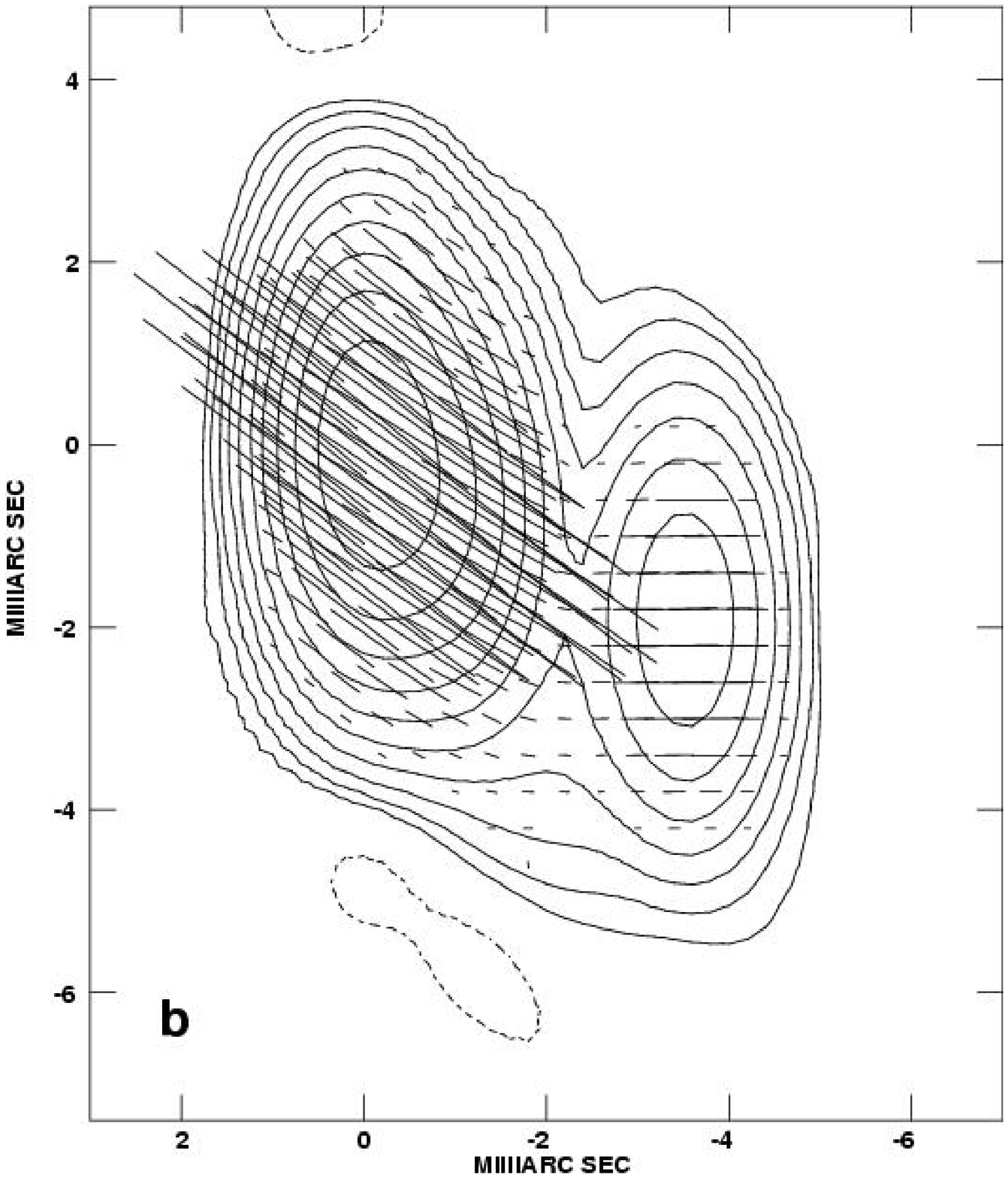}
\caption{a) Rotation measure image (color) for 3C\,279 overlaid on 
Stokes I contours at 15 GHz. The insets show plots of EVPA $\chi$ 
(deg) versus $\lambda^2$ (cm$^2$). (b) Electric vectors (1 mas = 
250 mJy beam$^{-1}$ polarized flux density) corrected for Faraday Rotation overlaid on
Stokes I contours. Contours start at 15 mJy beam$^{-1}$ and 
increase by factors of two.}
\label{3c279rm}
\end{figure}
\clearpage

\begin{figure}
\vspace{19.2cm}
\includegraphics{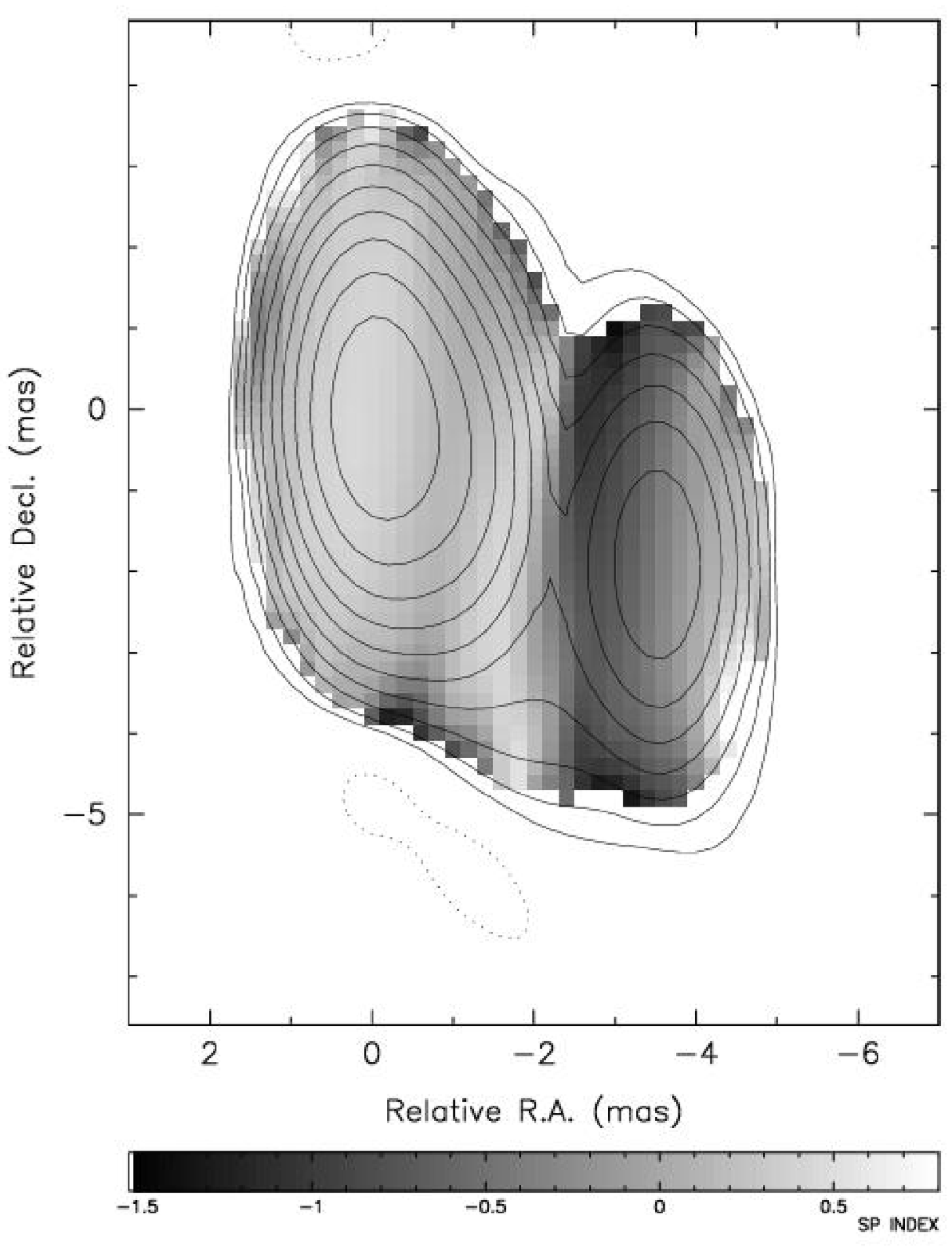}
\caption{Plot of spectral index $\alpha^{12.1}_{8.1}$ of the quasar 
3C\,279 overlaid on 15 GHz Stokes I contours. Contours start 
at 15 mJy beam$^{-1}$ and increase by factors of two.}
\label{3c279si}
\end{figure}
\clearpage

\begin{figure}
\vspace{19.2cm}
\includegraphics{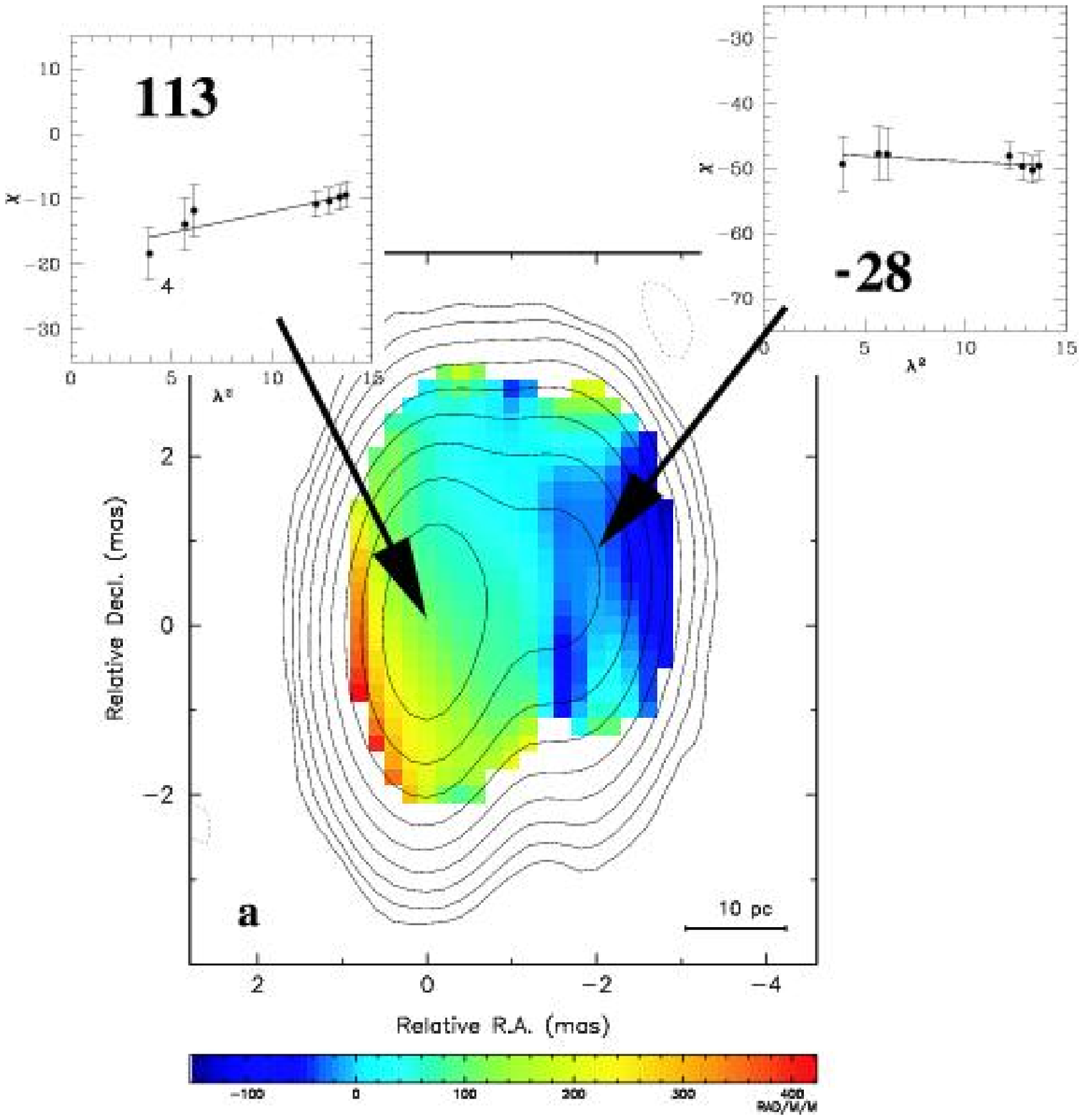}
\includegraphics{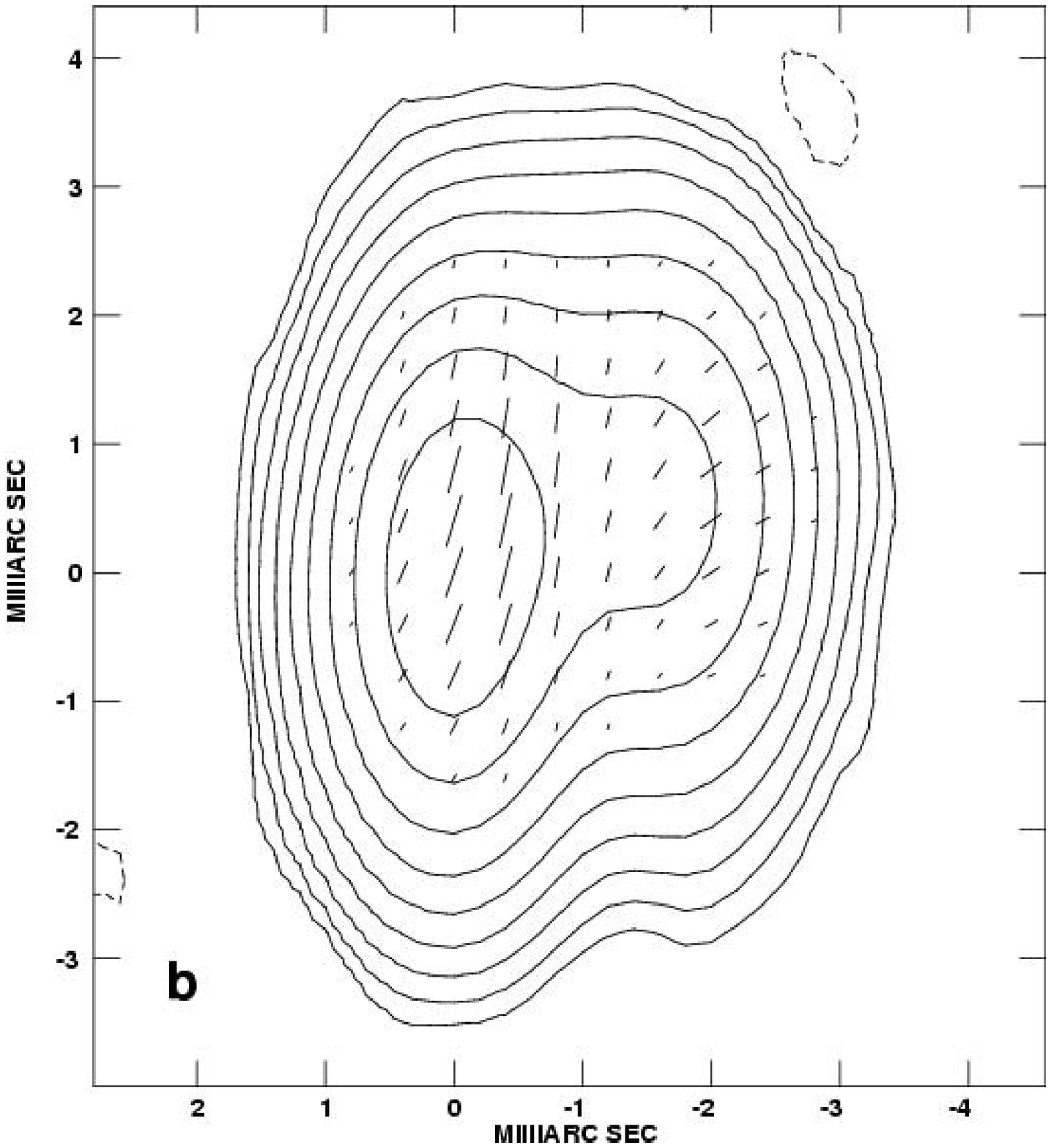}
\caption{a) Rotation measure image (color) for 1308+326 overlaid on 
Stokes I contours at 15 GHz. The insets show plots of EVPA $\chi$ 
(deg) versus $\lambda^2$ (cm$^2$). (b) Electric vectors (1 mas = 
50 mJy beam$^{-1}$ polarized flux density) corrected for Faraday Rotation 
overlaid on Stokes I contours. Contours start at 1.4 mJy beam$^{-1}$ and 
increase by factors of two.}
\label{1308rm}
\end{figure}
\clearpage

\begin{figure}
\vspace{19.2cm}
\includegraphics{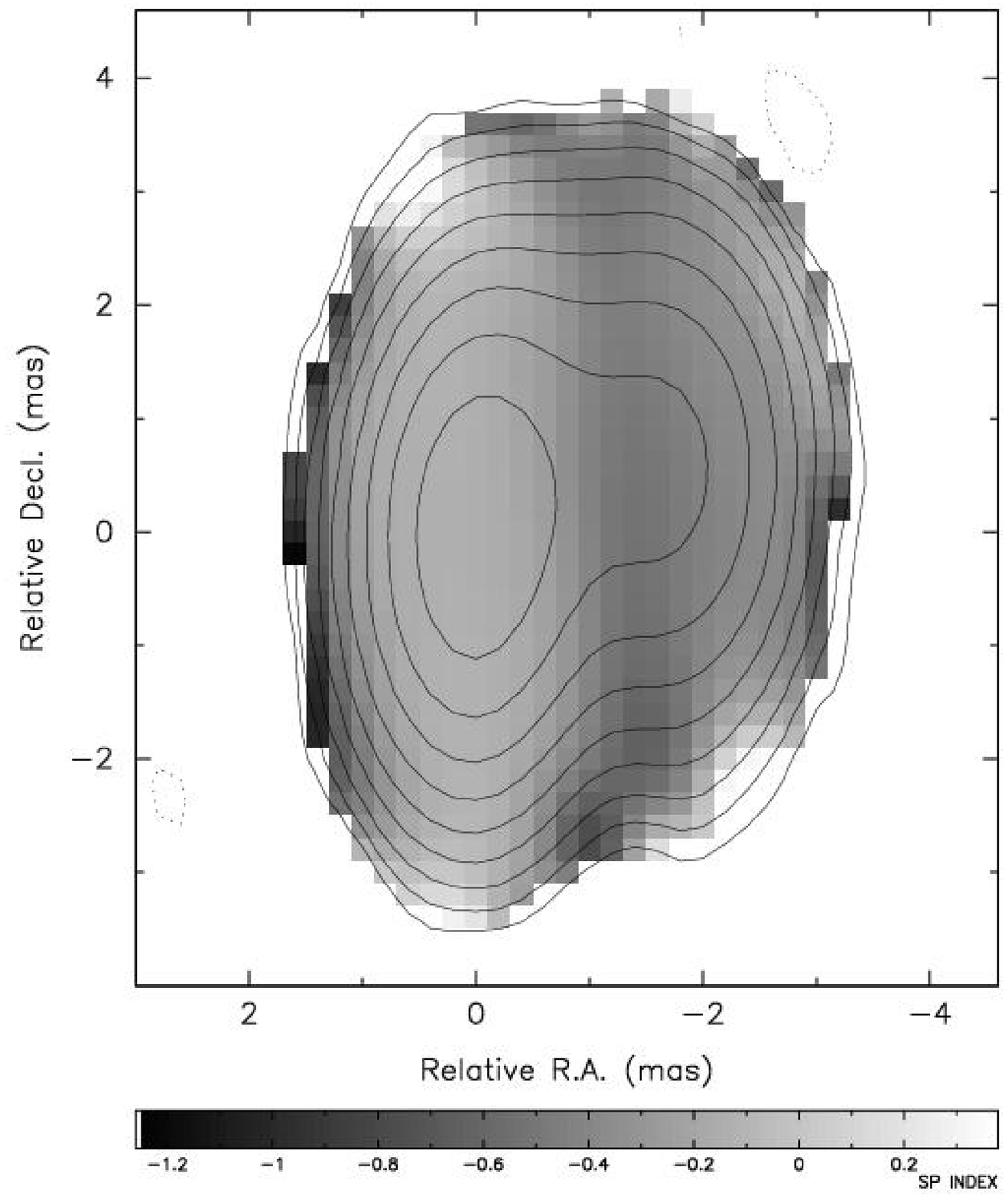}
\caption{Plot of spectral index $\alpha^{12.1}_{8.1}$ of the BL Lac 
object 1308+326 overlaid on 15 GHz Stokes I contours. Contours start 
at 1.4 mJy beam$^{-1}$ and increase by factors of two.}
\label{1308si}
\end{figure}
\clearpage

\begin{figure}
\vspace{19.2cm}
\includegraphics{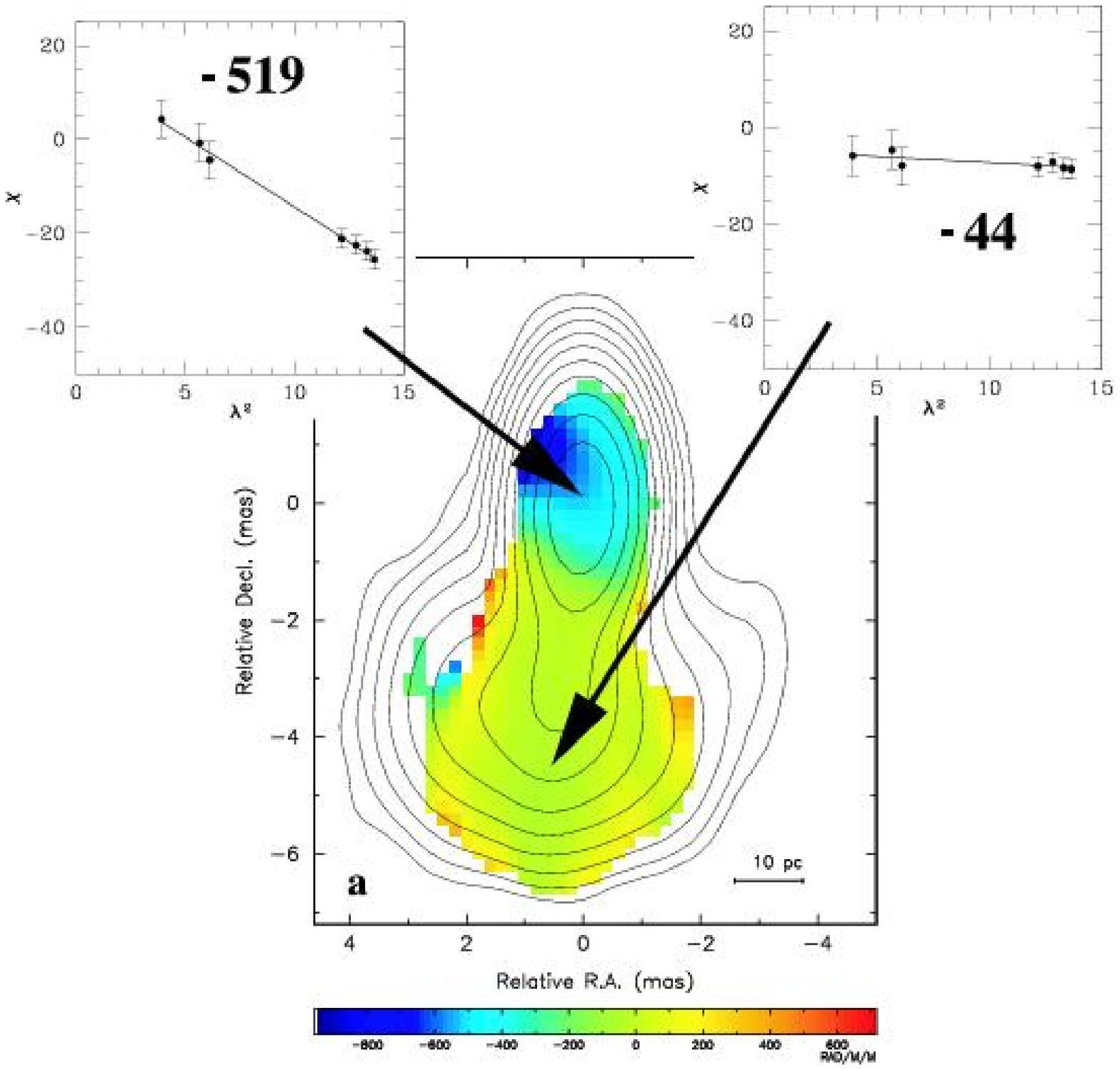}
\includegraphics{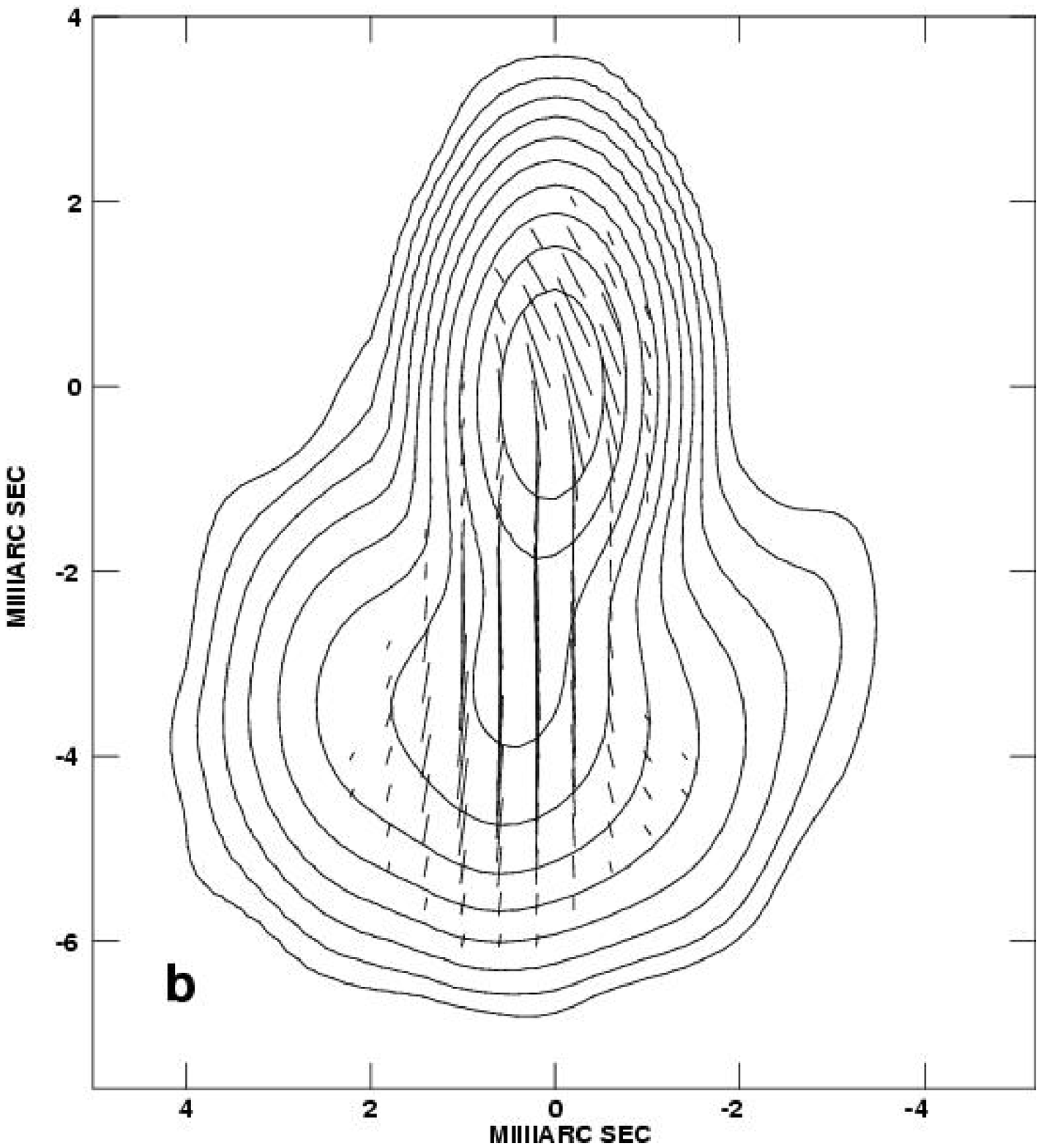}
\caption{(a) Rotation measure image (color) for 1611+343 overlaid on 
Stokes I contours at 15 GHz. The insets show plots of EVPA $\chi$ 
(deg) versus $\lambda^2$ (cm$^2$). (b) Electric vectors (1 mas = 
50 mJy beam$^{-1}$ polarized flux density) corrected for Faraday Rotation 
overlaid on Stokes I contours. Contours start at 2.7 mJy beam$^{-1}$ and 
increase by factors of two.}
\label{1611rm}
\end{figure}
\clearpage

\begin{figure}
\vspace{19.2cm}
\includegraphics{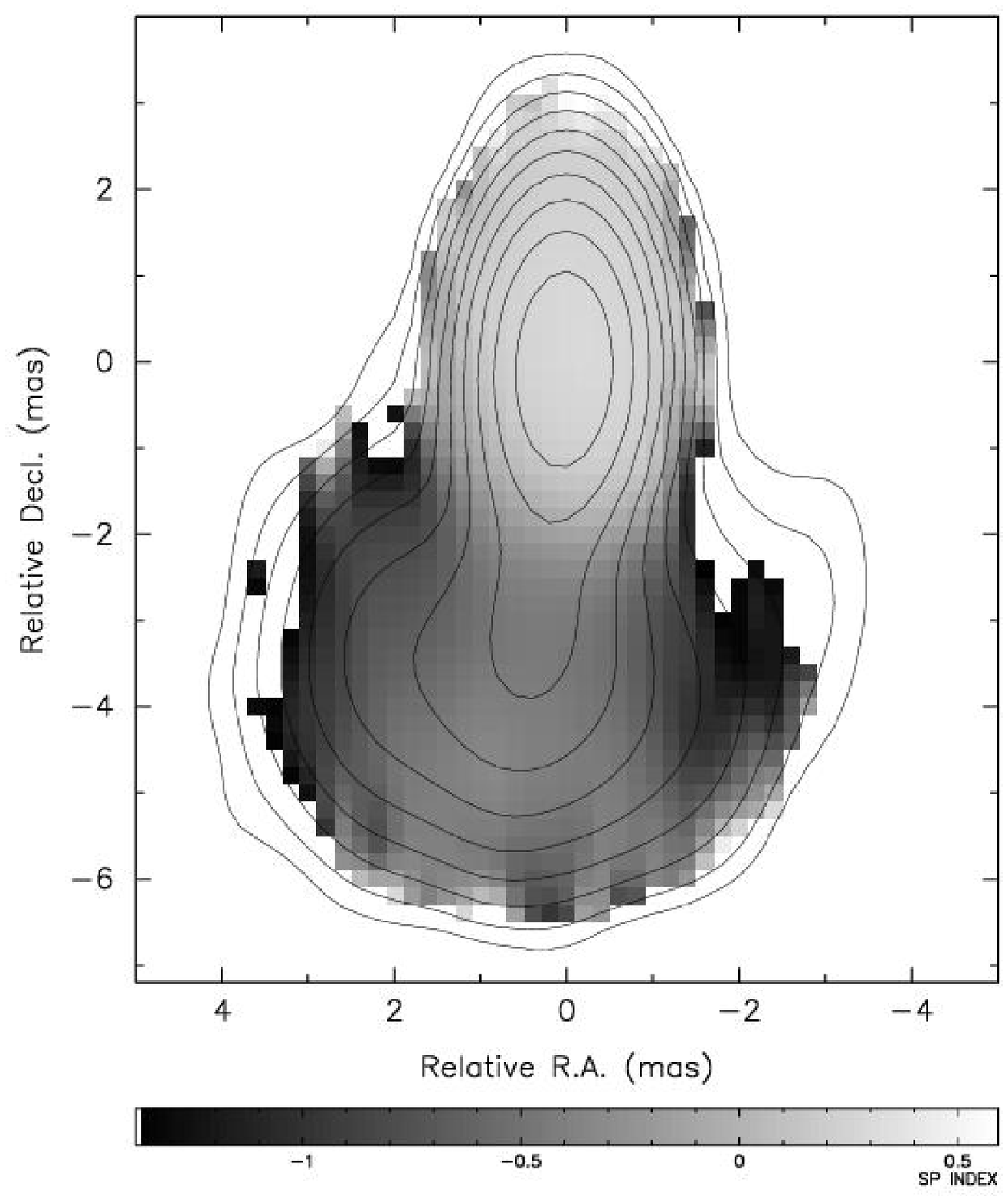}
\caption{Plot of spectral index $\alpha^{12.1}_{8.1}$ of the quasar 
1611+343 overlaid on 15 GHz Stokes I contours. Contours start 
at 2.7 mJy beam$^{-1}$ and increase by factors of two.}
\label{1611si}
\end{figure}
\clearpage

\begin{figure}
\vspace{19.2cm}
\includegraphics{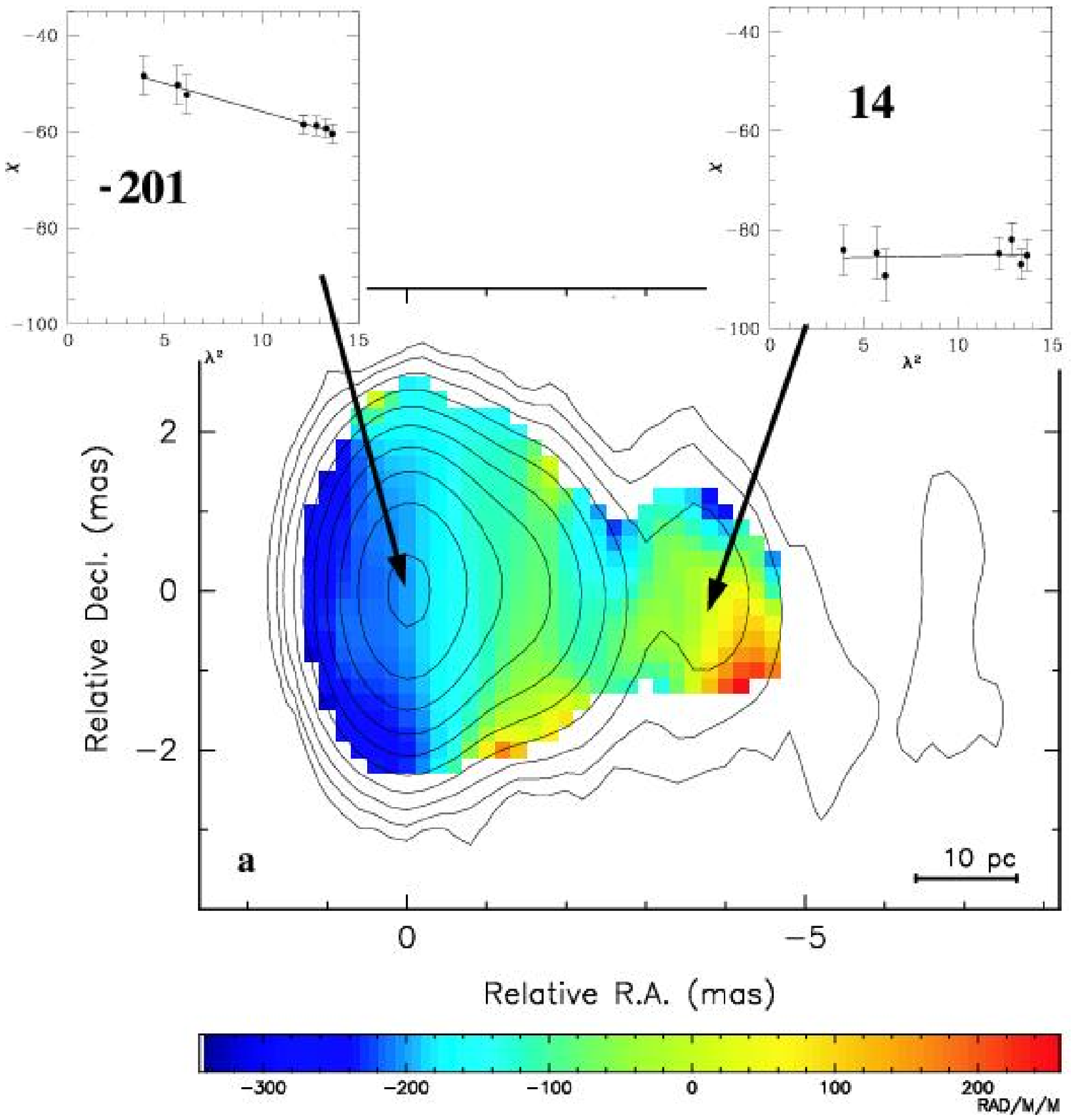}
\includegraphics{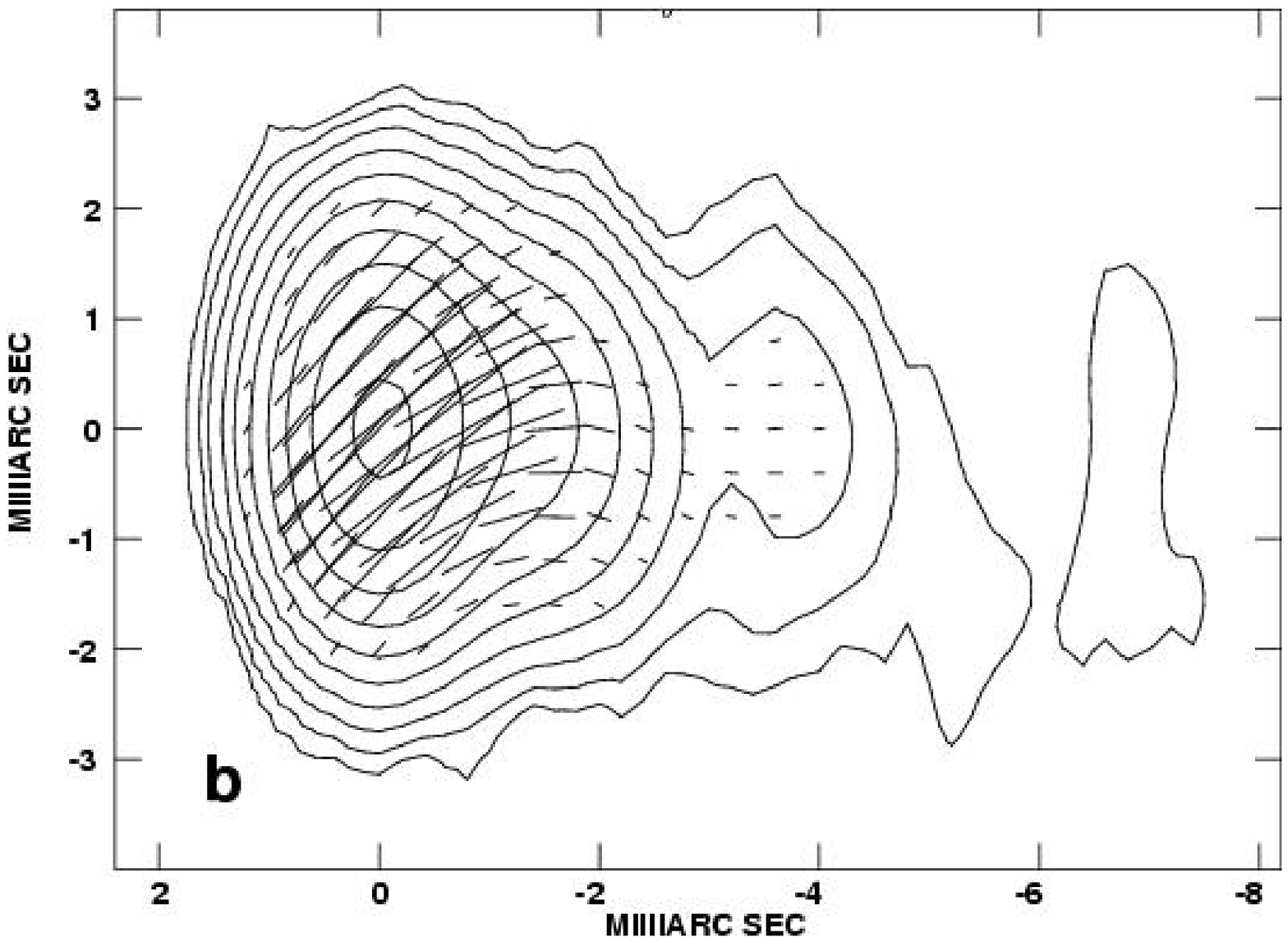}
\caption{a) Rotation measure image (color) for 1803+476 overlaid on 
Stokes I contours at 15 GHz. The insets show plots of EVPA $\chi$ 
(deg) versus $\lambda^2$ (cm$^2$). (b) Electric vectors (1 mas = 
33.3 mJy beam$^{-1}$ polarized flux density) corrected for Faraday Rotation 
overlaid on Stokes I contours. Contours start at 2.9 mJy beam$^{-1}$ and 
increase by factors of two.}
\label{1803rm}
\end{figure}
\clearpage

\begin{figure}
\vspace{19.2cm}
\includegraphics{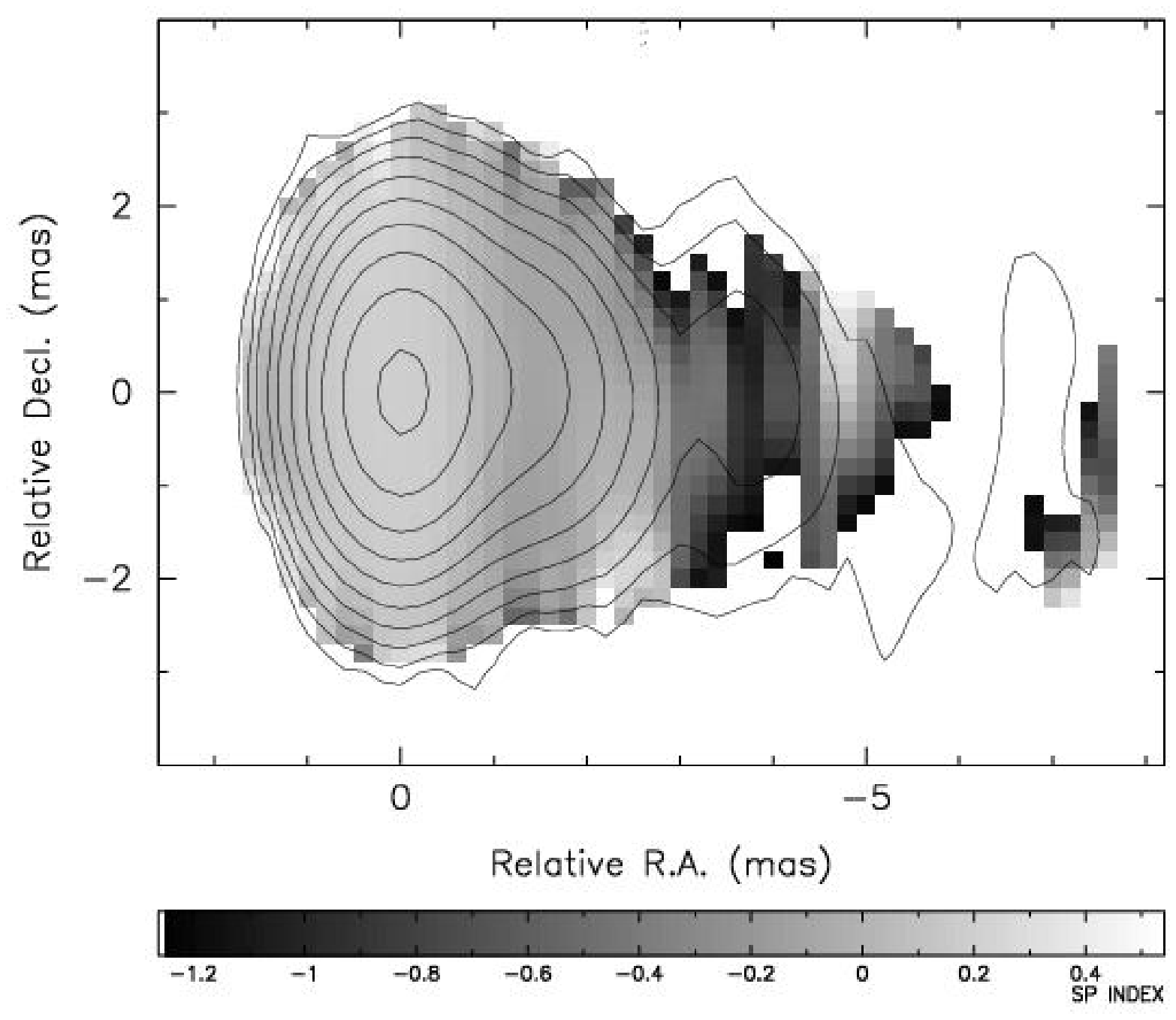}
\caption{Plot of spectral index $\alpha^{12.1}_{8.1}$ of the BL Lac 
object 1803+476 overlaid on 15 GHz Stokes I contours. Contours start 
at 2.9 mJy beam$^{-1}$ and increase by factors of two.}
\label{1803si}
\end{figure}
\clearpage

\begin{figure}
\vspace{19.2cm}
\includegraphics{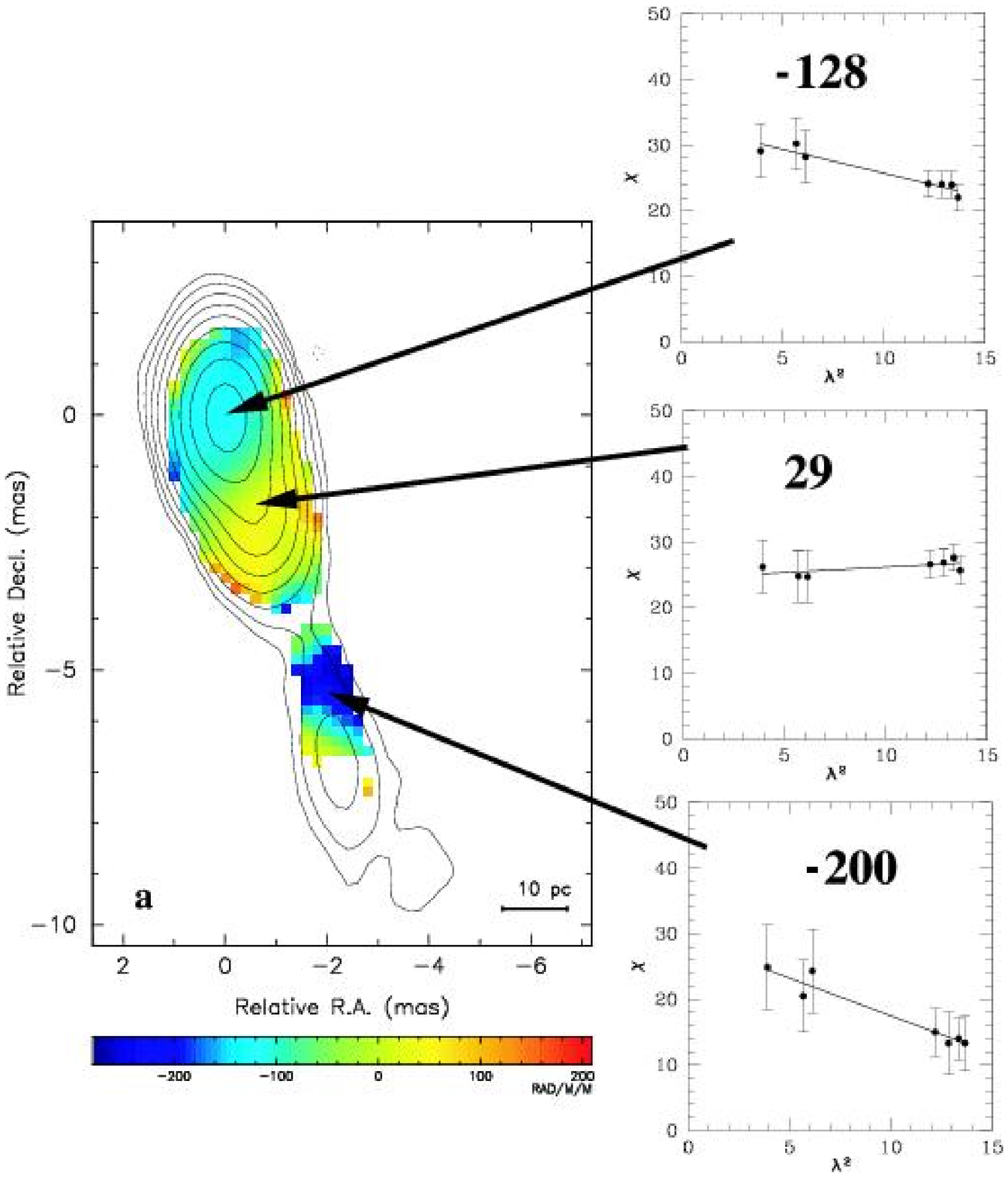}
\includegraphics{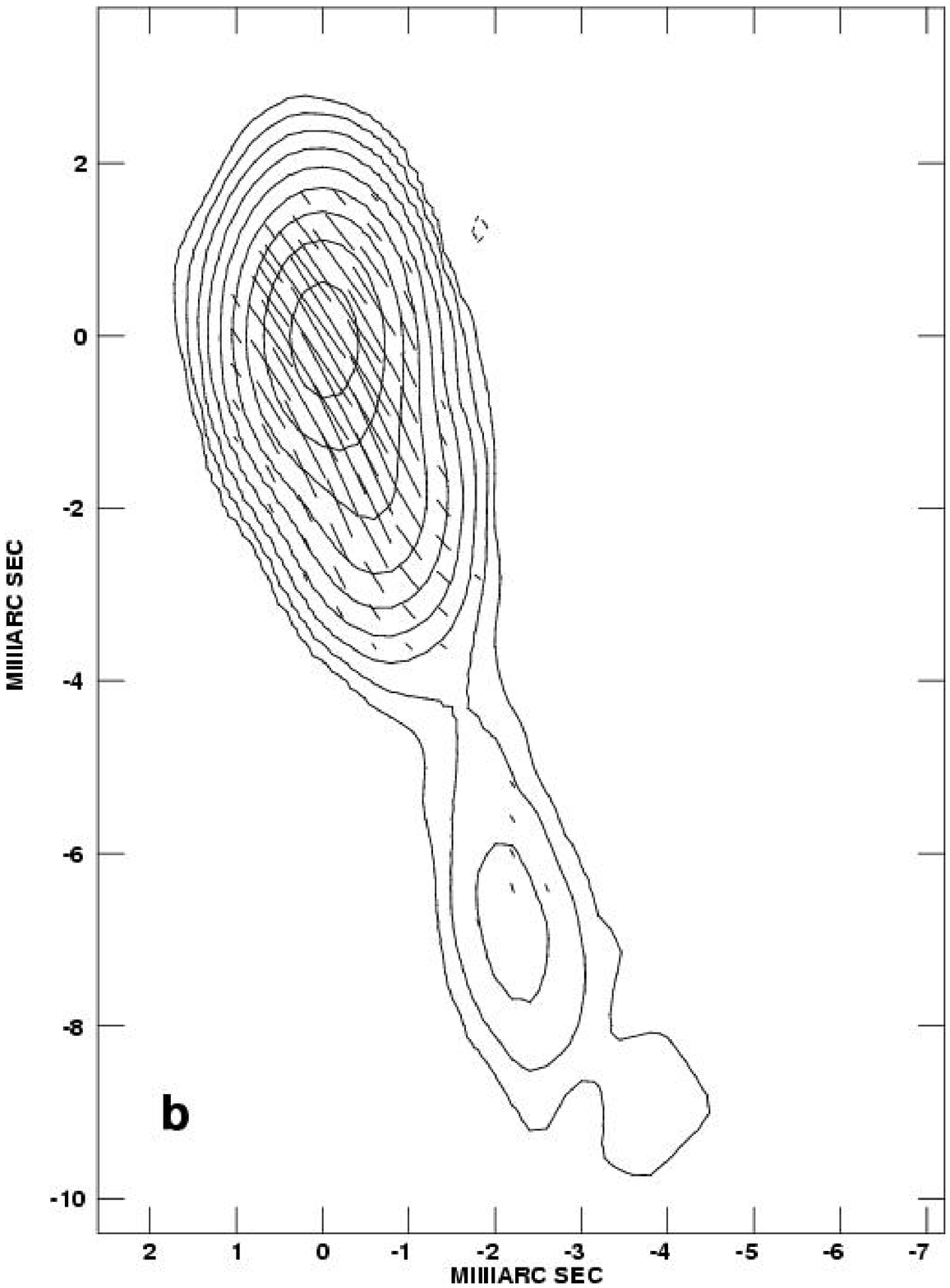}
\caption{(a) Rotation measure image (color) for 1823+568 overlaid on 
Stokes I contours at 15 GHz. The insets show plots of EVPA $\chi$ 
(deg) versus $\lambda^2$ (cm$^2$). (b) Electric vectors (1 mas = 
20 mJy beam$^{-1}$ polarized flux density) corrected for Faraday Rotation 
overlaid on Stokes I contours. Contours start at 1.8 mJy beam$^{-1}$ and 
increase by factors of two.}
\label{1823rm}
\end{figure}
\clearpage

\begin{figure}
\vspace{19.2cm}
\includegraphics{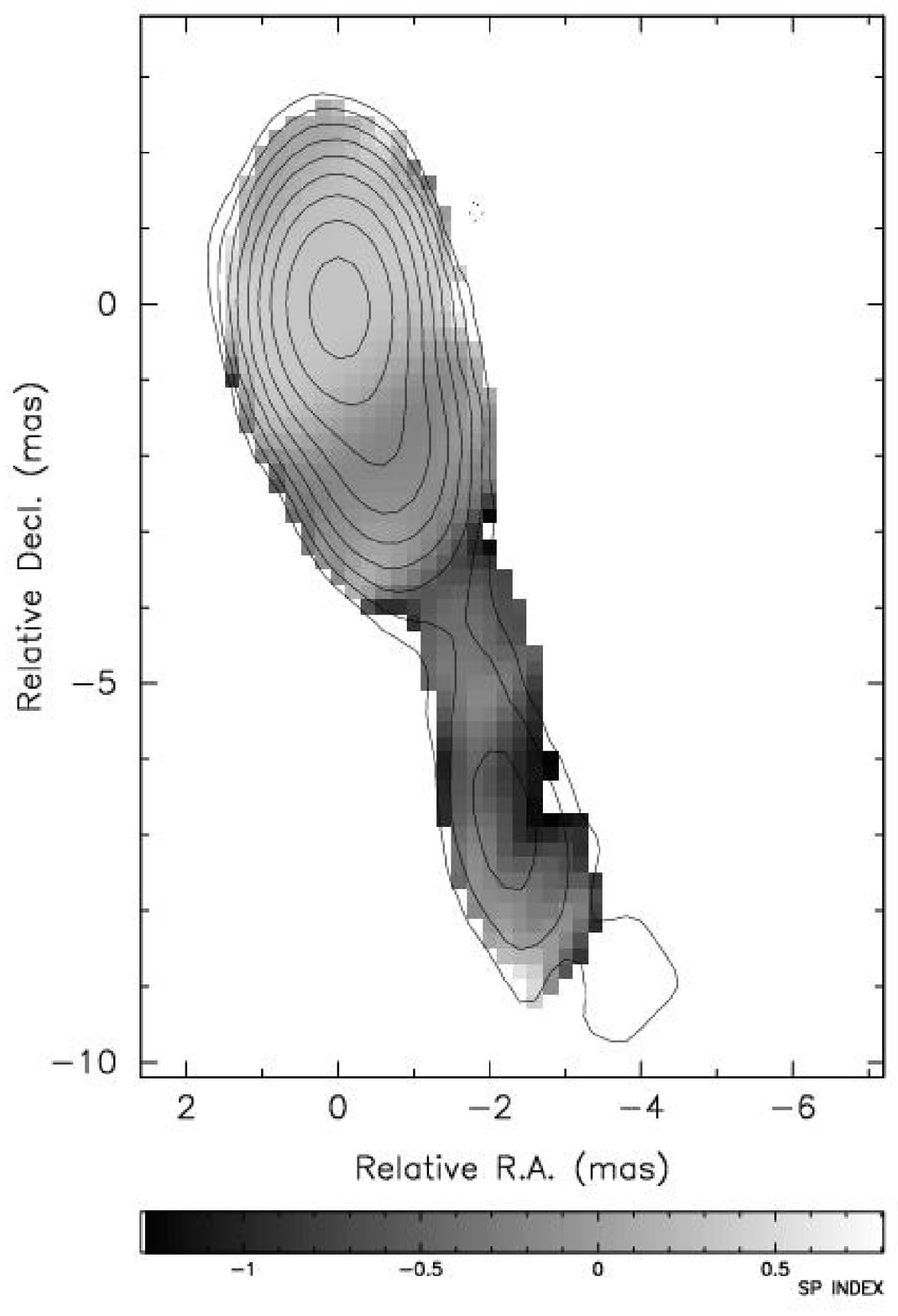}
\caption{Plot of spectral index $\alpha^{12.1}_{8.1}$ of the BL Lac 
object 1823+568 overlaid on 15 GHz Stokes I contours. Contours start 
at 1.8 mJy beam$^{-1}$ and increase by factors of two.}
\label{1823si}
\end{figure}
\clearpage

\begin{figure}
\vspace{19.2cm}
\includegraphics{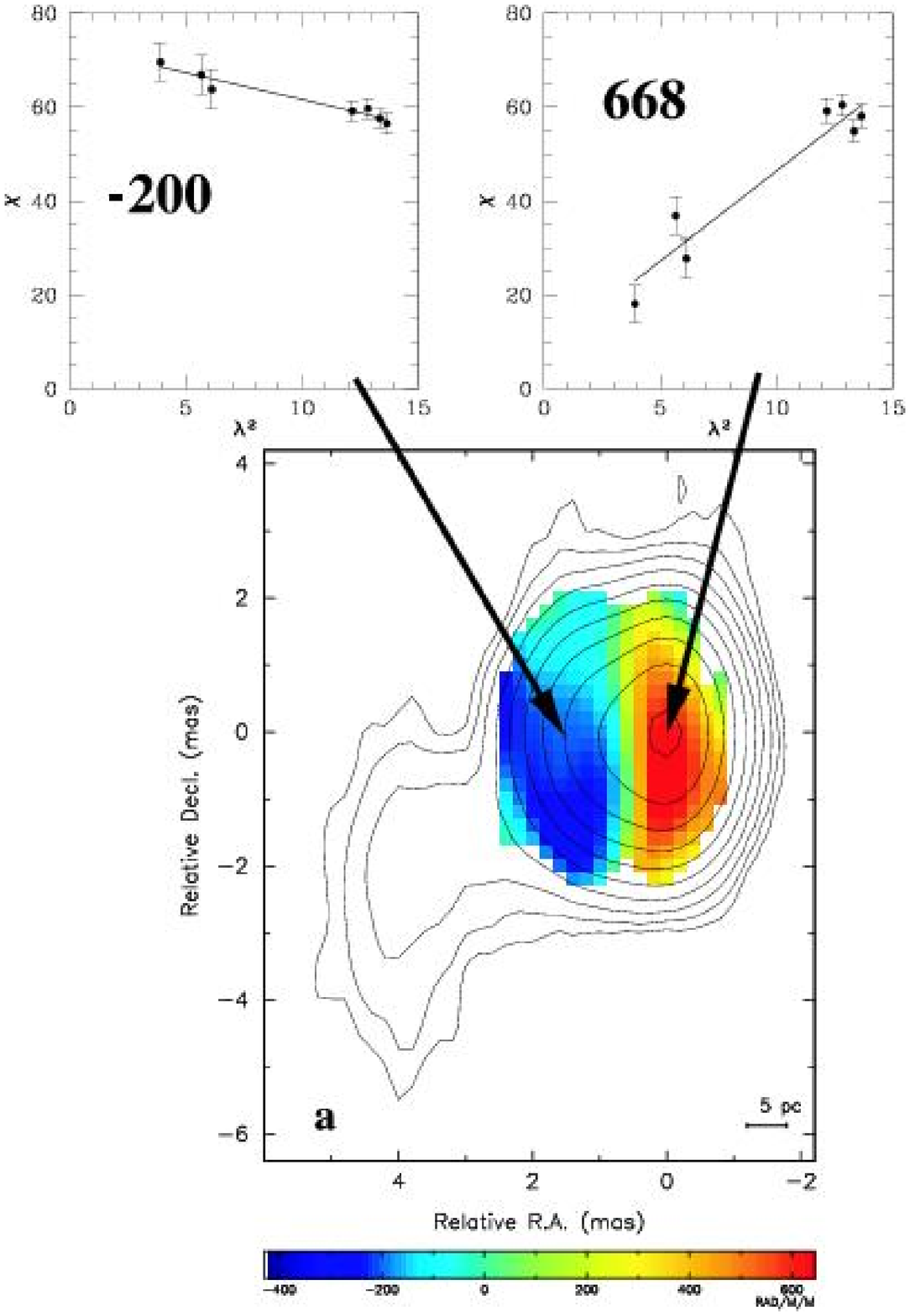}
\includegraphics{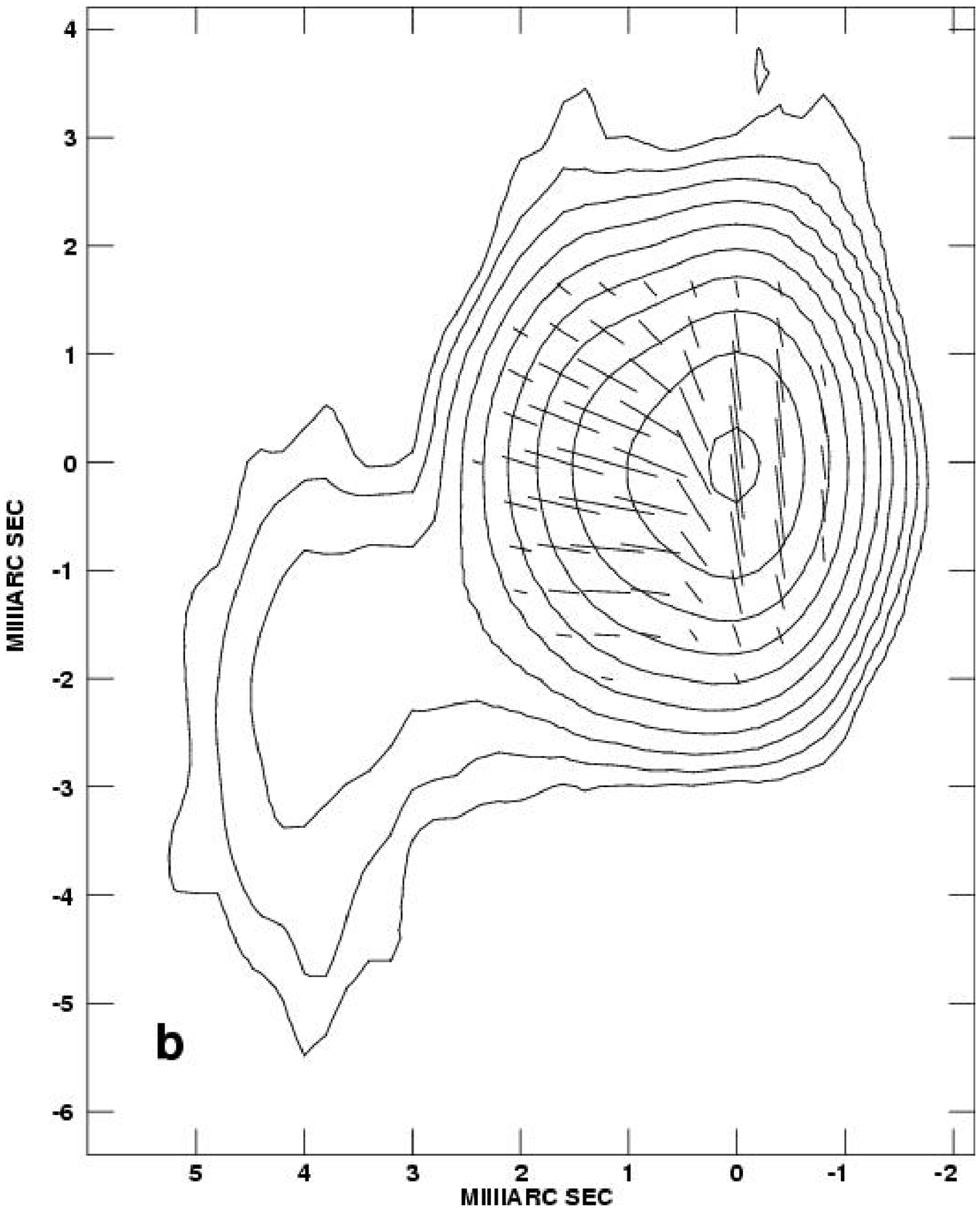}
\caption{(a) Rotation measure image (color) for 2005+403 overlaid on 
Stokes I contours at 15 GHz. The insets show plots of EVPA $\chi$ 
(deg) versus $\lambda^2$ (cm$^2$). (b) Electric vectors (1 mas = 
20 mJy beam$^{-1}$ polarized flux density) corrected for Faraday Rotation 
overlaid on Stokes I contours. Contours start at 2.5 mJy beam$^{-1}$ and 
increase by factors of two.}
\label{2005rm}
\end{figure}
\clearpage

\begin{figure}
\vspace{19.2cm}
\includegraphics{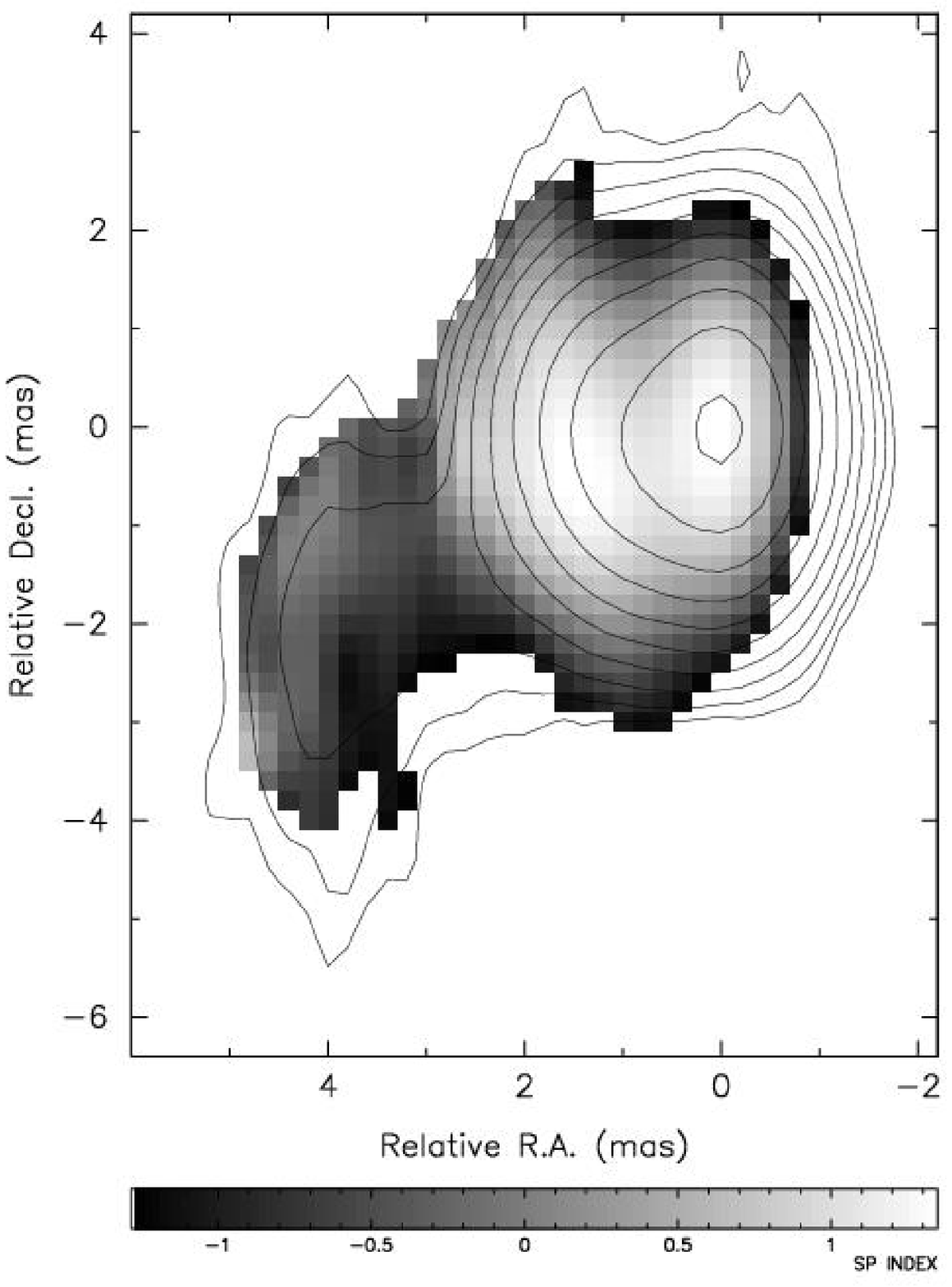}
\caption{Plot of spectral index $\alpha^{12.1}_{8.1}$ of the quasar 
2005+403 overlaid on 15 GHz Stokes I contours. Contours start 
at 2.5 mJy beam$^{-1}$ and increase by factors of two.}
\label{2005si}
\end{figure}
\clearpage

\begin{figure}
\vspace{19.2cm}
\includegraphics{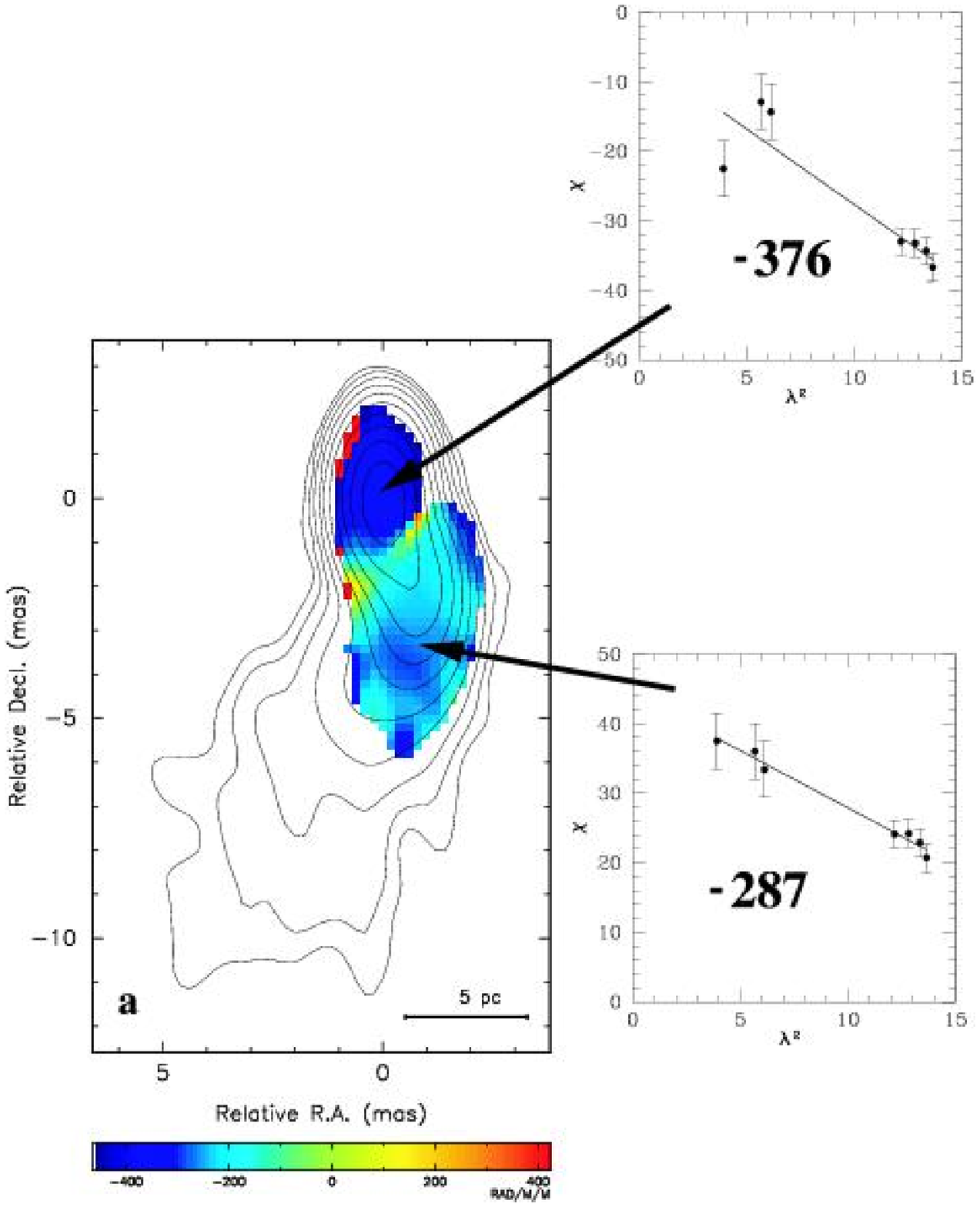}
\includegraphics{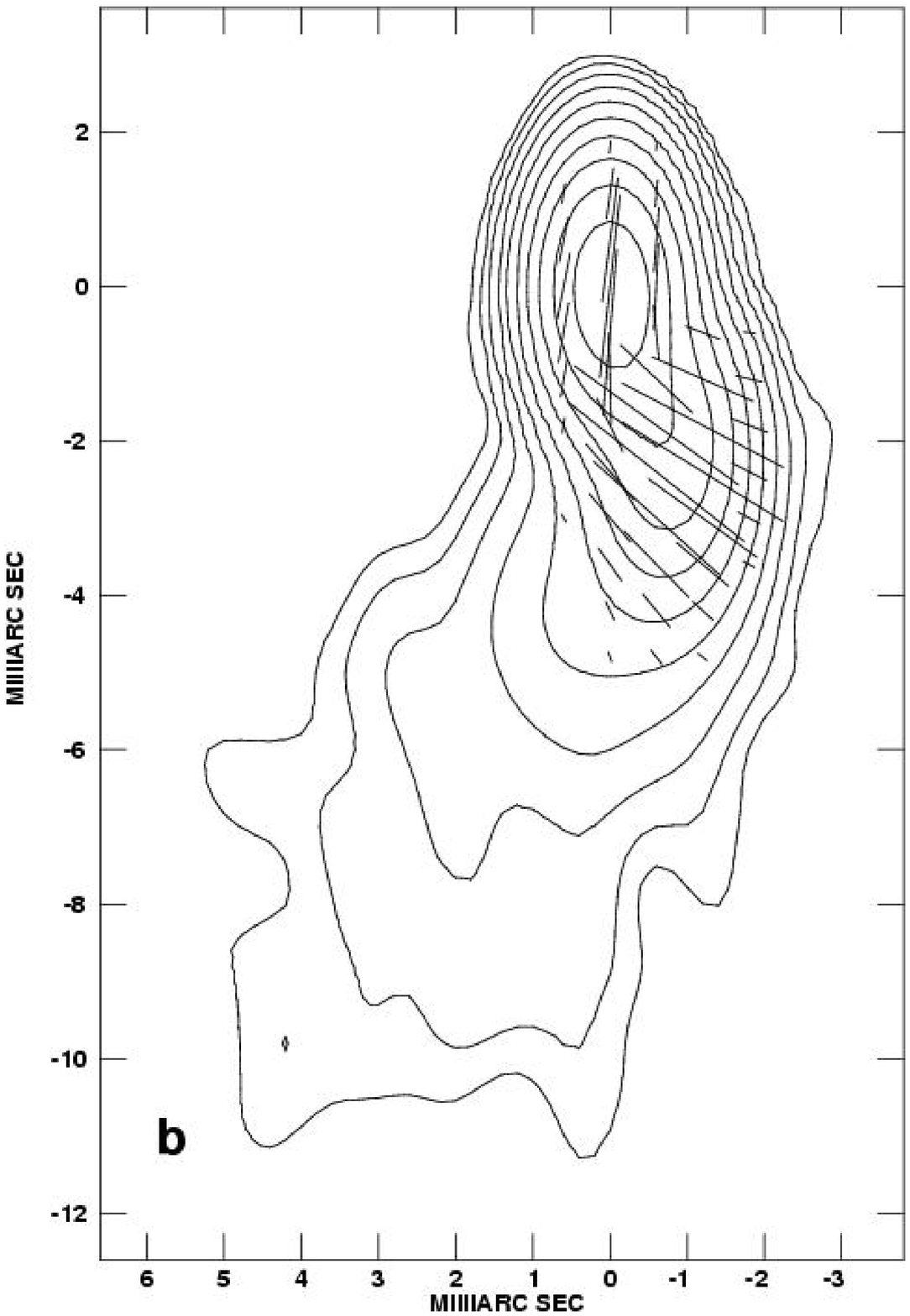}
\caption{a) Rotation measure image (color) for BL Lacertae overlaid on 
Stokes I contours at 15 GHz. The insets show plots of EVPA $\chi$ 
(deg) versus $\lambda^2$ (cm$^2$). (b) Electric vectors (1 mas = 
25 mJy beam$^{-1}$ polarized flux density) corrected for Faraday Rotation 
overlaid on Stokes I contours. Contours start at 2.2 mJy beam$^{-1}$ and 
increase by factors of two.}
\label{blrm}
\end{figure}
\clearpage

\begin{figure}
\vspace{19.2cm}
\includegraphics{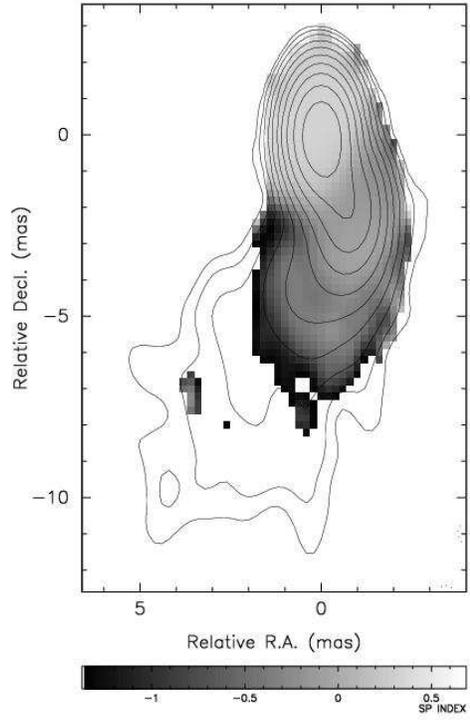}
\caption{Plot of spectral index $\alpha^{12.1}_{8.1}$ of BL Lacertae 
overlaid on 15 GHz Stokes I contours. Contours start 
at 2.2 mJy beam$^{-1}$ and increase by factors of two.}
\label{bllacsi}
\end{figure}
\clearpage

\begin{figure}
\vspace{19.2cm}
\includegraphics{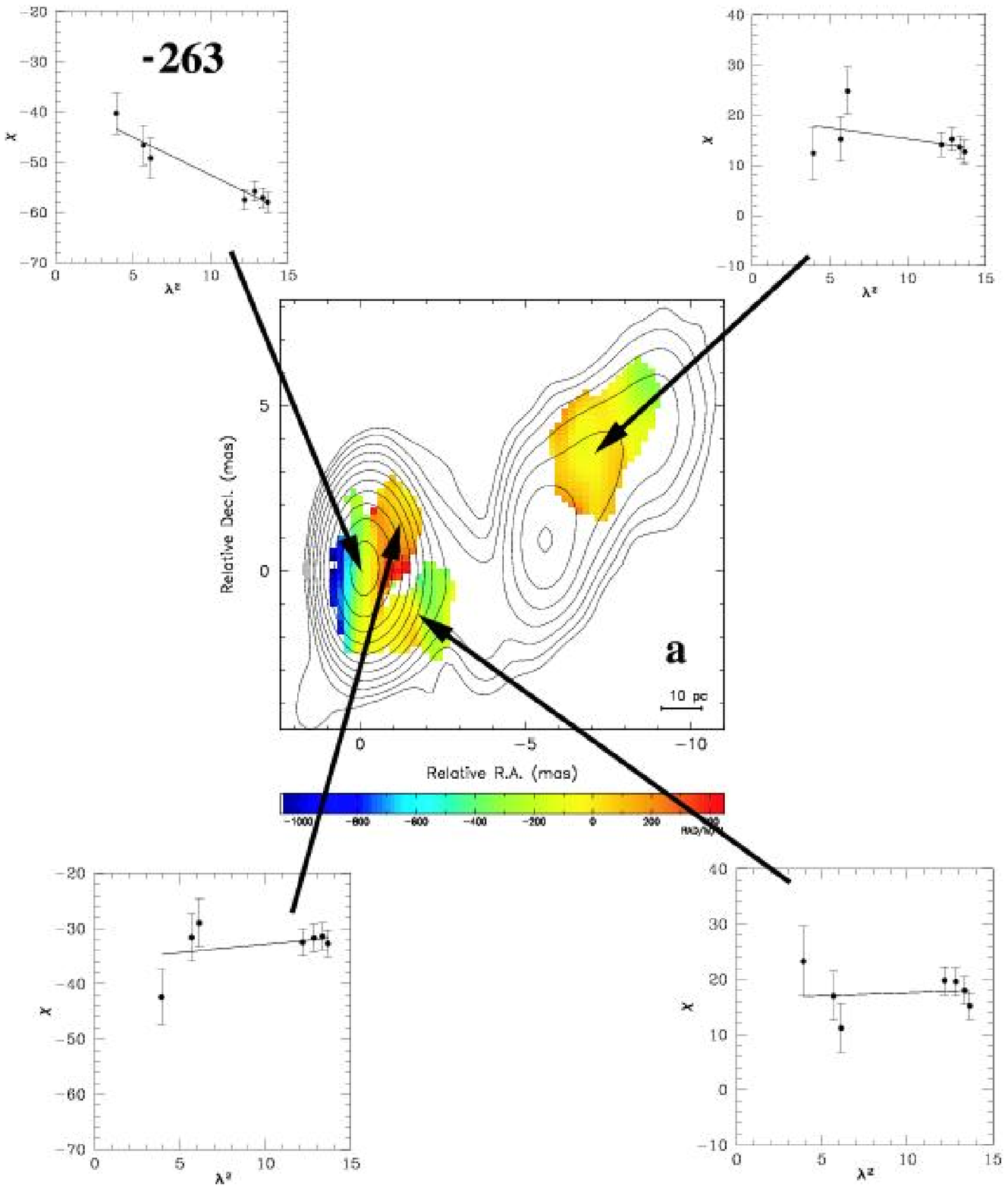}
\includegraphics{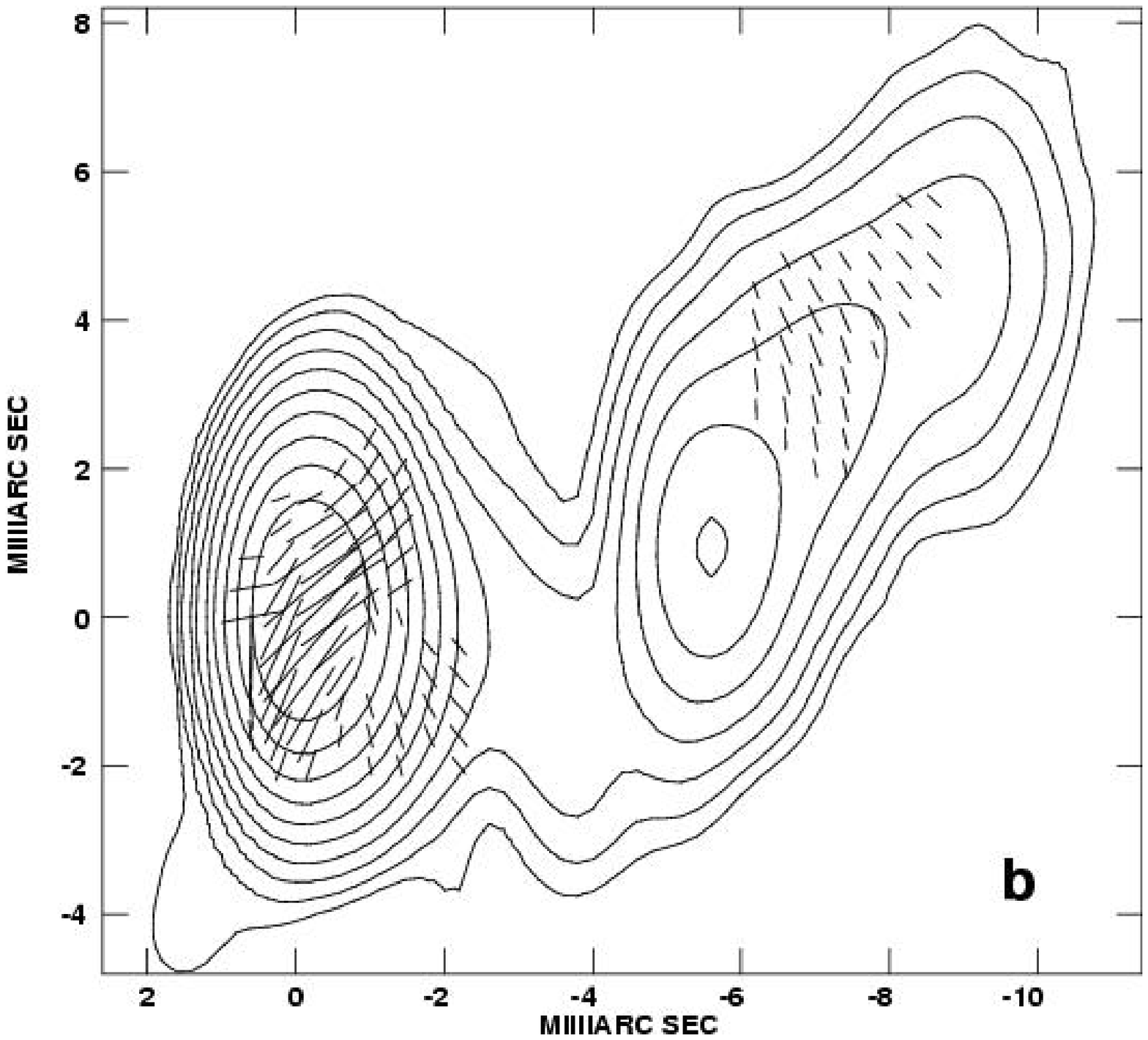}
\caption{(a) Rotation measure image (color) for 2251+158 overlaid on 
Stokes I contours at 15 GHz. The insets show plots of EVPA $\chi$ 
(deg) versus $\lambda^2$ (cm$^2$). (b) Electric vectors (1 mas = 
25 mJy beam$^{-1}$ polarized flux density) corrected for Faraday Rotation 
overlaid on Stokes I contours. Contours start at 4.0 mJy beam$^{-1}$ and 
increase by factors of two. One RM estimate only is placed
in the insets as that is the only reliable fit to a $\lambda^2$ 
law.}
\label{2251rm}
\end{figure}
\clearpage

\begin{figure}
\vspace{19.2cm}
\includegraphics{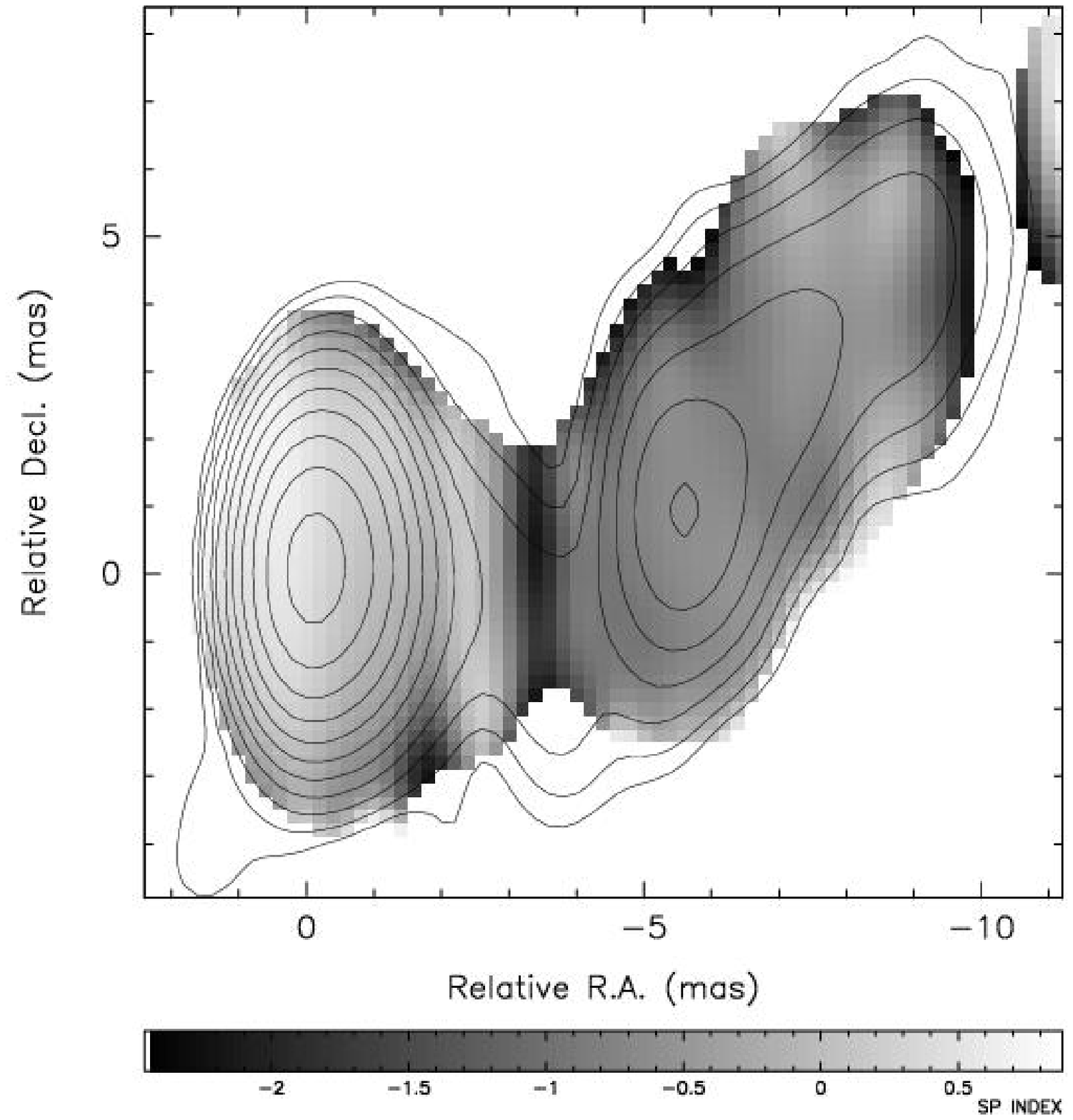}
\caption{Plot of spectral index $\alpha^{12.1}_{8.1}$ of the quasar 
2251+158 overlaid on 15 GHz Stokes I contours. Contours start 
at 4.0 mJy beam$^{-1}$ and increase by factors of two.}
\label{2251si}
\end{figure}
\clearpage


\begin{figure}
\vspace{19.2cm}
\includegraphics{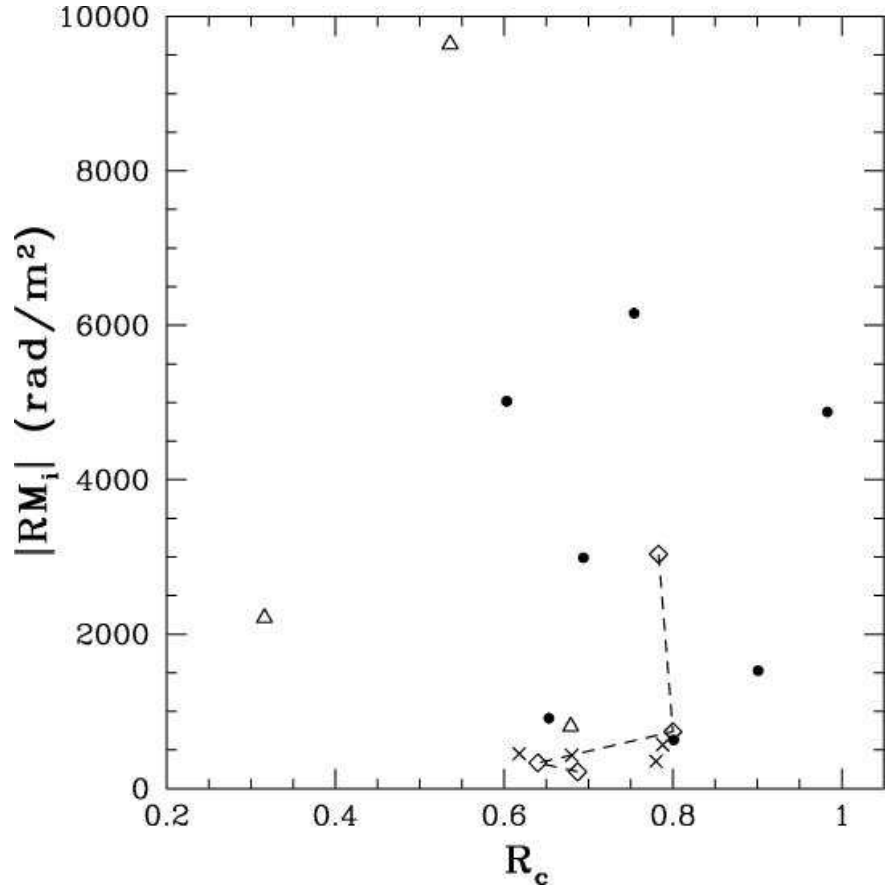}
\caption{Plot of maximum rotation measure (absolute value) in the rest 
frame versus core dominance at 15 GHz for the AGN presented in this paper. 
Dots are quasars, triangles are radio galaxies, and X's are BL Lac objects.
Multi-epoch data for 3C\,279 are diamonds connected by the dashed line.}
\label{rcrm}
\end{figure}
\clearpage

\begin{figure}
\vspace{19.2cm}
\includegraphics{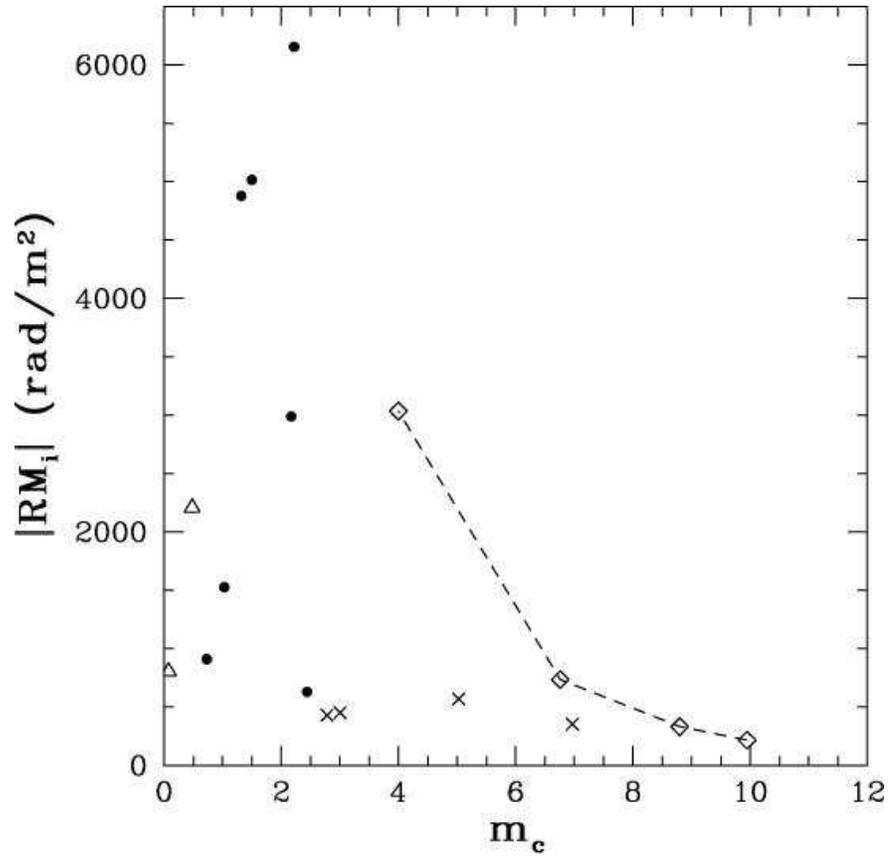}
\caption{Plot of the absolute value of the core
rotation measure in the rest frame versus 15 GHz core percent polarization  
for the sources listed in Table \ref{tbl-1}. Dots are quasars, triangles 
are radio galaxies, and X's are BL Lac objects. As in Fig.~\ref{rcrm} 
multi-epoch data for 3C\,279 are diamonds connected by a dashed line.}
\label{mcrm}
\end{figure}
\clearpage

\begin{figure}
\vspace{19.2cm}
\includegraphics{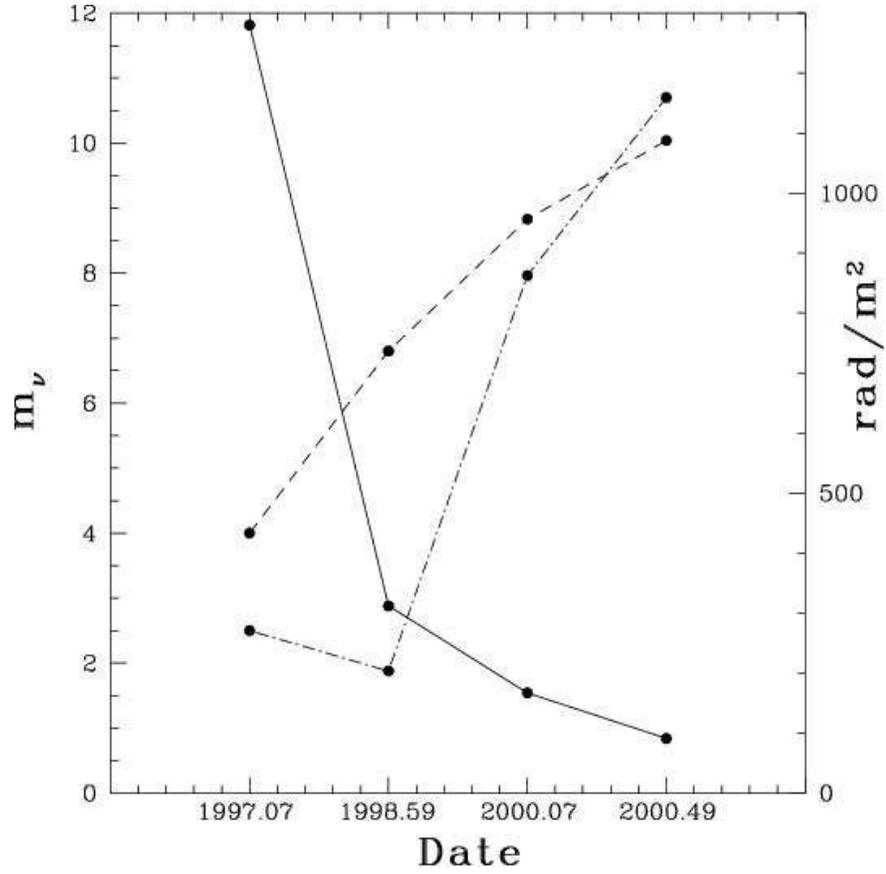}
\caption{A plot of core rotation measure (solid line), and percent 
polarization (dash-dot line = 8 GHz, dash line = 15 GHz) for the quasar 
3C\,279 over three years. 1997.07 data from \citet{tay98}, 
1998.59 data from \citet{tay00}, 2000.07 from \citet{zt01}, 
2000.49 this work. These epochs correspond to the diamonds 
shown in Fig.~\ref{rcrm}.}
\label{279time}
\end{figure}

\clearpage

\begin{figure}
\vspace{19.2cm}
\includegraphics{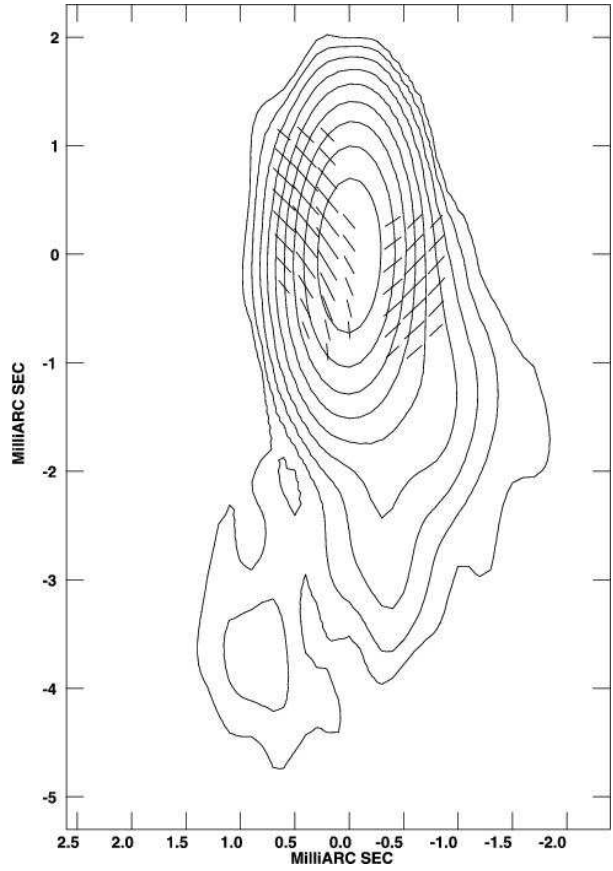}
\caption{15 GHz Stokes I contours of the quasar 0420$-$014 overlaid 
with E vectors (1 mas = 33.3 mJy beam$^{-1}$ uncorrected for 
Faraday rotation. This full-resolution image has a 
restoring beam 1.4$\times$ 0.6 mas at position angle -1.8\dg.}
\label{2pol}
\end{figure}
\clearpage

\begin{figure}
\vspace{19.2cm}
\includegraphics{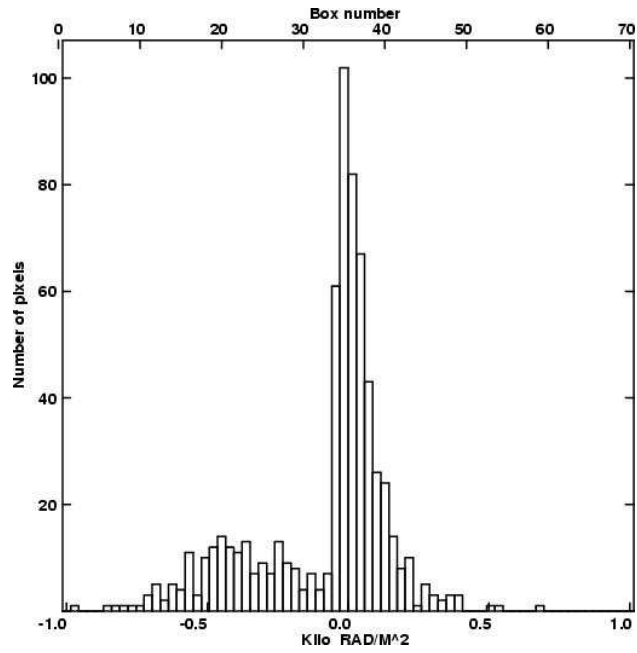}
\caption{Histogram of the rotation measure distribution 
for the quasar 1611+343. 70 boxes span a range of 
rotation measure of $\pm$ 1000 \radm.}
\label{1611dist}
\end{figure}
\clearpage

\end{document}